\documentclass[12pt,twoside,english]{extarticle}
\usepackage{lmodern}
\usepackage{helvet}
\usepackage[T1]{fontenc}
\usepackage[latin9]{inputenc}
\usepackage{color}
\usepackage{babel}
\usepackage{float}
\usepackage{mathrsfs}
\usepackage{varwidth}
\usepackage{amsmath}
\usepackage{amsthm}
\usepackage{amssymb}
\usepackage{graphicx}
\usepackage{geometry}
\usepackage{setspace}
\usepackage[pdfusetitle,
 bookmarks=true,bookmarksnumbered=false,bookmarksopen=false,
 breaklinks=true,pdfborder={0 0 0},pdfborderstyle={},backref=false,colorlinks=true]
 {hyperref}

\makeatletter

\providecommand{\tabularnewline}{\\}
\newenvironment{cellvarwidth}[1][t]
    {\begin{varwidth}[#1]{\linewidth}}
    {\@finalstrut\@arstrutbox\end{varwidth}}

\numberwithin{equation}{section}
\numberwithin{table}{section}
\numberwithin{figure}{section}
\usepackage[natbibapa]{apacite}
\theoremstyle{definition}
\newtheorem{defn}{\protect\definitionname}[section]
\theoremstyle{definition}
\newtheorem{example}{\protect\examplename}[section]
\theoremstyle{plain}
\newtheorem{assumption}{\protect\assumptionname}
\theoremstyle{plain}
\newtheorem{prop}{\protect\propositionname}[section]
\theoremstyle{plain}
\newtheorem{thm}{\protect\theoremname}[section]
\theoremstyle{plain}
\newtheorem{cor}{\protect\corollaryname}[section]
\newenvironment{lyxlist}[1]
	{\begin{list}{}
		{\settowidth{\labelwidth}{#1}
		 \setlength{\leftmargin}{\labelwidth}
		 \addtolength{\leftmargin}{\labelsep}
		 }}
	{\end{list}}
\theoremstyle{plain}
\newtheorem{lem}{\protect\lemmaname}[section]

\usepackage{graphicx}
\usepackage{amsmath}
\usepackage{dcolumn}
\usepackage{listings}
\usepackage{color}
\usepackage{setspace}
\usepackage[normalem]{ulem}  
\definecolor{hellgelb}{rgb}{1,1,0.8}
\definecolor{colKeys}{rgb}{0,0,1}
\definecolor{colIdentifier}{rgb}{0,0,0}
\definecolor{colComments}{rgb}{1,0,0}
\definecolor{colString}{rgb}{0,0.5,0}

\usepackage{lmodern}
\usepackage[T1]{fontenc}
\usepackage{geometry}
\usepackage{fancyhdr}
\usepackage{amsthm}
\usepackage{thmtools}
\usepackage{amsmath}
\usepackage{amssymb}
\usepackage{esint}
\usepackage{multirow}
\usepackage{mathrsfs} 
\usepackage{textgreek}
\usepackage{amsfonts}
\usepackage{amssymb}
\usepackage{mathrsfs} 

\usepackage[USenglish]{isodate}
\usepackage{chngcntr}

\numberwithin{equation}{section}
\numberwithin{table}{section}
\numberwithin{assumption}{section}
\counterwithout{figure}{section}
\counterwithout{table}{section}

\makeatother

  \providecommand{\assumptionname}{Assumption}

  \providecommand{\corollaryname}{Corollary}
  
  \providecommand{\definitionname}{Definition}
  \providecommand{\examplename}{Example}

  \providecommand{\lemmaname}{Lemma}

  \providecommand{\propositionname}{Proposition}

  \providecommand{\theoremname}{Theorem}
 \providecommand{\corollaryname}{Corollary}
 \providecommand{\theoremname}{Theorem}
 
\usepackage{thmtools}

\newtheoremstyle{MyTheoremstyle}
  {\topsep} 
  {\topsep} 
  {} 
  {} 
  {\bfseries} 
  {.} 
  {.90em} 
  {} 
\theoremstyle{MyTheoremstyle} 
\theoremstyle{MyTheoremstyle} 
\theoremstyle{MyTheoremstyle} 
\theoremstyle{MyTheoremstyle} 
\theoremstyle{MyTheoremstyle}

\declaretheoremstyle[
    headfont=\bfseries,
    notefont=\normalfont,
    bodyfont=\itshape,
    headpunct=\newline,
    headformat={%
        \makebox{\NAME\ \NUMBER\ }{\NOTE}%
    },
]{theorem}

\newlength{\spacelength}
\declaretheoremstyle[
    headfont=\bfseries,
    notefont=\normalfont,
    bodyfont=\itshape,
    headpunct=\newline,
    headformat={%
        \makebox[0pt][l]{\NAME\ \NUMBER\ }\hskip-\spacelength{\NOTE}%
    },
]{theore}

 \usepackage{fancyhdr}
\usepackage{booktabs}
\usepackage{float}
\usepackage{graphicx}
\usepackage{epstopdf}
\usepackage{morefloats}
\usepackage[referable]{threeparttablex}
\usepackage{footnote}

\usepackage{setspace} 
\usepackage{graphicx,color}

\usepackage{listings}
\RequirePackage[T1]{fontenc}
\RequirePackage{ae,fancyvrb}
\DefineVerbatimEnvironment{Sinput}{Verbatim}{fontshape=sl}
\DefineVerbatimEnvironment{Soutput}{Verbatim}{}
\DefineVerbatimEnvironment{Scode}{Verbatim}{fontshape=sl}

\author{
\textsc{\textcolor{MyBlue}{Alessandro Casini}}
\\
\small{\text{University of Rome Tor Vergata}}
\and
\textsc{\textcolor{MyBlue}{Adam McCloskey}}
\\
\small{\text{University of Colorado at Boulder}}
\and
\textsc{\textcolor{MyBlue}{Luca Rolla}}
\\
\small{\text{University of Rome Tor Vergata}}
\and
\textsc{\textcolor{MyBlue}{Raimondo Pala}}
\\
\small{\text{University of Rome Tor Vergata}}
}

\date{\small{\today} 
}

 \makeatletter

\numberwithin{equation}{section}
\makeatother

\usepackage{babel}
\addto\captionsenglish{%
}

\usepackage{color}
\usepackage{xcolor}

\renewcommand*{\thesection}{\arabic{section}}

\definecolor{MyRed}{rgb}{0.8,0,0}
\definecolor{MyBlue}{rgb}{0,0,0.7}
\definecolor{Green}{rgb}{0,0.5,0}
\definecolor{hellgelb}{rgb}{1,1,0.8}
\definecolor{colKeys}{rgb}{0,0,1}
\definecolor{colIdentifier}{rgb}{0,0,0}
\definecolor{colComments}{rgb}{1,0,0}
\definecolor{colString}{rgb}{0,0.5,0}

\definecolor{MyLightRed}{rgb}{2.2,0.2,0.4} 
\definecolor{MyLightRed2}{rgb}{0.6,0.2,0.3} 
\definecolor{MyLightRed2temp}{rgb}{0.6,0.2,0.3}
\definecolor{MyLightRed3}{rgb}{0.8,0.1,0.1} 
\definecolor{MyRed}{rgb}{0.7,0.0,0}

\definecolor{MyLigthBlue13}{rgb}{0,0.2,0.7}
 \definecolor{MyLigthBlack}{rgb}{0.2,0.25,0.3} 

\hypersetup{%
  bookmarks=true
  pdftitle = {Title Here},
  pdfsubject = {},
  pdfkeywords = {Keywords},
  pdfauthor = {Alessandro Casini},}
\hypersetup{ 
  colorlinks = {true}, 
  citecolor = {MyBlue},
  linkcolor = {MyRed},
  filecolor= {Green},	
  urlcolor = {MyBlue},
  hyperindex = {true},
  linktocpage = {true},
  linkbordercolor ={1 0 0},
 citebordercolor ={0 1 1},
 urlbordercolor ={0 1 1},
hypertexnames=false
}

\usepackage{bookmark}

\usepackage[bottom]{footmisc}
\usepackage{multibib}
\newcites{ReferencesSupp}{References}

\makeatother

\providecommand{\assumptionname}{Assumption}
\providecommand{\corollaryname}{Corollary}
\providecommand{\definitionname}{Definition}
\providecommand{\examplename}{Example}
\providecommand{\lemmaname}{Lemma}
\providecommand{\propositionname}{Proposition}
\providecommand{\theoremname}{Theorem}

\begin{document}
\pagebreak{}

\setcounter{page}{0}

\raggedbottom
\title{{\Large\textbf{Dynamic Local Average Treatment Effects in Time Series}}\textbf{\thanks{{\scriptsize We thank Isaiah Andrews, Josh Angrist, Jushan Bai, Elisa
Facchetti, Jean-Jacques Forneron, Eva Janssen, Hiro Kaido, Toru Kitagawa,
Sokbae (Simon) Lee, Daniel Lewis, Emi Nakamura, Serena Ng, Geert Mesters,
Pepe Montiel Olea, Aureo de Paula, Andrew Patton, Mounu Prem, Zhongjun
Qu, Bernard Salanié,  J\'{o}n Steinsson, Jim Stock, Stephen Terry,
Alex Torgovitsky and Kaspar W\"{u}thrich for useful comments. We
thank seminar participants at Bristol, BU, Cambridge, Columbia, Michigan,
Michigan State, Oxford and UCL and conference participants at the
NBER Summer Institute 2026. Casini acknowledges financial support
from EIEF through 2022 EIEF Grant and from the Italian Ministry of
Education and Research through FIS-2024-02854 grant. McCloskey acknowledges
support from the National Science Foundation under Grant SES-2341730.
The codes are available at \protect\href{https://github.com/alessandro-casini/DynamicLATE_ts}{https://github.com/alessandro-casini/DynamicLATE\_ts}.}
 }}}
\maketitle
\begin{abstract}
{\footnotesize This paper discusses identification, estimation, and
inference on dynamic local average treatment effects (LATEs) in instrumental
variables (IVs) settings. First, we show that compliers\textemdash observations
whose treatment status is affected by the instrument\textemdash can
be identified }{\footnotesize\emph{individually}}{\footnotesize{}
in time series data using smoothness assumptions and local comparisons
of treatment assignments. Second, we show that this result enables
not only better interpretability of IV estimates but also direct testing
of the exclusion restriction by comparing outcomes among identified
non-compliers across instrument values. Third, we document pervasive
weak identification in applied work using IVs with time series data
by surveying recent publications in leading economics journals. However,
we find that strong identification often holds in large subsamples
for which the instrument induces changes in the treatment. Motivated
by this, we introduce a method based on dynamic programming to detect
the most strongly-identified subsample and show how to use this subsample
to improve estimation and inference. We also develop new identification-robust
inference procedures that focus on the most strongly-identified subsample,
offering efficiency gains relative to existing full sample identification-robust
inference when identification fails over parts of the sample. Finally,
we apply our results to heteroskedasticity-based identification of
monetary policy effects. We find that about 60\% of observations are
compliers (i.e., cases where the variance of the policy shifts up
on FOMC announcement days), and we fail to reject the exclusion restriction.
Estimation using the most strongly-identified subsample helps reconcile
conflicting IV and GMM estimates in the literature. }{\footnotesize\par}
\end{abstract}
 {\footnotesize{\indent {\bf{JEL Classification}}: B41, C12, C32.}} \\  

\onehalfspacing
\thispagestyle{empty}

\pagebreak{}

\section{Introduction}

Economists work hard to extract plausibly exogenous variation in order
to identify causal effects. Many identification strategies used in
applied work either rely directly on instrumental variables (IVs)
or can be reframed in terms of IV identification. This holds also
in dynamic settings where, for example, external IVs may be constructed
using a narrative approach or heteroskedasticity is exploited to yield
additional identifying equations. Since \citet{imbens/angrist:1994},
it has been well-known that IV-based approaches identify the local
average treatment effect (LATE)\textemdash the average treatment effect
for the sub-population of compliers, i.e., those whose treatment status
is influenced by the policy intervention (the instrument). 

This paper makes four main contributions. First, we show that compliers
can be identified at the individual (time-period) level. Second, we
demonstrate that the exclusion restriction can be tested. Third, we
introduce a dynamic programming method to detect the most-strongly
identified subsample when instrument relevance is time-varying, and
we show how this subsample can be used to improve estimation. Fourth,
we develop identification-robust inference procedures based on the
strongly-identified subsample that are more efficient than existing
full sample procedures when identification fails over parts of the
sample. We discuss each contribution in turn.

In the LATE framework, the sub-population of compliers is unobserved.
This means that although a LATE can be identified, the specific sample
observations this effect represents is unknown. This limitation is
often described informally as the inability to observe an observation's
treatment status under both the intervention and non-intervention
scenarios. From a practical interpretability perspective, this presents
a challenge that has been widely discussed in the literature {[}see,
e.g., \citet{angrist/imbens/rubin:1996}, \citet{heckman:1996}, \citet{heckman/vytlacil:2005}
and \citet{imbens:2010}{]}. Some progress has been made by \citet{imbens/rubin:1997}
and \citet{abadie:2003} who show that the proportion of compliers
and some of their statistical characteristics can be identified provided
the treatment is binary.

We can list several reasons why identifying the compliers individually
is important for both policy analysis and empirical applications in
time series settings. A lack of individual complier identification
prevents researchers from determining when a policy transmission mechanism
is active. For example, researchers cannot determine whether compliance
occurs during recessions rather than expansions, nor can they identify
which recessionary episodes give rise to compliance. Likewise, they
cannot determine whether compliance occurs during periods of financial
distress or constrained monetary policy, or which specific episodes
generate compliance. In turn, this hinders the ability of policymakers
to target interventions toward similar time periods. Moreover, identifying
compliers can improve forecasting, as it allows one to predict ex
ante which time periods are most affected by the policy. Whereas \citet{abadie:2003}
identifies complier-specific characteristics  without identifying
which sample observations are compliers, we identify whether individual
time periods are compliers by using local comparisons to recover the
relevant counterfactual treatment assignment. Finally, as emphasized
by \citet{heckman/vitlacyl:2007}, it enhances external validity,
since the LATE may differ from the average treatment effect in the
broader population; without knowing who the compliers are, it is impossible
to assess how far the estimated causal effect can be generalized beyond
that subpopulation.

This paper considers IV identification in dynamic settings and shows
how compliers can be identified individually by exploiting the structure
of time series data. We first show that the notion of compliers can
be equivalently rewritten in terms of an inequality involving the
difference in means of the potential treatment under different instrument
values. Under assumptions of continuity over time in the mean of the
potential treatment assignment process\textemdash conditional on a
fixed hypothetical value of the instrument\textemdash it is possible
to recover counterfactual values by averaging observations in a neighborhood
around a given time point.

The economic intuition behind the continuity assumption is that the
policy transmission mechanism evolves gradually. For example, consider
heteroskedasticity-based identification of monetary policy effects
{[}cf. \citet{rigobon:2003} and \citet{nakamura/steinsson:2018}{]}
where the instrument indicates whether there was an FOMC announcement
on each date in the sample and the treatment variable is equal to
the variance of a short-term interest rate. Here compliers are defined
as observations for which the volatility of the policy variable (change
in short-term interest rate) increases if and only if there is an
FOMC announcement. Suppose there is an FOMC announcement on a particular
date. We then observe the treatment assignment under the realized
instrument value but do not observe the counterfactual treatment assignment
that would have occurred in the absence of the announcement. Continuity
implies that the mean treatment assignment under the counterfactual
state can be approximated using nearby non-announcement dates. More
generally, continuity allows information from neighboring periods
to recover the unobserved potential treatment assignments. Economically,
this assumption reflects the idea that, e.g., firms do not suddenly
stop responding to short-term interest rates, nor do institutional
features of financial markets abruptly change from one day to the
next. As a result, the complier status of the date in question can
be estimated and tested by comparing local means of the treatment,
one corresponding to nearby dates for which an announcement occurred
and the other corresponding to nearby dates for which it did not.\footnote{Our identification results immediately apply to cross-sectional settings
with spatial data provided that the temporal distance between observations
is interpreted as geographical distance, and analogous continuity
assumptions are imposed over space.} Applying our identification results and tests to the heteroskedasticity-based
identification of monetary policy, we find that about 60\% of observations
are compliers while the non-compliers are primarily concentrated in
the early zero lower bound period, when the central bank could no
longer lower interest rates and forward guidance was not aggressive.
Here identification of the complier dates allows one to determine
when the policy transmission mechanism is active and assess time variation
in policy effectiveness. 

Our identification argument is not confined to the case of FOMC announcements
with high-frequency data. We show that it also extends to narrative
identification approaches based on low-frequency data commonly used
in empirical macroeconomics, such as \citeauthor{ramey:2011}\textquoteright s
\citeyearpar{ramey:2011} identification of fiscal shocks, \citeauthor{mertens/ravn:2013}\textquoteright s
\citeyearpar{mertens/ravn:2013} identification of tax changes, and
\citeauthor{romer/romer:2004}\textquoteright s \citeyearpar{romer/romer:2004}
identification of monetary policy shocks. For recent methodological
work on narrative-based IV approaches see \citet{barnichon/mesters:2025}.

Identification of compliers is not only valuable in its own right.
The second main contribution of our paper is to show that this enables
us to test the exclusion restriction, a key condition for valid IV
estimation that is typically untestable in practice. By identifying
compliers, and thus also non-compliers, we show that the exclusion
restriction can be tested using a $t$-test that compares the average
outcomes of non-compliers across different instrument values. This
idea differs from that of \citet{kitagawa:2015}, who uses a nonnegativity
condition on conditional distributions of observables to construct
a Kolmogorov\textendash Smirnov test of instrument validity for a
binary treatment and a discrete instrument. Whereas Kitagawa's approach
is based on partial identification, our framework relies on exact
identification and does not require either a binary treatment or a
discrete instrument.

The third contribution of our paper is to establish new results and
introduce new methods on instrument relevance, a key condition for
LATE identification that requires a nontrivial correlation between
the instrument and the endogenous variable. We begin by analyzing
the problem of weak instruments, entailing low correlation between
the endogenous variable and instrument, in empirical work through
a survey of articles using IVs published from 2019 to 2022 in five
leading journals: American Economic Review, Econometrica, Journal
of Political Economy, Quarterly Journal of Economics, and Review of
Economic Studies. Our sample includes 1,560 specifications from 18
papers, with 199 involving time series and 1,361 involving panel data.\footnote{See the supplement for the full list of papers and inclusion criteria.}
The left panels of Figure \ref{Figure_Histogram} show histograms
of full sample first-stage $F$-statistics for the specifications
in our survey, truncated above 100 for visibility. Many $F$-statistics
concentrate around the $\chi_{1}^{2}$ critical values and fall below
the conventional thresholds of 10 and 23.1 suggested by \citet{staiger/stock:1997}
and \citet{montielolea/pflueger:2013}, raising serious concerns about
weak instruments. These findings align with those of \citet{andrews/stock/sun:2019},
who analyze cross-sectional studies. For example, we find that 75\%
of time series and 72\% of panel data specifications have first stage
$F$-statistics below 24. The median $F$-statistic is 12.63 for time
series and 9.29 for panel data.\footnote{For panel data specifications we consider each cross-sectional unit
individually to enable comparison to our proposed time series test
shown on the right panels.}  
\noindent\begin{flushleft}
\begin{center}
\begin{figure}[h]
\begin{centering}
\includegraphics[width=15cm,totalheight=7cm]{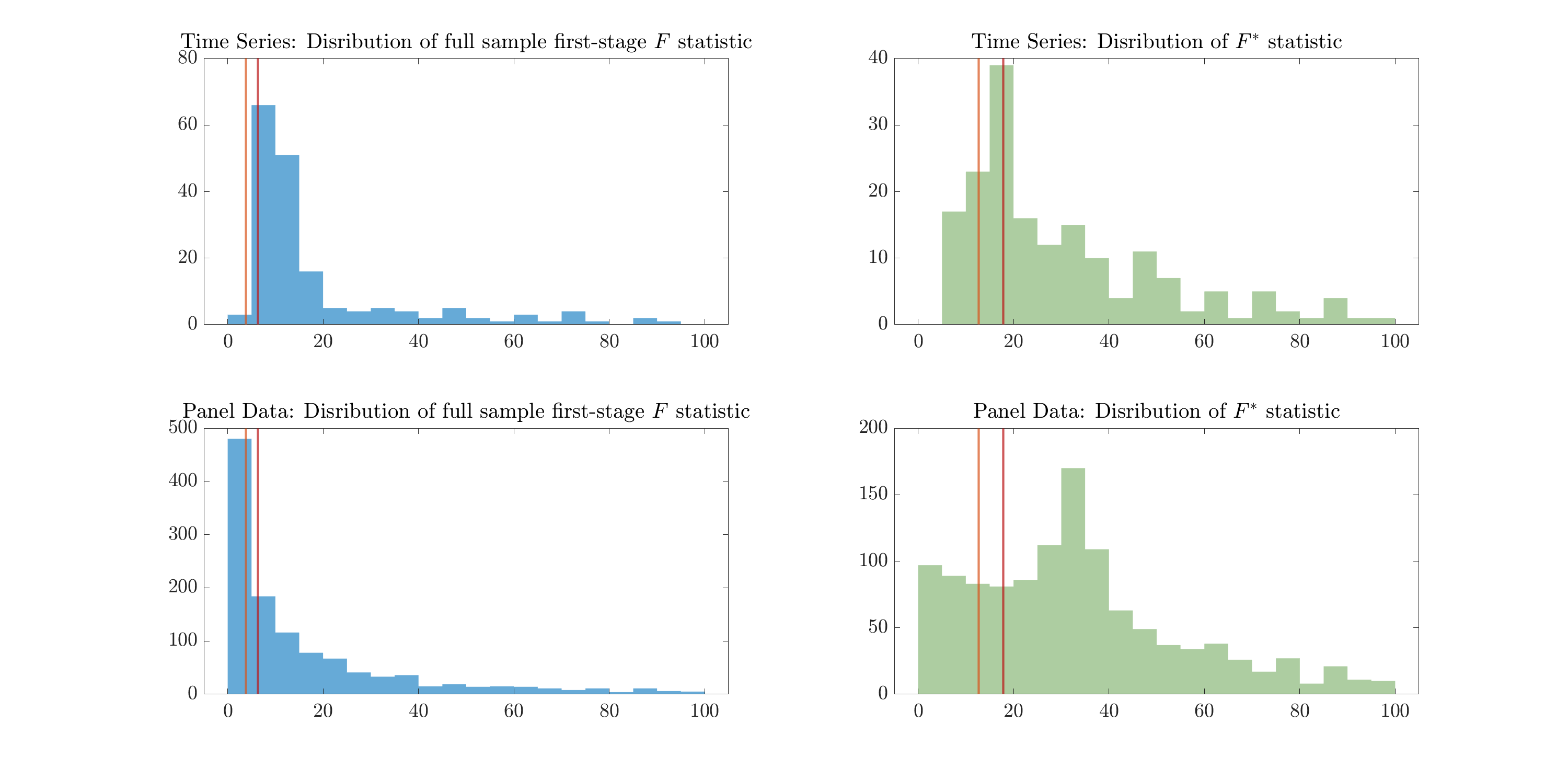}
\par\end{centering}
\raggedright{}\caption{\label{Figure_Histogram}{\scriptsize{} Distributions of the first-stage
$F$ (left panels) and $F^{*}$ statistics (right panels). The top
panels apply to time series specifications and the bottom panels apply
to panel data specifications. The orange and red vertical lines correspond
to the 5\% and 1\% level asymptotic critical values of the first-stage
$F$ ($\chi_{1}^{2}$ for left panels) and $F^{*}$ statistics (8.28
and 11.63) for right panels) under identification failure.}}
\end{figure}
\end{center}
\par\end{flushleft}

When identification fails or is weak, IV estimators can be severely
biased for LATEs and conventional inference methods are rendered invalid.
These problems have prompted extensive research on detecting weak
instruments and constructing identification-robust confidence sets.\footnote{See, e.g., \textcolor{MyBlue}{Andrews et al.} \citeyearpar{andrews/moreira/stock:2006},
\citet{kleibergen:2002}, \citet{moreira:2003} and \citet{staiger/stock:1997}.} However, there has been little work on estimation and inference in
a general LATE setting when identification may be stronger over subsamples.
We develop a framework for identification, estimation, and inference
on LATEs that accommodates time-varying instrument relevance. Within
this framework, we propose a first-stage $F$-test to detect whether
identification fails over all nontrivial subsamples. To solve the
computationally intensive problem of searching for maximal identification
strength among all possible sample partitions, we employ dynamic programming.
This optimization is more complex than that in the structural break
literature since evaluating identification strength requires more
than comparing parameter changes across regimes.

In an attempt to understand the sources of weak IVs we plot the histograms
of the $F^{*}$ statistic proposed in this paper (cf. Section \ref{Section: ID Failure Test})
in the right panels of Figure \ref{Figure_Histogram}. The statistic
$F_{}^{*}$ searches for the subsample with maximal identification
strength among all possible subsamples of size at least $\pi_{L}T$.\footnote{We set $\pi_{L}=0.6$ in Figure \ref{Figure_Histogram}. We discuss
the choice of $\pi_{L}$ below.} The idea is that while the IVs may appear weak in the full sample,
they may be strong in a possibly large subsample. Figure \ref{Figure_Histogram}
shows that this is indeed often the case. The red vertical lines in
Figure \ref{Figure_Histogram} mark the $95^{th}$ percentile of the
asymptotic distributions of the $F$ and $F^{*}$ statistics under
the null of identification failure. Although its quantiles are larger,
the $F^{*}$ statistics have substantially more mass in the upper
quantiles of their null distribution. This has at least three implications.
First, it confirms substantial time variation in the instruments'
strength. Second, strong identification appears to be frequently present
in a sizeable subsample even when the instruments appear weak in the
full sample. The median $F_{}^{*}$ is 27.22 for time series and 33.81
for panel data specifications. These are substantially higher than
their full sample counterparts and this difference cannot be simply
attributed to the different null asymptotic distributions of the two
test statistics given that the difference in the asymptotic critical
values is relatively small while the empirical distributions of the
two test statistics are markedly different. About one half of the
specifications that appear to suffer from weak IVs in the full sample
seem better characterized by strong IVs in the subsample with maximal
identification strength. Third, the subsamples where instruments appear
strong tend to be large. From an empirical perspective, this is encouraging:
although weak instruments in the full sample are common, researchers
can often succeed in identifying large subsamples where instruments
appear strong.

Motivated by this survey evidence, we construct consistent estimators
of LATEs when subsamples are strongly-identified. It is commonly believed
that if IVs are strong only in some portion of the sample, the full
sample IV estimator remains consistent for a LATE. We show that this
belief is unwarranted unless the LATE of interest is time-invariant
(i.e., homogeneous). If this condition fails, one can at best identify
a LATE corresponding to the strongly-identified subsample. Even when
the LATE is homogeneous, the full sample IV estimator may still be
severely biased if instruments are irrelevant over parts of the sample.

Our approach differs from that of \citet{magnusson/mavroeidis:2014}
and \citet{antonie/boldea:2018}, who use time variation in IV strength
to add moment conditions in a GMM context, enabling more efficient
inference and estimation. In contrast, we exploit this time variation
to identify the subsample where IVs are strongest and base our estimation
on this subsample. This insight allows for\textit{ consistent estimation
even when subsamples suffer from identification failure}.\footnote{Another major difference from \citet{magnusson/mavroeidis:2014} is
that we address the computational challenge for the case of multiple
breaks in the first-stage coefficient. \citet{magnusson/mavroeidis:2014}
did not attempt to address this issue and refer to it as ``computationally
demanding''.} If the parameter of interest is heterogeneous, our estimator remains
valid but is interpretable only within the strongly-identified sub-population.

We apply our methodology to the heteroskedasticity-based identification
strategy used to estimate the causal effects of monetary policy from
high-frequency data {[}e.g., \citet{nakamura/steinsson:2018}{]}.
The key identification condition for this strategy is that the volatility
of the daily changes in short-term interest rates is higher on FOMC
announcement days than on non-FOMC days. \citet{lewis:2020} provides
evidence of weak full sample identification and shows that IV and
GMM estimates even differ in sign. We find that identification is
substantially stronger over a subsample comprising 80\textendash 90\%
of the data, with the excluded subsample centered around the financial
crisis, during which volatility was high even on non-FOMC days. Estimation
using the most strongly-identified subsample yields IV and GMM estimates
that have the same sign and similar magnitudes. We recommend reporting
the most strongly-identified subsample estimates in addition to the
full sample estimates when strong full sample identification may be
in question.

Although our new methods are able to find the most strongly-identified
subsample, this subsample may still fail to be strongly-identified.
For our final theoretical contribution, we develop identification-robust
inference procedures using the most strongly-identified subsample.
We propose versions of the Anderson-Rubin, Lagrange Multiplier, and
conditional likelihood ratio tests, which depend only on this subsample.
These tests are more efficient than their full sample counterparts,
which include noise from regimes suffering from identification failure.
When instruments are strong throughout the sample, our tests coincide
with the conventional ones. When instruments are irrelevant over parts
of the sample, our tests achieve higher efficiency by focusing on
stronger segments. In the worst case, when IVs are weak everywhere,
our methods are no less efficient than existing ones. While there
is a trade-off between using fewer observations and more strongly
identified subsamples, simulations show that our tests have higher
power, indicating that the efficiency loss from a smaller sample size
is outweighed by the gain in identification strength.

The paper is organized as follows. Section \ref{Section: Statistical Framework for Identification of Causal Effects}
introduces the potential outcome framework and dynamic causal effects,
and presents identification results. Section \ref{Section: Application: Money Neutrality}
discusses issues pertaining to heteroskedasticity-based identification
of monetary policy. Section \ref{Section: ID Failure Test} presents
an $F$-test for full sample identification failure. Estimation and
inference robust to weak identification are discussed in Sections
\ref{Section: Estimation}-\ref{Section: Inference}. An empirical
application is considered in Section \ref{Section: Empirical Evidence}.
Section \ref{Section Conclusions} concludes. The supplements \citeauthor{casini/mccloskey/pala/rolla_Dynamic_Late_Supp_Not_Online}
(\citeyear{casini/mccloskey/pala/rolla_Dynamic_Late_Supp_Not_Online},
\citeyear{casini/mccloskey/pala/rolla_Dynamic_Late_Supp}) include
the Monte Carlo simulations, proofs and additional results.

\section{Identification of Dynamic Causal Effects\label{Section: Statistical Framework for Identification of Causal Effects}}

A growing literature in macroeconomics uses IVs to identify dynamic
causal effects when the policy variable of interest is endogenous.\footnote{See, e.g., \citet{gertler/karadi:2015}, \citet{mertens/ravn:2013},
\citet{plagborg-Moller/wolf:2022} and \citet{stock/watson:2018}.} Many existing identification approaches can be reframed in terms
of IVs, either derived from the modeling approach {[}e.g., heteroskedasticity-based
identification as in \citet{rigobon:2003} and \citet{nakamura/steinsson:2018}{]}
or through external IVs constructed using a narrative approach {[}cf.,
\citet{olea/stock/watson:2021}{]}. For example, \citet{romer/romer:1989}
study the FOMC minutes to pinpoint dates when monetary policy actions
were arguably exogenous. This allows the construction of exogenous
variables that can be interpreted as IVs for some structural shock
of interest.\footnote{See \citet{ramey:2011} for unanticipated defense spending shocks,
\citet{nakamura/steinsson:2018} and \citet{romer/romer:2004} for
monetary policy shocks, \citet{kanzig:2021} and \citet{killian:2009}
for oil market shocks, and \citet{romer/romer:2010} for tax shocks.}

We adopt a potential outcomes framework, as introduced by \citet{rubin:1974}
and extended to time series settings by \citet{angrist/kuersteiner:2011}
and \citet{rambachan/shephard:2021}. Let the stochastic process $V_{t}=(Y_{t},\,X_{t},\,D_{t},\,Z_{t})$
be defined on the probability space $\left(\Omega,\,\mathscr{F},\,\mathbb{P}\right)$,
where $Y_{t}$ is a vector of outcome variables, $D_{t}$ is a policy
variable, $X_{t}$ is a vector of other exogenous and/or lagged endogenous
variables, and $Z_{t}$ is a vector of instruments. Let $\vec{X}_{t}=\{\ldots,\,X_{t-1},\,X_{t},\}$
denote the covariate path up to time $t$, with analogous definitions
for $\vec{Y}_{t}$, $\vec{D}_{t}$ and $\vec{Z}_{t}$. Let the policy-relevant
information set at time $t$ denoted by $\mathscr{F}_{t}=\sigma(\widetilde{V}_{t})$
where $\sigma(\widetilde{V}_{t})$ is the $\sigma$-algebra generated
by the history of $V_{t}$, $\widetilde{V}_{t}=(\vec{Y}_{t-1},\,\vec{X}_{t},\,\vec{D}_{t-1},\,\vec{Z}_{t-1})$.

Policy decisions, $D_{t}$, depend on past observable variables and
the contemporaneous outcome through a systematic component and on
idiosyncratic information available to the policy-maker (i.e., the
random component). The systematic component is a time-varying non-stochastic
function of the observed random variables $\widetilde{V}_{t}$, contemporaneous
outcome $Y_{t}$, and the contemporaneous instrument $Z_{t}$. The
idiosyncratic information is represented by a scalar stochastic shock
$e_{t}$ that is not observed by the researcher. In a SVAR context,
$e_{t}$ is the structural shock to the policy variable $D_{t}.$
For example, if the monetary authority follows a simple Taylor rule
for the nominal interest rate, then $D_{t}$ is the policy interest
rate and $\widetilde{V}_{t}$ includes inflation, output and the natural
rate of interest.

We define two types of potential outcomes. The first, $Y_{t}\left(\left(\epsilon_{1:t}\right),\,\left(z_{1:t}\right)\right)$,
denotes the counterfactual values of $Y_{t}$ under hypothetical sequences
of the policy shocks $\epsilon_{1:t}$ and instruments $z_{1:t}$,
where $a_{1:t}=\left\{ a_{s}\right\} _{s=1}^{t}$. 
\begin{defn}
\label{Definition: Generalized Potential Outcome}A generalized potential
outcome, $Y_{t}\left(\left(\epsilon_{1:t}\right),\,\left(z_{1:t}\right)\right)$,
is defined as the value assumed by $Y_{t}$ if $e_{s}=\epsilon_{s}$
and $Z_{s}=z_{s}$ for $s=1,\ldots,\,t$. 
\end{defn}
This definition excludes dependence on future shocks or instruments.
The potential outcome process should not be confused with the observed
outcome $\left\{ Y_{t}\right\} _{t\geq1}=\left\{ Y_{t}\left(e_{1:t},\,Z_{1:t}\right)\right\} _{t\geq1}$.
For $h\geq0$ and any given $\epsilon$ and $z$, write the time-$t+h$
potential outcome along the path $\left(\left(e_{1:t-1},\,\epsilon,\,e_{t+1:t+h}\right),\,\left(Z_{1:t-1},\,z,\,Z_{t+1:t+h}\right)\right)$
as 
\begin{align*}
Y_{t,h}\left(\epsilon,\,z\right) & =Y_{t+h}\left(\left(e_{1:t-1},\,\epsilon,\,e_{t+1:t+h}\right),\,\left(Z_{1:t-1},\,z,\,Z_{t+1:t+h}\right)\right),
\end{align*}
where $Y_{t,h}\left(e_{t},\,Z_{t}\right)=Y_{t+h}$. Definition \ref{Definition: Generalized Potential Outcome}
captures the property that $Y_{t,h}\left(\epsilon,\,z\right)$ also
depends on policy shocks that occur between time $t+1$ and $t+h$.
The notation $Y_{t,h}\left(e,\,z\right)$ focuses on the effect of
a single policy shock on current and future outcomes akin to the idea
underlying an impulse response. When the potential outcomes do not
depend on the instruments, $Y_{t,h}\left(\epsilon,\,z\right)=Y_{t,h}\left(\epsilon\right)$,
and for $\epsilon\neq\epsilon'$, $Y_{t,h}\left(\epsilon\right)-Y_{t,h}\left(\epsilon'\right)$
for $h=0,\,1,\,\ldots$ are the dynamic causal effects of a policy
shock on the outcome. In a SVAR setting, one is often interested in
these dynamic causal effects which are in fact the impulse responses.

The second potential outcome that we discuss, $Y_{t}^{*}\left(\left(d_{1:t}\right),\,\left(z_{1:t}\right)\right)$,
is defined as the counterfactual values of $Y_{t}$ under hypothetical
sequences of treatments $d_{1:t}$ and instruments $z_{1:t}$. The
distinction with $Y_{t}\left(\epsilon_{1:t},\,z_{1:t}\right)$ is
that this formulation focuses on causal effects of the policy variable
$D$, not the policy shock $e$. For $t\geq1$, we assume that $d_{t}\in\mathbf{D}$,
$z_{t}\in\mathbf{Z}$ for some sets $\mathbf{D}$ and $\mathbf{Z}$.
In many applications outside SVARs, the causal effects of the policy
are of interest. Think about the slope of demand functions, price
elasticities, response coefficients or reaction functions of, for
example, asset prices to monetary policy, and so on. Typically these
causal effects are analyzed using event-studies, quasi-experiments,
IV regressions, etc. The recent literature on causal effects in time
series {[}e.g., \citet{rambachan/shephard:2021}{]} focuses on the
identification of causal effects of the structural shocks. In this
paper, we consider identification of causal effects of the policy
variable. We illustrate the difference between these two causal effects
and an application to SVAR using the following two examples. 
\begin{example}
\label{Example: Demand and Supply}Consider the following system of
simultaneous equations, 
\begin{align}
Y_{t} & =\beta D_{t}+\eta_{t}\qquad\mathrm{and}\qquad D_{t}=aY_{t}+e_{t},\label{Eq. (1) Demand Curve}
\end{align}
where the first equation is the demand curve, the second is the supply
curve, $Y_{t}$ and $D_{t}$ are the observed price and quantity,
and $\eta_{t}$ and $e_{t}$ are the structural shocks. The parameter
$\beta$ captures the slope of the demand function, which corresponds
to the causal effect $\partial Y_{t}^{*}\left(d\right)/\partial d=\beta$.
On the other hand, in a SVAR context one may be interested in the
impulse response of $Y_{t}$ given a shock to supply $e_{t}$. Solving
for the reduced-form of \eqref{Eq. (1) Demand Curve}, 
\begin{align*}
Y_{t} & =\frac{\beta}{1-a\beta}e_{t}+\frac{1}{1-a\beta}\eta_{t},
\end{align*}
shows that the lag-0 impulse response is $\mathrm{d}Y_{t}\left(e\right)/\mathrm{d}e=\beta/\left(1-a\beta\right)$,
which differs from $\beta$. 
\end{example}
\begin{example}
\label{Example: SVAR, Potential Outcomes}Consider the following reduced-form
VAR, 
\begin{align*}
V_{t} & =A_{1}V_{t-1}+A_{2}V_{t-2}+\ldots+A_{p}V_{t-p}+u_{t},
\end{align*}
where $V_{t}=(D_{t},\,Y_{t}')'$ is $n\times1$, $D_{t}$ is a scalar,
and $u_{t}$ is a vector of reduced-form VAR innovations. The latter
are related to structural shocks, $\varepsilon_{t}=\left(e_{t},\,\eta'_{t}\right)'$,
via $u_{t}=B_{0}\varepsilon_{t}$ where $B_{0}$ is a non-singular
matrix. Under suitable conditions, $V_{t}$ admits a moving-average
representation $V_{t}=\sum_{j=0}^{\infty}C_{j}\left(A\right)B_{0}\varepsilon_{t-j}$,
where $C_{j}\left(A\right)=\sum_{i=1}^{j}C_{j-i}\left(A\right)A_{i}$
for $j=1,\,2,\ldots$ with $C_{0}\left(A\right)=I_{n}$ and $A_{i}=0$
for $i>p$. Then, the outcome variable admits a moving-average representation,
\begin{align*}
Y_{t} & =\sum_{j=0}^{\infty}c_{ye,j}e_{t-j}+\sum_{j=0}^{\infty}c_{y\eta,j}\eta_{t-j},
\end{align*}
where $c_{ye,j}$ and $c_{y\eta,j}$ are blocks of $C_{j}\left(A\right)B_{0}$
partitioned conformably to $Y_{t}$, $e_{t}$ and $\eta_{t}$. If
$e_{t}$ is the policy shock, the potential outcomes here are defined
as 
\begin{align*}
Y_{t,h}\left(\epsilon\right)= & Y_{t,h}\left(\epsilon,\,z\right)=\sum_{j=0,j\neq h}^{\infty}c_{ye,j}e_{t+h-j}+\sum_{j=0}^{\infty}c_{y\eta,j}\eta_{t+h-j}+c_{ye,h}\epsilon.
\end{align*}
The potential outcome $Y_{t,h}^ {}\left(\epsilon\right)$ tells us
what $Y_{t+h}$ would be if $e_{t}=\epsilon$ and it does not depend
upon $z$ since the instrument $Z_{t}$ is excluded from the VAR.
Here the absence of causal effects means that $c_{ye,h}=0$ for all
$h$, coinciding with the canonical condition that the impulse responses
are identically equal to zero.

The potential outcome framework is useful because it allows the study
of nonparametric conditions such that common statistical estimands
(e.g., impulse responses) have a causal interpretation. \citet{olea/stock/watson:2021}
show how to use the instrument $Z_{t}$ to identify the impulse response
coefficient $\phi_{r,e,h}=\partial Y_{t+h}^{\left(r\right)}/\partial e_{t}$
(the effect of $e_{t}$ on the $r$th variable in $Y_{t+h}$). From
the moving-average representation we have $\phi_{r,e,h}=\iota'_{r+1}C_{h}\left(A\right)B_{0}\iota{}_{1}$
where $\iota_{s}$ denotes the $s$-th standard basis vector. This
shows that $\phi_{r,e,h}$ depends on the $A$'s and the first column
of $B_{0}$. The following assumptions are needed for the identification
of $\phi_{r,e,h}$: (i) $\mathbb{E}(Z_{t}e_{t})=\theta\neq0$ (instrument
relevance) and (ii) $\mathbb{E}(Z_{t}\eta_{t})=0$ (instrument exogeneity).
By (i)-(ii), $B_{0}^{\left(:,1\right)}=B_{0}\iota_{1}$ is identified
up to scale by the covariance between $Z_{t}$ and the reduced-form
innovations $u_{t}$:~$\Gamma=\mathbb{E}(Z_{t}u_{t})=\mathbb{E}(Z_{t}B_{0}\varepsilon_{t})=\theta B_{0}^{\left(:,1\right)}$.
Using the scale normalization $B_{0}^{\left(1,1\right)}=1$ {[}see
\citet{stock/watson:2018} for a discussion{]} we have $\Gamma^{\left(1,1\right)}=\mathbb{E}(Z_{t}e_{t})=\theta$
and $B_{0}^{\left(:,1\right)}=\Gamma/\Gamma^{\left(1,1\right)}=\Gamma/\iota'_{1}\Gamma$.
It follows that $\phi_{r,e,h}$ is identified since $\phi_{r,e,h}=\iota'_{r+1}C_{h}\left(A\right)\Gamma/\iota'_{1}\Gamma$,
where $A$ can be estimated consistently from the reduced-form VAR
and $\Gamma$ can be estimated consistently by using the VAR residuals
$\widehat{u}_{t}$ in place of $u_{t}$. On the other hand, identifying
the causal effects of the policy $D_{t}$ here would require additional
identification restrictions.

\citet{olea/stock/watson:2021} use shortfalls in OPEC oil production
associated with wars and civil disruptions as an instrument for the
oil supply shock in the SVAR of \citet{killian:2009} who investigates
the effect of oil supply and demand shocks on oil production and prices.
This variable is plausibly correlated with the oil supply shock and,
because the shortfalls are associated with political events such as
wars in the Middle East, it is plausibly uncorrelated with the demand
shocks. Using the analog of the nonparametric conditions we provide
below, applied to the shock $e_{t}$ rather than the policy $D_{t}$,
permits a causal interpretation of the impulse response even when
$\mathbb{E}(Z_{t}e_{t})=0$ for a sub-population.
\end{example}
In the following, we discuss identification of causal effects of the
policy via IV estimands.

\subsection{\label{Subsection: Identification-Conditions}Identification Conditions}

We explicitly allow for endogeneity and rely on IVs. We define $D_{t}(z)$
as the potential treatment assignment at time $t$ when $Z_{t}$
is set equal to $z\in\mathbf{Z}$. The instrument $Z_{t}$ is assumed
to be (conditionally) independent of the potential outcomes $Y_{t,j}^{*}\left(d,\,z\right)$
and treatments $D_{t}(z)$ but correlated with the observed treatment
$D_{t}$. 
\begin{assumption}
\label{Assumption: Independence}$($Independence$)$ For all $d\in\mathbf{D}$,
$z\in\mathbf{Z}$ and $t\geq1$, we have 
\begin{align}
\left\{ \left\{ Y_{t,h}^{*}\left(d,\,z\right)\right\} _{h\geq0},\,D_{t}\left(z\right)\right\}  & \bot Z_{t}|\,\widetilde{V}_{t}.\label{Eq: Assumption Independence}
\end{align}
\end{assumption}
Assumption \ref{Assumption: Independence} states that, given $\widetilde{V}_{t}$,
the instrument is as good as randomly assigned. In studies of monetary
policy such as \citet{nakamura/steinsson:2018}, $Z_{t}=1$ if there
is an FOMC announcement on day $t$, and $Z_{t}=0$ otherwise. Since
FOMC announcements are scheduled in advance, $Z_{t}$ is deterministic,
and therefore Assumption \ref{Assumption: Independence} holds.\footnote{Unscheduled FOMC meetings that occur during emergencies are typically
excluded from the sample. } 

The second assumption is that potential outcomes $Y_{t,h}^{*}\left(d,\,z\right)$
are a function of $d$ but not of $z$. In the case of FOMC announcements,
this means that potential realizations of expected output growth respond
to changes in the monetary policy variable in the same way regardless
of whether the change is associated with an FOMC announcement or not.
\begin{assumption}
\label{Assumption: Exclusion }$($Exclusion$)$ For all $d\in\mathbf{D},\,t\geq1$
and $h\geq0$, we have 
\begin{align}
\left\{ Y_{t,h}^{*}\left(d,\,z\right)=Y_{t,h}^{*}\left(d,\,z'\right)\right\}  & |\,\widetilde{V}_{t},\qquad\mathrm{for}\,\mathrm{all}\,z,\,z'\in\mathbf{Z}.\label{Eq: Assumption Exclusion Restriction}
\end{align}
\end{assumption}
In a dynamic simultaneous equations model (e.g., a SVAR) the exclusion
restriction requires the instrument not to appear in the causal equation
of interest. In Example \ref{Example: SVAR, Potential Outcomes},
Assumption \ref{Assumption: Exclusion } corresponds to condition
(ii), i.e., $\mathbb{E}(Z_{t}\eta_{t})=0$ where $\eta_{t}$ is composed
of the structural shocks other than $e_{t}$. Under Assumption \ref{Assumption: Exclusion }
we write $Y_{t,h}^{*}\left(d,\,z\right)=Y_{t,h}^{*}\left(d\right)$.

Identification based on IVs requires instrument relevance or ``existence
of a first-stage''. The latter means that $\mathbb{E}(D_{t}\left(z\right)|\,\widetilde{V}_{t})$
is a non-trivial function of $z$. In cross-sectional settings, the
existence of a first-stage is typically assumed to hold for all units
to guarantee strong identification. Strong identification of this
form often fails to hold in applications involving time series data
due to temporary misspecification, bad luck, rare events or parameter
instability. The analysis based on articles in five leading journals
that we report earlier suggests that there are time periods for which
the first-stage exists (strong identification) and others for which
it does not (identification failure). Standard first-stage $F$-tests
are then likely to indicate weak identification since they are based
on averaging these two sub-populations.

We provide a theoretical framework to address this identification
problem by assuming that there are two sub-populations. One comprises
a fraction $\pi_{0}\in\left[0,\,1\right]$ of the overall population
for which the first-stage exists. For the second sub-population, which
comprises a fraction $1-\pi_{0}$ of the population, the first-stage
does not exist. This leads to a new notion of LATE, which we name
$\pi$-LATE, the LATE for the (unknown) $\pi_{0}$ fraction of the
population for which the first-stage exists. If $\pi_{0}=1,$ then
one recovers LATE.

Denote by $|\mathbf{S}_{0,T}|$ the cardinality of $\mathbf{S}_{0,T}$
(i.e., the number of indices in $\mathbf{S}_{0,T})$.
\begin{assumption}
\label{Assumption: First-Stage}$($Partial first-stage$)$ Assume
there exists $\mathbf{S}_{0,T}\subseteq\left\{ 1,\,\ldots,\,T\right\} $
such that $|\mathbf{S}_{0,T}|=\left\lfloor \pi_{0}T\right\rfloor $
with $\pi_{0}\in(0,\,1]$ and for $t\in\mathbf{S}_{0,T}$, $\mathbb{E}(D_{t}\left(z\right)|\,\widetilde{V}_{t})$
is a non-trivial function of $z$.\footnote{We assume that all expectations exist.}
\end{assumption}
Assumption \ref{Assumption: First-Stage} implies that there are two
sub-populations: one for which the first-stage exists and one for
which it does not. An average treatment effect can only be identified
via IVs for the fraction $\pi_{0}$ of the population for which a
first-stage exists.

The next assumption is monotonicity which, under heteroskedasticity-based
identification of monetary policy (see Section \ref{Section: Application: Money Neutrality}),
means that while for some days the FOMC announcement does not coincide
with higher volatility in the policy variable, all of those days in
which the announcement affects the volatility of the policy variable,
volatility is shifted up.
\begin{assumption}
\label{Assumption: Monotonicity}$($Monotonicity$)$ $\mathbf{D}\subseteq\mathbb{R}$.
For all $t=1,\ldots,\,T$ and $z,\,z'\in\mathbf{Z}$, either $D_{t}\left(z\right)\geq D_{t}\left(z'\right)$
or $D_{t}\left(z'\right)\geq D_{t}\left(z\right)$ with probability
one.  
\end{assumption}
If $\pi_{0}=1$ (so $|\mathbf{S}_{0,T}|=T$), the condition reduces
to that in \citet{imbens/angrist:1994}. 

Following \citet{kolesar/plagborgmoller:2025}, we impose the following
assumption.
\begin{assumption}
\label{Assumption KP 2024}For all $t\geq1$ and $h\geq0,$ (i) $Y_{t,h}^{*}\left(\cdot\right)$
is locally absolutely continuous on $\mathbf{D}$ and (ii) $\mathbb{E}\left[\left.\int_{\mathbf{D}}|\partial Y_{t,h}^{*}\left(d\right)/\partial d|\mathrm{d}d\right|\widetilde{V}_{t}\right]<\infty$.
\end{assumption}
Assumption \ref{Assumption KP 2024} allows $D_{t}$ to be either
discrete, continuous or mixed. When $D_{t}$ is discrete or mixed,
it is implicitly assumed that to deal with the gaps in the support
of $D_{t}$ one extends $Y_{t,h}^{*}\left(\cdot\right)$ to $\mathbf{D}$
such that the extension is locally absolutely continuous. The support
of $D_{t}$ is allowed to be unbounded. These conditions are weaker
than counterparts imposed in the recent literature {[}cf. \citet{casini/mccloskey:2024}
and \citet{rambachan/shephard:2021}{]}, in particular local absolute
continuity replaces differentiability of $Y_{t,h}^{*}\left(\cdot\right)$
plus bounded support of $D_{t}$. It allows the application of the
fundamental theorem of calculus to $Y_{t,h}^{*}\left(\cdot\right)$
without requiring the support of $D_{t}$ to be bounded.

\subsection{\label{Subsection: Identification-Results}Identification Results}

\subsubsection{\label{Subsubsection: Identification of Causal Effects}Identification
of Causal Effects}

We first discuss the case of a discrete instrument. When the first-stage
does not exist for all $t$, it is useful to define an IV estimand
corresponding to the sub-population for which it does. Let the generalized
Wald estimand be defined for all $z',\,z\in\mathbf{Z}$ by 
\begin{align}
\beta_{\pi,t,h}\left(\widetilde{v}\right) & =\frac{\mathbb{E}\left(Y_{t+h}|\,Z_{t}=z',\,\widetilde{V}_{t}=\widetilde{v}\right)-\mathbb{E}\left(Y_{t+h}|\,Z_{t}=z,\,\widetilde{V}_{t}=\widetilde{v}\right)}{\mathbb{E}\left(D_{t}|\,Z_{t}=z',\,\widetilde{V}_{t}=\widetilde{v}\right)-\mathbb{E}\left(D_{t}|\,Z_{t}=z,\,\widetilde{V}_{t}=\widetilde{v}\right)},\qquad\text{for }t\in\mathbf{S}_{0,T},\label{Eq. (beta_j(v)) pi-LATE}
\end{align}
where $\widetilde{v}\in\mathbf{V}$. This is the ratio of a reduced-form
generalized impulse response to a first-stage generalized impulse
response for $t\in\mathbf{S}_{0,T}$. We show that for $t\in\mathbf{S}_{0,T}$,
the estimand $\beta_{\pi,t,h}\left(\widetilde{v}\right)$ identifies
a weighted average of causal effects for the compliers. Recall that
$t\in\mathbf{S}_{0,T}$ and $\pi_{0}$ are related by $|\mathbf{S}_{0,T}|=\left\lfloor \pi_{0}T\right\rfloor $.
When $\pi_{0}=1$ and there is no conditioning on $\widetilde{V}_{t}=\widetilde{v}$,
$\beta_{1,t,h}$ reduces to the Wald estimand considered by \citet{rambachan/shephard:2021}.
For $t\notin\mathbf{S}_{0,T}$, $\beta_{\pi,t,h}$ does not identify
a causal effect because the denominator of \eqref{Eq. (beta_j(v)) pi-LATE}
is equal to zero.

We show that for $t\in\mathbf{S}_{0,T}$, the generalized Wald estimand
is equal to a weighted average of marginal effects where the latter
are the derivatives $\partial Y_{t,\,h}^{*}\left(d\right)/\partial d$.
\begin{prop}
\label{Proposition: Continuous pi-LATE}($\pi$-LATE) Let Assumptions
\ref{Assumption: Independence}-\ref{Assumption KP 2024} hold. For
$t\in\mathbf{S}_{0,T}$, $h\geq0$, $\widetilde{v}\in\mathbf{V}$
and $z',\,z\in\mathbf{Z}$, we have 
\begin{align}
\beta_{\pi,t,h}\left(\widetilde{v}\right) & =\int_{\mathbf{D}}\mathbb{E}\left[\left.\frac{\partial Y_{t,\,h}^{*}\left(d\right)}{\partial d}\right|\,D_{t}\left(z\right)\leq d\leq D_{t}\left(z'\right),\,\widetilde{V}_{t}=\widetilde{v}\right]w_{t}\left(d|\,\widetilde{v}\right)\mathrm{d}d,\quad\mathrm{where}\label{Eq. pi-LATE}\\
w_{t}\left(d|\,\widetilde{v}\right) & =\frac{\mathbb{P}\left(D_{t}\left(z\right)\leq d\leq D_{t}\left(z'\right)|\,\widetilde{V}_{t}=\widetilde{v}\right)}{\int_{\mathbf{D}}\mathbb{P}\left(D_{t}\left(z\right)\leq r\leq D_{t}\left(z'\right)|\,\widetilde{V}_{t}=\widetilde{v}\right)\mathrm{d}r}\geq0\quad\mathrm{and}\quad\int_{\mathbf{D}}w_{t}\left(d|\,\widetilde{v}\right)\mathrm{d}d=1.\nonumber 
\end{align}
\end{prop}
Proposition \ref{Proposition: Continuous pi-LATE} shows that $\beta_{\pi,t,h}\left(\widetilde{v}\right)$
identifies a weighted average of causal effects for compliers, characterized
by $D_{t}(z')>D_{t}(z)$, for observations with a first-stage, with
weights $w_{t}\left(d|\,\widetilde{v}\right)$ determined by the (conditional)
likelihood that $D_{t}\left(z\right)\leq d\leq D_{t}\left(z'\right)$.
We refer to the average treatment effect on the right-hand side of
\eqref{Eq. pi-LATE} as the time-$t$ $\pi$-LATE since it is the
LATE for the observations in this sub-population, which is a fraction
$\pi_{0}$ of the whole population. In practice, the IV estimand $\beta_{\pi,t,h}\left(\widetilde{v}\right)$
is characterized by two types of averaging. First, there is averaging
over time. For any treatment $d$, the average involves only those
observations whose treatment variable can be induced to change by
a change in the instrument and is computed only over those observations
that satisfy the first-stage (i.e., $t\in\mathbf{S}_{0,T}$). The
second averaging is over different treatment values $d$ at the same
date $t$. This is reflected in the weight $w_{t}\left(\cdot\right)$,
which is proportional to the conditional probability that $D_{t}\left(z\right)\leq d\leq D_{t}\left(z'\right)$.

Stationarity of the conditional joint distribution of the average
potential outcome and treatment assignment functions for observations
with a first-stage lends further interpretability to the generalized
Wald estimand. Specifically, if $\{Y_{t,h}^{*}\left(d\right),D_{t}(z)\}|\widetilde{V}_{t}$
is identically distributed across $t$ for all $t\in\mathbf{S}_{0,T}$,
$d\in\mathbf{D}$ and $z\in\mathbf{Z}$, Proposition \ref{Proposition: Continuous pi-LATE},
immediately implies that $\beta_{\pi,t,h}$ is equal for all $t\in\mathbf{S}_{0,T}$.
Given this, we can write $\beta_{\pi,t,h}=\beta_{\pi,h}$, making
explicit that the generalized Wald estimand \eqref{Eq. (beta_j(v)) pi-LATE}
equals a weighted average of causal effects for members of the sub-population
with a first-stage, which represents a $\pi_{0}$-sized fraction of
the total population. Under this assumption, we refer to the average
causal effect inside of the integral as $\pi$-LATE since it is a
LATE for a member of the $\mathbf{S}_{0,T}$ sub-population whose
treatment variable can be induced to change by a change in the instrument.

The sample counterpart to the generalized Wald estimand \eqref{Eq. (beta_j(v)) pi-LATE}
involves replacing the conditional expectations with sample estimates
based upon observations $t\in\mathbf{S}_{0,T}$, yielding an estimator
of a causal effect. When Assumption \ref{Assumption: First-Stage}
holds with $\pi_{0}\in\left(0,\,1\right)$, the full sample estimand,
i.e., the ratio of the time averages of the numerator and denominator
of \eqref{Eq. (beta_j(v)) pi-LATE}, is a poor representative of the
full sample average treatment effects because it includes observations
for which the instrument is not relevant in the averaging. We caution
that the usual practice of estimating the conditional expectations
in \eqref{Eq. (beta_j(v)) pi-LATE} with full sample estimates will
not estimate the full sample LATE, but $\pi$-LATE.

\citet{angrist/graddy/imbens:2000}, \citet{kolesar/plagborgmoller:2025}
and \citet{rambachan/shephard:2021} consider related results. The
difference here is that we do not require $D_{t}$ to be continuous
or that the first-stage holds for all $t.$\footnote{\citet{sojitra/syrgkanis:2025} focus is on the causal effect of treatment
histories on long term outcomes, rather than of one-time shocks or
single policy shifts on outcomes at horizon $h$.}

A connection to program evaluation with binary policy actions arises
when we map a dynamic problem with continuous variables into one with
binary policy actions and instruments. For example, consider the analysis
of causal effects of monetary policy using heteroskedasticity-based
identification {[}cf. \citet{nakamura/steinsson:2018} and \citet{rigobon/sack:2003}{]}.
Define a binary instrument $Z_{t}$ with $Z_{t}=1$ if there is a
scheduled announcement on day $t$ and $Z_{t}=0$ otherwise. The policy
$\Delta i_{t}$ typically reflects changes in short-term interest
rates. Identification relies on higher volatility in $\Delta i_{t}$
during announcement days (policy sample) compared to non-announcement
days (control sample). Think about mapping $|\Delta i_{t}|$ into
a binary treatment such that $D_{t}=1$ if $|\Delta i_{t}|\geq\delta$
for some threshold $\delta>0$ and $D_{t}=0$ if $|\Delta i_{t}|<\delta$
{[}cf. \citet{rigobon/sack:2003}{]}. Here $\pi$-LATE captures the
average treatment effect for the sub-population whose interest rate
changes exceed $\delta$ only when there is an announcement (i.e.,
when $Z_{t}=1$). Observations where $|\Delta i_{t}|<\delta$ regardless
of announcements are ``never-takers,'' while those with $|\Delta i_{t}|\geq\delta$
regardless of announcements are ``always-takers.'' Under monotonicity,
these groups form the non-compliers, whose responses are driven by
idiosyncratic factors other than announcement-specific effects. In
Section \ref{Section: Application: Money Neutrality} we document
regimes where the volatility of $\Delta i_{t}$ is high even in the
absence of announcements.

\subsubsection{\label{Subsubsec: Identification of Compliers}Identification of
Compliers and Exclusion Restriction }

A practical challenge for the $\pi$-LATE framework, and LATE frameworks
in general, is that the sub-population of compliers is unknown. However,
in time series settings with binary instruments, we show below that
one can identify the compliers individually, i.e., to determine whether
each observation $t$ is a complier. In this section, we consider
a binary instrument, e.g., $Z_{t}=1$ if $t$ is an FOMC meeting day
and $Z_{t}=0$ otherwise. Under Assumption \ref{Assumption: Monotonicity},
assume without loss of generality that $D_{t}(1)\geq D_{t}(0)$ for
all $t$. Then, observation $t_{0}$ is a complier if and only if
$D_{t_{0}}\left(1\right)>D_{t_{0}}\left(0\right)$ with probability
one\textemdash if the treatment changes in response to the instrument.

We begin with the following assumption which states that each observation
is either a complier or a non-complier with certainty.
\begin{assumption}
\label{Assumption: Pure behavior}(Deterministic complier status)
For each $t$ either $\mathbb{P}\left(D_{t}\left(1\right)>D_{t}\left(0\right)\right)=1$
or $\mathbb{P}\left(D_{t}\left(1\right)>D_{t}\left(0\right)\right)=0$.
\end{assumption}
Assumption \ref{Assumption: Pure behavior} rules out cases where
$\mathbb{P}\left(D_{t}\left(1\right)>D_{t}\left(0\right)\right)=p$
for some $p\in\left(0,\,1\right)$. A non-complier cannot be characterized
by $\mathbb{P}\left(D_{t}\left(1\right)>D_{t}\left(0\right)\right)>0.$
The latter probability must be zero. It allows for continuous treatment
variables. For example, $D_{t}\left(0\right)$ may be an arbitrary
continuous random variable, and $D_{t}\left(1\right)=D_{t}\left(0\right)+c$
where $c\geq0.$ Under Assumption \ref{Assumption: Pure behavior},
Lemma \ref{Lemma: D1>D0 E(D1)>E(D0)} in the supplement shows that
$\mathbb{P}\left(D_{t_{0}}\left(1\right)>D_{t_{0}}\left(0\right)\right)=1$
is equivalent to $\mathbb{E}\left(D_{t_{0}}\left(1\right)\right)>\mathbb{E}\left(D_{t_{0}}\left(0\right)\right)$.
This equivalence implies that compliers can be identified by comparing
the expected treatment values under different instrument values.\footnote{Note that this result is different from that in Lemma 2.1 in \citet{abadie:2003}
who shows that under several assumptions the proportion of compliers
can be identified by $\mathbb{E}\left(D_{i}\left(1\right)\right)-\mathbb{E}\left(D_{i}\left(0\right)\right)$
in a cross-sectional setting. He uses this lemma to show that any
statistical characteristic that can be defined in terms of moments
of the joint distribution of $\left(Y_{i},\,D_{i},\,Z_{i}\right)$
is identified for compliers. He then remarks that it is not possible
to identify compliers individually under these assumptions.} Under mild smoothness assumptions that we discuss below, the latter
two expected values can be estimated consistently from the sample
so that we can determine whether $t_{0}$ is a complier in large samples
by looking at the corresponding inequality based on sample quantities.

Let $\mathbf{P}\subset\{1,\ldots,T\}$ denote the ``policy sample'',
the set of observations for which $Z_{t}=1$ so that $D_{t}=D_{t}(1)$
for all $t\in\mathbf{P}$, and let $\mathbf{C}=\{1,\ldots,T\}\setminus\mathbf{P}$
denote the ``control sample'', where $D_{t}=D_{t}(0)$. It is reasonable
to assume that, for a given value of the instrument, the potential
treatment assignments vary smoothly over time. Suppose we wish to
determine whether an observation $t_{0}\in\mathbf{P}$ is a complier.
Since $D_{t_{0}}\left(0\right)$ is not observed, under time-smoothness
we approximate $\mathbb{E}\left(D_{t_{0}}\left(0\right)\right)$ by
averaging nearby observations in the control sample. Letting $N_{0}(t_{0})$
denote the $n_{0}$ largest indices $s\in\mathbf{C}$ such that $s\leq t_{0}-1,$
this implies 
\[
\overline{D}_{C,t_{0}-1,n_{0}}\equiv n_{0}^{-1}\sum_{s\in N_{0}(t_{0})}D_{s}\overset{\mathbb{P}}{\rightarrow}\mathbb{E}\left(D_{t_{0}-1}\left(0\right)\right)
\]
as $n_{0}\rightarrow\infty$ with $n_{0}/|\mathbf{C}|\rightarrow0$
under mild conditions. This a standard nonparametric condition: the
window size $n_{0}$ must grow, but slowly relative to the size of
$\mathbf{C}$, so that the averaging neighborhood becomes dense on
the $\mathbf{C}$ while remaining asymptotically local. In addition,
it follows that $\mathbb{E}\left(D_{t_{0}-1}\left(0\right)\right)$
is close to $\mathbb{E}\left(D_{t_{0}}\left(0\right)\right)$. A similar
argument can be applied to $\mathbb{E}\left(D_{t_{0}}\left(1\right)\right)$
using adjacent days in the policy sample: we have $\overline{D}_{P,t_{0},n_{1}}\overset{\mathbb{P}}{\rightarrow}\mathbb{E}\left(D_{t_{0}}(1)\right)$
as $n_{1}\rightarrow\infty$ with $n_{1}/|\mathbf{P}|\rightarrow0$,
where $\overline{D}_{P,t_{0},n_{1}}=n_{1}^{-1}\sum_{s\in N_{1}(t_{0})}D_{s}$
and $N_{1}(t_{0})$ denotes the $n_{1}$ largest indices $s\in\mathbf{P}$
such that $s\leq t_{0}$. Thus, observation $t_{0}\in\mathbf{P}$
is a complier if and only if $\overline{D}_{P,t_{0},n_{1}}-\overline{D}_{C,t_{0}-1,n_{0}}\overset{\mathbb{P}}{\rightarrow}c$
as $n_{0},n_{1}\rightarrow\infty$ with $n_{0}/|\mathbf{C}|,n_{1}/|\mathbf{P}|\rightarrow0$
for $c>0$.

Intuitively, even though $D_{t_{0}}\left(0\right)$ is not observed
when $t_{0}\in\mathbf{P}$, observations close to $t_{0}$ characterized
by no FOMC announcement provide information about what $\mathbb{E}\left(D_{t_{0}}(0)\right)$
would have been in the absence of an FOMC announcement.\footnote{One could also use the observations to the right of $t_{0}$ to construct
$\overline{D}_{C,t_{0}+1,n}$, i.e., $D_{t_{0}+1},\ldots,\,D_{t_{0}+n}$.} There are about six weeks in between any two FOMC meetings, and so
$n_{0}\approx30$. Alternatively, following \citet{nakamura/steinsson:2018}
the control sample could include all Tuesdays and Wednesdays that
are not FOMC meeting days. Nevertheless, one can skip the observation
that pertains to the previous meeting, say $D_{t_{-1}}\left(0\right)$,
whose realization is not observed, and continue averaging using the
observations prior to that meeting as well to construct the average
$\overline{D}_{C,t_{0}-1,n_{0}}$ possibly applying down-weighting
for observations further in time from $t_{0}$, i.e., use $\ldots,D_{t_{-1}-1},\,D_{t_{-1}+1},\,D_{t_{-1}+2}\ldots,\,D_{t_{0}-2},\,D_{t_{0}-1}$.
Similarly, observations in $\mathbf{P}$ close to $t_{0}$ provide
information about what $\mathbb{E}\left(D_{t_{0}}\left(1\right)\right)$
would have been, though here the successive observations are separated
chronologically by the observations in the control sample $\mathbf{C}$.

The economic intuition behind this continuity assumption is that the
latent determinants of treatment take-up evolve gradually over time.
Firms do not suddenly stop responding to short-term interest rates,
and households do not suddenly become insensitive to borrowing costs.
Consequently, the propensity to receive treatment is expected to change
smoothly.

We now present the formal result for identification of the compliers.
The following two assumptions can be justified in large samples when
the mean (potential) treatment assignments in both the control and
policy samples vary smoothly over time. Under an infill asymptotic
embedding where the original observations indexed by $t=1,\ldots,\,T$
are mapped into the unit interval $[0,\,1]$ via $u=t/T$, if $\lim_{T\rightarrow\infty}\mathbb{E}(D_{Tu}(z))$
is continuous in $u$ under a fixed instrument value $z\in\mathbf{Z}$,
the following assumptions hold. This type of continuity accommodates
general forms of smoothly time-varying means but not abrupt breaks
in mean.\footnote{However, breaks in the mean of the assignment process can be estimated
under some conditions as we explain below. Then, time-smoothness is
required to hold only in regimes defined by successive break dates.}
\begin{assumption}
\label{Assumption: Local LLN}(i) For any $t\in\mathbf{C},$ $\overline{D}_{C,t,n}\overset{\mathbb{P}}{\rightarrow}\mathbb{E}\left(D_{t}\right)$
as $n\rightarrow\infty$ with $n/|\mathbf{C}|\rightarrow0.$ (ii)
For $t\in\mathbf{P}$ $\mathbb{E}(D_{t-1}\left(0\right))=\mathbb{E}(D_{t}\left(0\right))$.
\end{assumption}
\begin{assumption}
\label{Assumption Local LLN Treatment Sample}(i) For any $t\in\mathbf{P}$
$\overline{D}_{P,t,n_{}}\overset{\mathbb{P}}{\rightarrow}\mathbb{E}\left(D_{t}\right)$
as $n_{}\rightarrow\infty$ with $n/|\mathbf{P}|\rightarrow0$. (ii)
For $t\in\mathbf{C}$ $\mathbb{E}(D_{t}\left(1\right))=\mathbb{E}(D_{s^{*}\left(t\right)}\left(1\right))$
where $s^{*}\left(t\right)=\mathrm{argmin}_{s\in\mathbf{P}}|t-s|$
is the closest index in $\mathbf{P}$ to $t.$ 
\end{assumption}
Assumptions \ref{Assumption: Local LLN}-\ref{Assumption Local LLN Treatment Sample}
impose smoothness restrictions ensuring that the mean of the potential
treatment assignment evolves gradually over time under both the no-announcement
and announcement trajectories. Under an infill asymptotic embedding,
these conditions guarantee that local averages of the observed treatment
variable consistently recover the underlying conditional means. Assumption
\ref{Assumption: Local LLN}(i) requires a law of large numbers to
apply to the rolling-window sample average of $D_{t}$ at the points
of continuity of $\mathbb{E}\left(D_{t}\right)$. It is a minimal
technical assumption. Assumption \ref{Assumption: Local LLN}(ii)
strengthens part (i) a bit by requiring that for $t\in\mathbf{P}$
the potential treatment assignment under the trajectory $Z_{t}=0$
has a locally constant mean. Assumption \ref{Assumption Local LLN Treatment Sample}(i)
adapts Assumption \ref{Assumption: Local LLN}(i) to the observations
in $\mathbf{P}$. This is a stronger assumption since two successive
observations in the policy sample are separated by several observations
in the control sample. Assumption \ref{Assumption Local LLN Treatment Sample}(ii)
requires that $\mathbb{E}\left(D_{t}\left(1\right)\right)$ for $t\in\mathbf{C}$
is equal to the mean of the potential treatment assignment at the
closest date in the policy sample $s^{*}\left(t\right)$. This is
a first moment constancy assumption on the potential treatment assignment
under the trajectory $Z_{t}=1$. Assumption \ref{Assumption: Local LLN}
is used to identify the compliers in $\mathbf{P}$, while Assumption
\ref{Assumption Local LLN Treatment Sample} is used to identify the
compliers in $\mathbf{C}$. 

 Assumptions \ref{Assumption: Local LLN}-\ref{Assumption Local LLN Treatment Sample}
are implied by stationarity and local stationarity conditions commonly
imposed in the analysis of economic time series. Together, these assumptions
formalize the idea that observations close in time provide reliable
information about the counterfactual evolution of the treatment process.
This, in turn, enables identification of compliers through local comparisons
between control and policy subsamples.
\begin{thm}
\label{Theorem: Compliers}Let Assumptions \ref{Assumption: Pure behavior}-\ref{Assumption Local LLN Treatment Sample}
hold and $n_{0},n_{1}\rightarrow\infty$ with $n_{0}/|\mathbf{C}|,n_{1}/|\mathbf{P}|\rightarrow0$.
Then:\\
 (i) $t\in\mathbf{P}$ is a complier if and only if $\overline{D}_{P,t,n_{1}}-\overline{D}_{C,t-1,n_{0}}\overset{\mathbb{P}}{\rightarrow}c$
where $c>0$. \\
 (ii) $t\in\mathbf{C}$ is a complier if and only if $\overline{D}_{P,s^{*}\left(t\right),n_{1}}-\overline{D}_{C,t,n_{0}}\overset{\mathbb{P}}{\rightarrow}\widetilde{c}$
where $\widetilde{c}>0$. 
\end{thm}
Theorem \ref{Theorem: Compliers} shows that the compliers can be
identified individually through local comparisons of realized treatment
assignments between the policy sample $\mathbf{P}$ and the control
sample $\mathbf{C}$. For a policy-sample date $t\in\mathbf{P}$,
a complier corresponds to the treatment assignment at $t$ exceeding,
in probability, the counterfactual treatment level inferred from nearby
control-sample observations. This is captured by a strictly positive
limiting gap between a local average taken within $\mathbf{P}$ and
the corresponding local average within $\mathbf{C}.$ For non-intervention
dates $t\in\mathbf{C}$, a complier is characterized by the existence
of a nearby policy-sample date whose local treatment assignment exceeds
the local control-sample average at $t.$ Economically, the theorem
formalizes that a complier is a date where the policy shock genuinely
shifts the path of the treatment variable relative to what would have
occurred absent the shock. Because the assumptions guarantee smooth
evolution of the counterfactual treatment path, local averages in
the control sample provide valid proxies for the counterfactual $D_{t}\left(z\right)$.
Thus, identifying compliers reduces to detecting persistent, sign-consistent
deviations of the observed treatment path from its locally inferred
counterfactual trajectory.

To the best of our knowledge, there is no equivalent result in the
cross-sectional setting. The assumptions of the theorem are easily
satisfied in time series applications. Using Theorem \ref{Theorem: Compliers}
is straightforward: one computes the difference between two sample
averages and checks whether it is greater than zero. Given the sampling
uncertainty associated with the two averages, one can conduct inference
using a $t$-statistic for the null hypothesis $\mathbb{E}\left(D_{t_{0}}\left(1\right)\right)-\mathbb{E}\left(D_{t_{0}}\left(0\right)\right)=0$
($t_{0}$ is not a complier) versus the alternative hypothesis that
$\mathbb{E}\left(D_{t_{0}}\left(1\right)\right)-\mathbb{E}\left(D_{t_{0}}\left(0\right)\right)>0$
($t_{0}$ is a complier).

In the supplement we extend Theorem \ref{Theorem: Compliers} to the
case of continuous IV $Z_{t}$. This identification argument is quite
general and applies well beyond the setting of FOMC announcements.
In particular, it covers any narrative identification approach in
which $Z_{t}$ is interpreted as a shock constructed from narrative
records (e.g., military spending shocks, tax shocks, oil-supply shocks,
natural disasters, etc.). We explicitly demonstrate in the supplement
how Theorem \ref{Theorem: Compliers} extends to \citeauthor{ramey:2011}'s
\citeyearpar{ramey:2011} identification fiscal multipliers based
on military spending news and to \citeauthor{romer/romer:2004}'s
\citeyearpar{romer/romer:2004} narrative measure of monetary policy
shocks. Finally, our framework also applies to cross-sectional settings
based on spatial data by replacing continuity over time with continuity
over geographical space.

An additional challenge specific to the $\pi$-LATE framework is that
the set of observations with a first stage $\mathbf{S}_{0,T}$ is
also unknown. However, as the following result states, under Assumption
\ref{Assumption: Monotonicity}, in the absence of covariates $\widetilde{V}_{t}$,
$\mathbf{S}_{0,T}$ is equal to the (identified) set of compliers
\textemdash i.e., observations for which the first-stage holds individually.
\begin{prop}
\label{Proposition: Compliers FS indivdually}Suppose $Z_{t}$ is
binary and let Assumptions \ref{Assumption: First-Stage} without
conditioning on $\widetilde{V}_{t}$, \ref{Assumption: Monotonicity}
and \ref{Assumption: Pure behavior} hold. Then, the set of compliers
coincide with $\mathbf{S}_{0,T}$.
\end{prop}
Knowledge of the compliers sub-population (and hence of the non-compliers
sub-population) can be used to test the exclusion restriction (cf.
Assumption \ref{Assumption: Exclusion }) by comparing the mean outcomes
of groups of non-compliers across different values of the instrument.
Intuitively, for a non-complier the treatment status is not affected
by changes in the instrument. One can divide any large subset of non-compliers
into two groups according to their assignment status. If one can reject
the hypothesis that the average outcomes in these two groups is the
same, then the exclusion restriction cannot hold.

Under Assumption \ref{Assumption: Monotonicity} with $D_{t}(1)\geq D_{t}(0)$,
the set of non-compliers is $\mathcal{NC}=\{t\in\{1,\ldots,T\}:D_{t}(1)=D_{t}(0)=D_{t}\}$.
Let $\mathcal{NC}^{s}$ be any non-empty subset of $\mathcal{NC}$
such that $\mathcal{NC}_{\mathbf{P}}^{s}=\mathcal{NC}^{s}\cap\mathbf{P}\neq\emptyset$
and $\mathcal{NC}_{\mathbf{C}}^{s}=\mathcal{NC}^{s}\cap\mathbf{C}\neq\emptyset$.
We can test the exclusion restriction in Assumption \ref{Assumption: Exclusion }
under the following assumption on the subsets $\mathcal{NC}_{\mathbf{P}}^{s}$
and $\mathcal{NC}_{\mathbf{C}}^{s}$. 
\begin{assumption}
\label{Assumption: Non-Complier LLN}(i) $\mathbb{E}[Y_{t}^{*}(D_{t},z)|t\in\mathcal{NC}_{\mathbf{P}}^{s}]=\mathbb{E}[Y_{r}^{*}(D_{r},z)|r\in\mathcal{NC}_{\mathbf{C}}^{s}]$
for all $t,r\geq1$, and $z\in\mathbf{Z}$. (ii) For $\mathbf{R}=\mathbf{C}$
or $\mathbf{P}$, $|\mathcal{NC}_{\mathbf{R}}^{s}|^{-1}\sum_{t\in\mathcal{NC}_{\mathbf{R}}^{s}}Y_{t}\overset{\mathbb{P}}{\rightarrow}\mathbb{E}\left[Y_{t}|t\in\mathcal{NC}_{\mathbf{R}}^{s}\right]$
as $|\mathcal{NC}_{\mathbf{R}}^{s}|\rightarrow\infty$.
\end{assumption}
Condition (i) states that the potential outcome for non-compliers
is mean-stationary and the mean is the same across control and policy
subsamples. Condition (ii) states that a law of large numbers holds
for non-compliers observations in both the control and policy subsamples.
As long as the policy sample does not tend to contain systematic different
values of the policy variable $D_{t}$ among non-compliers than the
control sample, these are relatively mild conditions.\footnote{The assumption is stated for the case $h=0$. However, it can be immediately
generalized to any $h>0$. }
\begin{prop}
\label{Proposition: exclusion-restriction}Suppose $Z_{t}$ is binary
and let Assumptions \ref{Assumption: Monotonicity} and \ref{Assumption: Non-Complier LLN}
hold. If Assumption \ref{Assumption: Exclusion } holds, then as $|\mathcal{NC}_{\mathbf{P}}^{s}|,|\mathcal{NC}_{\mathbf{C}}^{s}|\rightarrow\infty$,
\[
|\mathcal{NC}_{\mathbf{P}}^{s}|^{-1}\sum_{t\in\mathcal{NC}_{\mathbf{P}}^{s}}Y_{t}-|\mathcal{NC}_{\mathbf{C}}^{s}|^{-1}\sum_{t\in\mathcal{NC}_{\mathbf{C}}^{s}}Y_{t}\overset{\mathbb{P}}{\rightarrow}0.
\]
\end{prop}
Using Proposition \ref{Proposition: exclusion-restriction} to test
Assumption \ref{Assumption: Exclusion } is simple: since non-compliers
can be identified individually using Theorem \ref{Theorem: Compliers},
one can immediately compute the sample averages specified in Proposition
\ref{Proposition: exclusion-restriction} and conduct inference using
a $t$-statistic for the null hypothesis that the population mean
of $t\in\mathcal{NC}_{\mathbf{P}}^{s}$ is equal to that of $t\in\mathcal{NC}_{\mathbf{C}}^{s}$.
The researcher has the ability to choose the subset of non-compliers
$\mathcal{NC}^{s}$ when implementing this test. The simplest choice
is to set $\mathcal{NC}^{s}=\mathcal{NC}$, however, the researcher
also has the ability to direct the power of the test toward particular
types of non-compliers they may suspect of being more likely to violate
the exclusion restriction. 

\section{\label{Section: Application: Money Neutrality}High-Frequency Identification
of Monetary Policy Effects}

To study the effects of monetary policy on real variables, a large
literature has relied on high-frequency identification. This exploits
the fact that at the time of an FOMC meeting a large amount of economic
news is revealed. Here we discuss \citeauthor{rigobon:2003}'s \citeyearpar{rigobon:2003}
heteroskedasticity identification approach which uses a 1-day window
{[}see, e.g., \citet{nakamura/steinsson:2018}{]}, and can be reformulated
as IV-based identification. In Section \ref{Subsection: Heteroske Ident}
we explain when the resulting reduced-form estimands have a causal
meaning within the potential outcome framework of Section \ref{Section: Statistical Framework for Identification of Causal Effects}.
In Section \ref{Subsection: Weak-or-Lack of Identification and the pi-LATE}
we discuss the weak identification problem of current approaches and
show how the $\pi$-LATE framework can be used to strengthen identification.

\subsection{\label{Subsection: Heteroske Ident}Heteroskedasticity-Based Identification}

Consider the following system of equations: 
\begin{align}
\tilde{Y}_{t} & =\beta_{0}\tilde{D}_{t}+\eta_{t},\qquad\mathrm{and}\qquad\tilde{D}_{t}=a\tilde{Y}_{t}+e_{t},\label{Eq. (1) NS and RS, Outcome variable}
\end{align}
where $\tilde{Y}_{t}$ is the (demeaned) daily change in an outcome
variable, (e.g., an asset price or a bond yield) and $\tilde{D}_{t}$
is the (demeaned) daily change in the unexpected component of a short-term
interest rate or policy news (e.g., $\Delta i_{t}$ as discussed after
Proposition \ref{Proposition: Continuous pi-LATE}), $\eta_{t}$ is
a shock to $\tilde{Y}_{t}$, $e_{t}$ is the monetary policy shock
and $a$ and $\beta_{0}$ are scalar parameters. The errors $\eta_{t}$
and $e_{t}$ have no serial correlation and are mutually uncorrelated.
The parameter of interest is $\beta_{0}$ which represents the causal
effect of monetary policy on the outcome variable. The model in \eqref{Eq. (1) NS and RS, Outcome variable}
could arise from a bivariate VAR. In fact, one could add a vector
$X_{t}$ of exogenous variables to the model in \eqref{Eq. (1) NS and RS, Outcome variable}.
However, to focus on the main intuition, we follow \citet{nakamura/steinsson:2018}
and we omit $X_{t}$ and lagged terms of $\tilde{Y}_{t}$ and $\tilde{D}_{t}$.
See \citet{casini/mccloskey:2024} for a detailed discussion of why
the lags can be omitted in this setting.

The model in \eqref{Eq. (1) NS and RS, Outcome variable} is a special
case of the generalized framework studied in Section \ref{Section: Statistical Framework for Identification of Causal Effects}.
It is useful because it directly motivates a particular IV estimand.
However, we study the causal interpretation of this estimand in the
general case for which the linear model with stable parameters is
not the correct specification.

Heteroskedasticity-based identification requires that the variance
of the monetary shock increases in the days of FOMC announcements,
while the variance of other shocks is unchanged. Let $T_{P}$ denote
the number of days containing an FOMC announcement (policy sample),
and let $T_{C}$ denote the number of days that do not contain an
FOMC announcement (control sample). Let $\sigma_{e,P}^{2}=T_{P}^{-1}\sum_{t\in\mathbf{P}}\mathbb{E}\left(e_{t}^{2}\right)$
and $\sigma_{e,C}^{2}=T_{C}^{-1}\sum_{t\in\mathbf{C}}\mathbb{E}\left(e_{t}^{2}\right)$
be the average variance of the monetary policy shock in the policy
and control samples. Define $\sigma_{\eta,P}^{2}$ and $\sigma_{\eta,C}^{2}$
similarly. Then, the identification conditions are 
\begin{align}
\sigma_{e,P} & >\sigma_{e,C}\qquad\mathrm{and}\qquad\sigma_{\eta,P}=\sigma_{\eta,C}.\label{Eq. Vol_P >Vol_C}
\end{align}
The condition $\sigma_{e,P}>\sigma_{e,C}$ is the relevance condition
while $\sigma_{\eta,P}=\sigma_{\eta,C}$ is the exclusion restriction.
 In the supplement we show that the resulting Wald estimand is 
\begin{align}
\beta_{\pi,t,0}^{*} & =\frac{\mathbb{E}\left(\tilde{D}_{t}\tilde{Y}_{t}|\,Z_{t}=1\right)-\mathbb{E}\left(\tilde{D}_{t}\tilde{Y}_{t}|\,Z_{t}=0\right)}{\mathbb{E}\left(\tilde{D}_{t}^{2}|\,Z_{t}=1\right)-\mathbb{E}\left(\tilde{D}_{t}^{2}|\,Z_{t}=0\right)},\label{Eq. (beta0*)}
\end{align}
which corresponds to the Wald estimand \eqref{Eq. (beta_j(v)) pi-LATE}
for $h=0$, $Y_{t}=\tilde{D}_{t}\tilde{Y}_{t}$, $D_{t}=\tilde{D}_{t}^{2}$
and no conditioning variable $\widetilde{V}_{t}$.  The following
corollary of Proposition \ref{Proposition: Continuous pi-LATE} presents
the causal meaning of $\beta_{\pi,t,0}^{*}$ under the general setting
of Section \ref{Section: Statistical Framework for Identification of Causal Effects}.
\begin{cor}
\label{Corollary: pi-LATE HET}$($LATE in heteroskedasticity-based
identification$)$ Let Assumptions \ref{Assumption: Independence}-\ref{Assumption KP 2024}
hold for $Y_{t}=\tilde{D}_{t}\tilde{Y}_{t}$ and $D_{t}=\tilde{D}_{t}^{2}$
with $\tilde{D}_{t}(1)^{2}\geq\tilde{D}_{t}(0)^{2}$. For $t\in\mathbf{S}_{0,T}$,
we have 
\begin{align}
\beta_{\pi,t,0}^{*} & =\frac{\int_{\mathbf{D}}\mathbb{E}\left(\left.\frac{\partial\left(\tilde{d}\tilde{Y}_{t,0}^{*}\left(\tilde{d}^ {}\right)\right)}{\partial(\tilde{d}^{2})}\right|\tilde{D}_{t}(1)^{2}\geq\tilde{d}^{2}\geq\tilde{D}_{t}(0)^{2}\right)\mathbb{P}\left(\tilde{D}_{t}(1)^{2}\geq\tilde{d}^{2}\geq\tilde{D}_{t}(0)^{2}\right)\mathrm{d}(\tilde{d}^{2})}{\int_{\mathbf{D}}\mathbb{P}\left(\tilde{D}_{t}(1)^{2}\geq\tilde{d}^{2}\geq\tilde{D}_{t}(0)^{2}\right)\mathrm{d}(\tilde{d}^{2})}.\label{Eq. (beta0*) LATE}
\end{align}
\end{cor}
Corollary \ref{Corollary: pi-LATE HET} shows that the Wald estimand
in \eqref{Eq. (beta0*)} has a causal meaning because it is the ratio
of a reduced-form generalized impulse response of $\tilde{D}_{t}\tilde{Y}_{t}$
to a first-stage generalized impulse response of $\tilde{D}_{t}^{2}$.
More specifically, $\beta_{\pi,t,0}^{*}$ identifies a weighted average
of the derivative of the product between the potential outcome and
policy variable for compliers. Hence, contrary to popular belief,
the causal interpretation of the heteroskedasticity-based estimator
(i.e., Rigobon's estimator) is not the same as that of a standard
IV estimator\textemdash though it remains local in nature as it averages
over compliers.\footnote{We also note that the interpretation becomes even more complex over
the region when $\widetilde{D}_{t}$ changes sign. The interpretation
of the marginal causal effect becomes an average over the two signs
of $\widetilde{D}_{t}$. } We continue to refer to it as LATE with the understanding that it
is a LATE for $\tilde{D}_{t}\tilde{Y}_{t}$, not $\tilde{Y}_{t}$
itself.

Here the compliers are the observations for which the announcement
induces a higher volatility of the policy $\tilde{D}_{t}$. In contrast,
the non-compliers are characterized by idiosyncratic or general equilibrium
factors that dominate the news specific to the announcement. That
is, regimes where $\tilde{D}_{t}^{2}$ remains low regardless of the
presence of an announcement correspond to ``never-takers,'' while
regimes where $\tilde{D}_{t}^{2}$ remains high even in the absence
of an announcement correspond to ``always-takers.'' Noting that
$D_{t}=\tilde{D}_{t}^{2}$ in this context, we can apply Theorem \ref{Theorem: Compliers}
to identify the compliers individually. We do so in the empirical
application in Section \ref{Section: Empirical Evidence}.

It is important to consider how the interpretation of the causal effect
identified by ${\beta}_{\pi,t,0}^{*}$ in Corollary \ref{Corollary: pi-LATE HET}
varies with the functional relationship between $\tilde{Y}_{t}$ and
$\tilde{D}_{t}$. Let us begin with the linear case with stable parameters
as in \eqref{Eq. (1) NS and RS, Outcome variable}. From \eqref{Eq. (beta0*)},
simple algebra shows that ${\beta}_{\pi,t,0}^{*}$ reduces to $\beta_{0}$
when the denominator of \eqref{Eq. (beta0*) LATE} is nonzero, which
means that Rigobon's estimator identifies the causal effect of the
policy (i.e., the slope coefficient in \eqref{Eq. (1) NS and RS, Outcome variable}).
This result does not generally extend to the case where $\beta_{0}$
is time-varying or the first-stage is zero. At most one could identify
a $\pi$-LATE. We will return to this in Section \ref{Subsection: Weak-or-Lack of Identification and the pi-LATE}.

Let us turn to analyzing the consequences of nonlinearities. When
$D_{t}$ and the shock $\eta_{t}$ are additively separable (i.e.,
$Y_{t}=\varphi_{D}(D_{t})+\varphi_{\eta}\left(\eta_{t}\right)$ for
some nonlinear functions $\varphi_{D}\left(\cdot\right)$ and $\varphi_{\eta}\left(\cdot\right)$),
\citet{kolesar/plagborgmoller:2025} show that the estimand resulting
from a regression of $Y_{t}$ on $D_{t}$ using $Z_{t}=(W_{t}-\mathbb{E}\left(W_{t}\right))D_{t}$
as an instrument for which $\mathrm{Cov}\left(D_{t}^{2},\,W_{t}\right)\neq0$
identifies a weighted average of marginal effects of the policy shock
$e_{t}$ with weights that are not guaranteed to be positive. As a
result, the researcher may infer an incorrect sign for the marginal
effects. Thus, this estimand is not weakly causal {[}cf. \citet{blandhol/bonney/mogstad/torgovitsky:2025}.\footnote{\citet{kitagawa/wang/xu:2015} show how a reasonable economic interpretation
can potentially be restored when the weights are negative.} The authors also note that for the case $Y_{t}=e_{t}\varphi_{\eta}\left(\eta_{t}\right)$
with $\mathbb{E}[\varphi_{\eta}\left(\eta_{t}\right)]=0$ and $e_{t}\bot\eta_{t}$
the estimand is nonzero while the true causal effect of the policy
shock is zero since $\mathbb{E}\left[Y_{t}|\,e_{t}\right]=0$.

Corollary \ref{Corollary: pi-LATE HET} provides even more negative
news about the effect of nonlinearities for heteroskedasticity-based
identification than that shown by \citet{kolesar/plagborgmoller:2025}:
in a general nonparametric model, Corollary \ref{Corollary: pi-LATE HET}
implies that Rigobon's Wald estimand ${\beta}_{\pi,t,0}^{*}$, which
is in general different from the IV estimand examined by \citet{kolesar/plagborgmoller:2025},
does not necessarily equal a weighted average of marginal effects.
The intuition is that the instrument affects $\mathrm{Var}\left(D_{t}\right)$
and not $\mathbb{E}\left(D_{t}\right)$, so variation in the instrument
induces exogenous variation in $D_{t}^{2}$, which has a causal effect
on $D_{t}Y_{t}$ not just $Y_{t}$. In short, it is generally difficult
to interpret ${\beta}_{\pi,t,0}^{*}$ when the true model is nonlinear.
Thus, we concur with the recommendation of \citet{kolesar/plagborgmoller:2025}
that the linearity assumption should be checked carefully when using
heteroskedasticity-based identification. This is likely even more
important in the context of SVARs and local projections than in the
 current event study setting since the former aggregates data over
a month or a quarter while the latter uses relatively higher frequency
data (e.g., a 30-minute or 1-day change in policy and outcome variables
around an announcement), where linearity may be a more credible assumption
since a nonlinear function can be locally well approximated by a linear
one.

\subsection{\label{Subsection: Weak-or-Lack of Identification and the pi-LATE}Weak
or Lack of Identification and the Usefulness of $\pi$-LATE}

The key identification condition that the volatility of the policy
variable is higher during FOMC announcement days appears reasonable
in principle, since each announcement day is likely to be associated
with substantial monetary news. However, the volatility of monetary
policy variables can be high for other reasons. There are multi-year
periods during which the volatility of several macroeconomic variables
is elevated. In this case, general equilibrium factors dominate the
news specific to the announcement. For example, during the 2007-09
financial crisis and the Covid-19 pandemic, volatility was high across
many macroeconomic and financial variables. These facts pose serious
challenges for identification, as the first-stage condition may not
hold for all $t$. To see this, examine the denominator in \eqref{Eq. (beta0*)}.
If the first-stage does not hold for all $t$, we may have 
\begin{align}
T_{P}^{-1}\sum_{t\in\mathbf{P}}\mathrm{Var}\left(\tilde{D}_{t}\right)-T_{C}^{-1}\sum_{t\in\mathbf{C}}\mathrm{Var}\left(\tilde{D}_{t}\right) & \thickapprox0,\label{Eq. VarP-VarC =00003D00003D 0}
\end{align}
which would render the estimate of the average treatment effect highly
imprecise.

\begin{center}
\begin{figure}[H]
\includegraphics[width=16cm,totalheight=8cm]{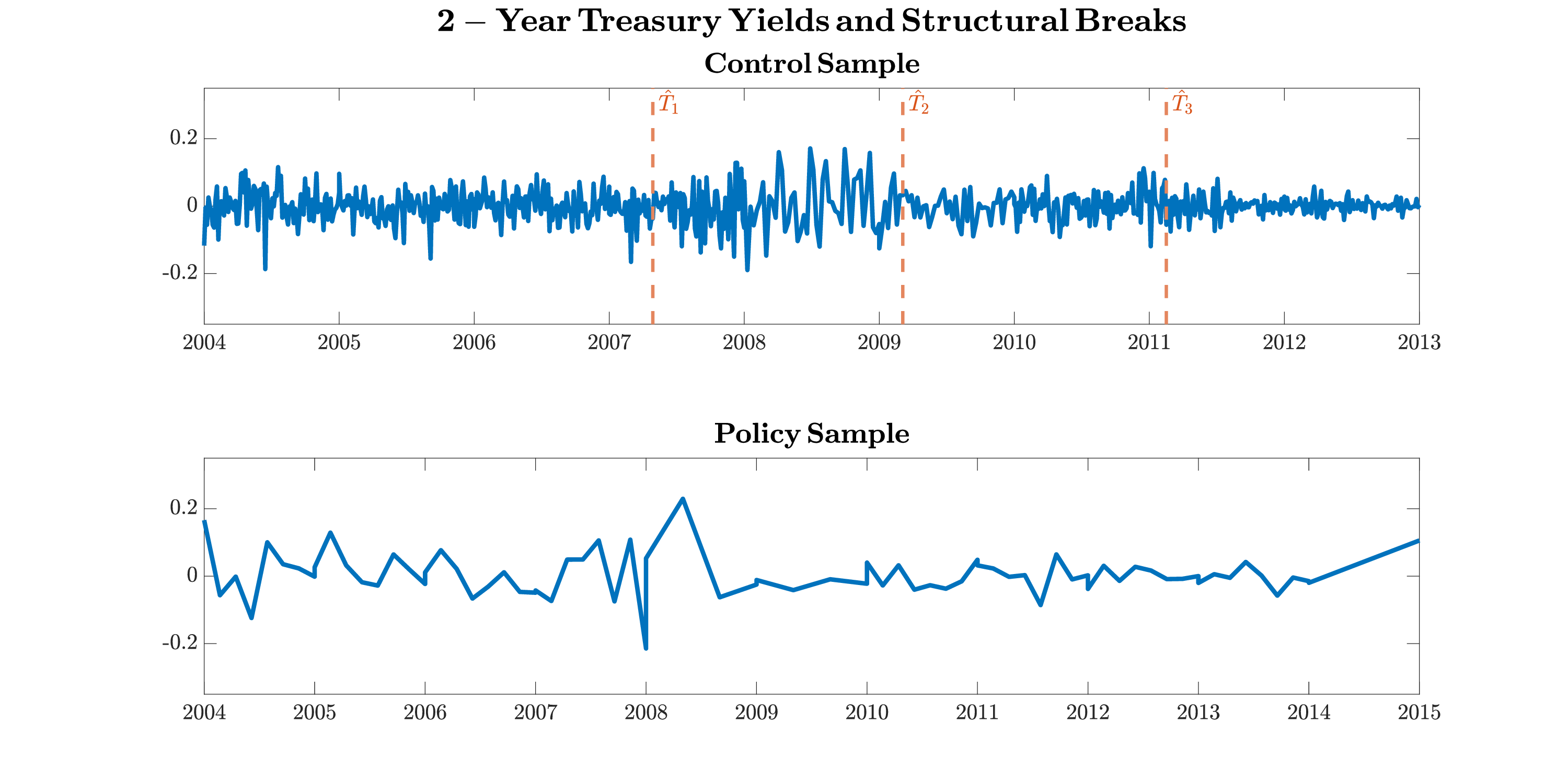}

\caption{{\small\label{Figure_Plot_Dt}}{\scriptsize Plot of 2-years Treasury
yields in control (top panel) and policy sample (bottom panel). Vertical
broken lines are the estimated break dates using \citeauthor{casini/perron:change-point-spectra}\textquoteright s
\citeyearpar{casini/perron:change-point-spectra} test.}}
\end{figure}
\end{center}

Using an $F$-test for weak identification, \citet{lewis:2020} shows
that the monetary policy effects based on a 1-day window in \citet{nakamura/steinsson:2018}
appear to be weakly-identified. We show that this arises from significant
time variation in the volatility of the policy variable within both
policy and control samples. Figure \ref{Figure_Plot_Dt} plots $\tilde{D}_{t}$
(2-Year Treasury yields) for the control and policy samples. The policy
sample includes all regularly scheduled FOMC meeting days from 1/1/2000
to 3/19/2014. The control sample includes all Tuesdays and Wednesdays
that are not FOMC meeting days between 1/1/2000 and 12/31/2012.

There appear to be multiple volatility regimes. Using the structural
break test from \citet{casini/perron:change-point-spectra}, which
allows for stable or smoothly varying volatility under the null and
abrupt breaks under the alternative, we detect three breaks in the
control sample. The first break (April 24, 2007) marks the start of
the 2007-09 financial crisis. The second (July 28, 2009) captures
the crisis period itself, characterized by the highest volatility.
Afterward, volatility returns to pre-crisis levels until the third
break (February 2, 2011), which aligns with the zero lower bound (ZLB)
period and the start of unconventional monetary policy. The final
regime shows the lowest volatility, reflecting initial policy effects
and stabilization.\footnote{We do not test for breaks in the policy sample due to small size ($T_{P}=74$),
treating it as a single regime.}

These findings show significant time variation in $\mathrm{Var}(\tilde{D}_{t})$.
In the second regime, control-sample volatility is close to the policy-sample
average, contributing to the weak identification in \eqref{Eq. VarP-VarC =00003D00003D 0}.
\citet{nakamura/steinsson:2018} find their estimates imprecise and
not economically meaningful for some of the interest rates they use
as outcome variables. \citet{lewis:2020} reports a first-stage $F$-statistic
of 8.11\textemdash well below the 23 critical value\textemdash suggesting
weak identification.

We propose to focus on $\pi$-LATE. The fraction $\pi_{0}$ of the
sample (i.e., all $t\in\mathbf{S}_{0,T}$) that has a first-stage
corresponds to the regimes in the control sample where $\mathrm{Var}(\tilde{D}_{t})$
is low (relative to its average level). For example, it is likely
that the regime $[\widehat{T}_{1}+1,\,\widehat{T}_{2}]$ does not
belong to $\mathbf{S}_{0,T}$ since $\mathrm{Var}(\tilde{D}_{t})$
within this regime appears close to the average volatility in the
policy sample. By construction, it is easier to identify $\pi$-LATE
than full sample LATE. The usefulness of $\pi$-LATE depends on the
magnitude of $\pi_{0}$: a small $\pi_{0}$ implies that identification
is achievable only in a small portion of the population, whereas a
large $\pi_{0}$ indicates that the identified $\pi$-LATE is representative
of a substantial part of the population.\footnote{It is possible that in practice the $\pi_{0}$ fraction of the sample
contains a mixture of strong and weak identification. We discuss weak
identification in the context of $\pi$-LATE formally in Section \ref{Section: Inference}.}

The $\pi$-LATE parameter is the same as the LATE parameter in Section
\ref{Subsection: Heteroske Ident} but instead of supposing that a
first-stage exists, only uses observations with a nonzero first-stage.
Let $T_{P,S}$ denote the number of days in $\mathbf{S}_{0,T}$ that
contain an FOMC announcement, and let $T_{C,S}$ the number of days
in $\mathbf{S}_{0,T}$ that do not contain an FOMC announcement. This
means $T_{P,S}+T_{C,S}=\pi_{0}T$.\footnote{For notational simplicity we assume that $\pi_{0}T$ is an integer
so that we avoid using the notation $\left\lfloor \pi_{0}T\right\rfloor $,
where $\left\lfloor \cdot\right\rfloor $ denotes the largest smaller
integer function.} Let $\mathbf{P}_{\mathbf{S}}=\mathbf{P}\cap\mathbf{S}_{0,T}$ and
$\mathbf{C}_{\mathbf{S}}=\mathbf{C}\cap\mathbf{S}_{0,T}$. The Wald
estimand is 
\begin{align}
\widetilde{\beta}_{\pi,t,0}^{*} & =\frac{\mathbb{E}\left(\tilde{D}_{t}\tilde{Y}_{t}|t\in\mathbf{P}_{\mathbf{S}}\right)-\mathbb{E}\left(\tilde{D}_{t}\tilde{Y}_{t}|t\in\mathbf{C}_{\mathbf{S}}\right)}{\mathbb{E}\left(\tilde{D}_{t}^{2}|t\in\mathbf{P}_{\mathbf{S}}\right)-\mathbb{E}\left(\tilde{D}_{t}^{2}|t\in\mathbf{C}_{\mathbf{S}}\right)},\label{Eq. (beta0*) pi}
\end{align}
 to which Corollary \ref{Corollary: pi-LATE HET} immediately applies
without the (now redundant) qualifier ``for $t\in\mathbf{S}_{0,T}$.''
 $\pi$-LATE is the average treatment effect for the sub-population
for which a first-stage holds: observations for which $\tilde{D}_{t}^{2}$
is induced to be higher by the announcement (i.e., the sub-population
of compliers in $\mathbf{S}_{0,T})$.

If the treatment effect is constant across the population {[}e.g.,
as in \eqref{Eq. (1) NS and RS, Outcome variable}{]}, then the $\pi$-LATE
for the sub-population $\mathbf{S}_{0,T}$ is equal to both the LATE
and ATE in the full population. To determine which treatment effect
is identified, we must determine which parts of the sample belong
to $\mathbf{S}_{0,T}$. We discuss this in Sections \ref{Section: ID Failure Test}-\ref{Section: Estimation}.

\section{\label{Section: ID Failure Test}Testing for Full Population Identification
Failure}

In this section, we introduce a test of the null hypothesis that no
subpopulation exists for which a LATE can be identified, even weakly.
In other words, the test assesses whether identifying a sub-population
LATE is possible at all. However, we strongly caution against using
this as a pretest before estimation or inference, as doing so may
introduce pretest bias and invalidates standard inference unless the
inference method is modified to account for the pretest {[}see, e.g.,
\citet{andrews:2018}{]}. Instead, the test should be viewed as a
diagnostic tool for evaluating whether there is evidence of identifiable
sub-population LATEs in a given application. We apply it for this
purpose to several existing studies that appear to face identification
challenges. Notably, such a pretest is unnecessary for conducting
identification-robust inference on sub-population LATEs, which we
discuss in Section \ref{Section: Inference}.

In accord with the analysis of Section \ref{Section: Statistical Framework for Identification of Causal Effects},
consider an IV regression model with a single endogenous variable
and multiple instruments. In matrix format, the structural equation
is 
\begin{equation}
Y=D\beta+X\gamma_{1}+u,\qquad t=1,\ldots,\,T,\label{Eq. Structural eq. IV model}
\end{equation}
where $Y$ is a $T\times1$ vector of outcome variables, $D$ is $T\times1$
vector of endogenous variables, $X$ is a $T\times p$ matrix of $p$
exogenous regressors, $u$ is a $T\times1$ vector of error terms,
and $\beta\in\mathbb{R}$ and $\gamma_{1}\in\mathbb{R}^{p}$ are unknown
parameters. The reduced-form equation is 
\begin{align}
D_{t} & =Z'_{t}\theta\mathbf{1}\{t\in\mathbf{S}_{0,T}\}+X'_{t}\gamma_{2}+e_{t},\label{Eq. Reduced Form Eq. IV model}
\end{align}
where $Z_{t}$ is a $q\times1$ vector of instruments, $e_{t}$ is
an error term, and $\theta\in\mathbb{R}^{q}$ and $\gamma_{2}\in\mathbb{R}^{p}$
are unknown parameters. For $t\notin\mathbf{S}_{0,T}$, the instrument
$Z_{t}$ is irrelevant. For $t\in\mathbf{S}_{0,T}$, the instrument
$Z_{t}$ is relevant if $\theta\neq0$. We assume that $|\mathbf{S}_{0,T}|=\pi_{0}T$
for some $\pi_{0}\in(0,1]$, noting that this is without loss of generality
since it does not rule out complete identification failure which occurs
when $\theta=0$ for any $\pi_{0}\in(0,1]$. The hypothesis testing
problem is 
\[
H_{\theta,0}:\,\theta=0\quad\mathrm{versus}\quad H_{\theta,1}:\,\theta\neq0.
\]
We discuss both the cases for which the sub-population $\mathbf{S}_{0,T}$
is known and unknown. For the sake of the exposition, we focus on
homogeneous $\theta$ in $\mathbf{S}_{0,T}$.\footnote{We could allow for $\theta_{t}\neq0$ for $t\in\mathbf{S}_{0,T}$
at the expense of additional notation and longer proofs, though the
key insights would not change. Actually, the computational procedures
we develop to implement our methods allow $\theta_{t}\neq0$ for $t\in\mathbf{S}_{0,T}$.}

Consider the $\left(\pi T\times T\right)$ selection matrix $S_{T}$
that selects the $\pi T$ rows of a matrix corresponding to the indices
in $\mathbf{S}_{T}$. That is, for an arbitrary $T\times k$ matrix
$A$, $S_{T}A$ is the $\left(\pi T\times k\right)$ matrix whose
elements are the rows of $A$ that correspond to the indices in $\mathbf{S}_{T}$.
For example, if $\mathbf{S}_{T}=\left\{ 1,\ldots,\,0.25T,\,0.75T+1,\ldots,\,T\right\} $,
\[
S_{T}A=\left[A^{\left(1,:\right)\prime}:\cdots:A^{\left(0.25T,:\right)\prime}:A^{\left(0.75T+1,:\right)\prime}:\cdots:A^{\left(T,:\right)\prime}\right]',
\]
where $A^{\left(r,:\right)}$ denotes the $r^{th}$ row of the matrix
$A$. Using the standard projection matrix notation, $P_{A}=A(A^{\prime}A)^{-1}A^{\prime}$
and $M_{A}=I-P_{A}$, let $\widetilde{A}(S_{T})=M_{S_{T}X}S_{T}A$
for any arbitrary $T\times k$ matrix $A$. The following $F$ test
statistic is useful for testing whether $\theta=0$ in the regression
\eqref{Eq. Reduced Form Eq. IV model} when the sub-population $\mathbf{S}_{0,T}$
is known: 
\begin{align*}
F_{T}\left(\mathbf{S}_{T}\right) & =\frac{\widetilde{D}\left(S_{T}\right)'\widetilde{Z}\left(S_{T}\right)\widehat{J}(S_{T})^{-1}\widetilde{Z}\left(S_{T}\right)'\widetilde{D}\left(S_{T}\right)}{q\left(\pi T-p-q\right)},
\end{align*}
for $\mathbf{S}_{T}=\mathbf{S}_{0,T}$ and $Z=[Z_{1}:\cdots:Z_{T}]^{\prime}$
and $\widehat{J}(S_{T})$ a consistent estimate of the long-run variance
$\lim_{T\rightarrow\infty}(T\pi)^{-1}\mathrm{Var}(\widetilde{Z}(S_{T})'S_{T}e)$
with $e=[e_{1}:\cdots:e_{T}]^{\prime}$. HAC or DK-HAC estimators
can be used to estimate the long-run variance {[}cf. \citet{andrews:91},
\citet{casini_hac} and \citet{newey/west:87}{]}.

For the case of an unknown sub-population, we follow the structural
break literature and search for maximal identification strength over
all sub-populations of minimal size $\pi_{L}T$ that can be partitioned
into $m$ distinct smaller sub-populations, where $\pi_{L}>0$ and
$1\leq m\leq m_{+}$ for some upper bound on the number of regimes
$m_{+}>0$: 
\[
F_{T}^{*}=\sup_{\pi\in[\pi_{L},\,1]}\max_{1\leq m\leq m_{+}}\sup_{\mathbf{S}_{T}\in\Xi_{\epsilon,\pi,m,T}}F_{T}\left(\mathbf{S}_{T}\right),
\]
where $\Xi_{\epsilon,\pi,m,T}$ denotes the set of all possible partitions
of a fraction $\pi$ of $\{1,\ldots,T\}$ that involve $m$ regimes
$\left(\left(\lambda_{L,1}T,\,\lambda_{R,1}T\right),\ldots,\,\left(\lambda_{L,m}T,\,\lambda_{R,m}T\right)\right)$
for $\lambda_{L,i},\,\lambda_{R,i}\in\left[0,\,1\right]$ such that
(i) $\lambda_{L,i}<\lambda_{R,i}$ for all $i$, (ii) $\lambda_{R,i}<\lambda_{L,i+1}$
for $i=1,\ldots,\,m-1$, (iii) $\left|\lambda_{R,i}-\lambda_{L,i}\right|\geq\epsilon$
for all $i$ and some (small) $\epsilon>0$ and (iv) $\sum_{i=1}^{m}(\lambda_{R,i}-\lambda_{L,i})=\pi.$
Conditions (i) and (ii) correspond to $T\lambda_{L,i}$ ($T\lambda_{R,i}$)
denoting the start (end) date of regime $i$ within the sub-population
$\mathbf{S}_{T}$ while condition (iii) implies that each regime involves
a non-negligible fraction of the sample. The statistic $F_{T}^{*}$
thus implicitly searches for maximal identification strength over
all possible sub-populations of size $\pi_{L}T$ and larger with less
than $m_{+}$ distinct regimes that are at least an $\epsilon$ fraction
of the overall sample size.

The tuning parameters $\pi_{L}$ and $\epsilon$ determine the types
of sub-populations for which the test can detect identification: smaller
values of $\pi_{L}$ allow detection in smaller sub-populations, while
smaller values of $\epsilon$ enable detection in sub-populations
with shorter regimes. The choice of these lower bounds should be guided
by the empirical context, reflecting the smallest sub-population and
regime sizes for which LATE inference remains meaningful in the application.\footnote{In the structural break literature, common recommendations for $\epsilon$
are 0.05, 0.10 and 0.15. See \citet{casini/perron_Oxford_Survey}
for a review.} In our simulations and empirical applications we set $\pi_{L}=0.6$
and $\epsilon=0.05$.

For $X_{t}^{\prime}$ the $t^{th}$ row of $X$, let $w_{t}=(X'_{t},\,Z'_{t})'$
and $W_{r}\left(\cdot\right)$ denote a $r$-vector of independent
Wiener processes on $\left[0,\,1\right]$. We derive the asymptotic
null distributions of $F_{T}\left(\mathbf{S}_{T}\right)$ and $F_{T}^{*}$
under the following standard high-level assumptions that permit both
heteroskedastic and serially correlated errors. Sufficient conditions
for them can be found in the supplement.
\begin{assumption}
\label{Assumption: w ULLN}$T^{-1}\sum_{t=1}^{\left\lfloor Ts\right\rfloor }w_{t}w'_{t}\overset{\mathbb{P}}{\rightarrow}sQ$,
uniformly in $s\in\left[0,\,1\right]$ for some p.d. matrix $Q$.
\end{assumption}
\begin{assumption}
\label{Assumption: we FCLT}$T^{-1/2}\sum_{t=1}^{\left\lfloor Ts\right\rfloor }w_{t}e_{t}\Rightarrow\Omega_{we}^{1/2}W_{p+q}\left(s\right)$
for some p.d. variance matrix $\Omega_{we}$. 
\end{assumption}
\begin{assumption}
\label{Assumption: hat-J uniform consistency}$\widehat{J}(S_{T})$
is p.d. for all $T,$ $\mathbf{S}_{T}\in\Xi_{\epsilon,\pi,m,T}$ and
$\widehat{J}(S_{T})\overset{\mathbb{P}}{\rightarrow}\lim_{T\rightarrow\infty}\left(T\pi\right)^{-1}\mathrm{Var}($
$e^{\prime}S_{T}^{\prime}\widetilde{Z}(S_{T}))$ uniformly in $\mathbf{S}_{T}\in\Xi_{\epsilon,\pi,m,T}$.
\end{assumption}
\begin{thm}
\label{Theorem: Asymptotic Distribution Sup Fstar test}Let Assumptions
\ref{Assumption: w ULLN}-\ref{Assumption: hat-J uniform consistency}
hold. Under $H_{\theta,0}$, 
\begin{align*}
F_{T}\left(\mathbf{S}_{T}\right)\Rightarrow F\left(\mathbf{S}\right)\quad\mathrm{if}\quad\mathbf{S}_{T}\in\Xi_{\epsilon,\pi,m,T},\qquad\mathrm{and}\qquad & F_{T}^{*}\Rightarrow\sup_{\pi\in[\pi_{L},\,1]}\max_{1\leq m\leq m_{+}}\sup_{\mathbf{S}\in\Xi_{\epsilon,\pi,m}}F\left(\mathbf{S}\right),
\end{align*}
where $\mathbf{S}=\lim_{T\rightarrow\infty}T^{-1}\mathbf{S}_{T}$,
$\Xi_{\epsilon,\pi,m}=\lim_{T\rightarrow\infty}T^{-1}\Xi_{\epsilon,\pi,m,T}$
and 
\[
F\left(\mathbf{S}\right)=\frac{1}{q\pi}\left\Vert \sum_{i=1}^{m}\left(W_{q}\left(\lambda_{R,i}\right)-W_{q}\left(\lambda_{L,i}\right)\right)\right\Vert ^{2}.
\]
\end{thm}
When $\pi=1$ ($\pi_{L}=1$ and $m_{+}=1$), $F_{T}\left(\mathbf{S}_{T}\right)$
($F_{T}^{*}$) reduces to the usual first-stage $F$-statistic for
$\theta=0$ in \eqref{Eq. Reduced Form Eq. IV model}. For $\pi\in(0,\,1)$
($\pi_{L}\in(0,\,1)$), the consistency of tests against $H_{\theta,1}$
using $F_{T}\left(\mathbf{S}_{0,T}\right)$ ($F_{T}^{*}$) follows
from similar arguments as for the $\pi_{0}=1$ case. The asymptotic
null distributions of both $F\left(\mathbf{S}\right)$ and $F_{T}^{*}$
are free of nuisance parameters. The critical values are obtained
via simulations and reported in Table \ref{Table CVs of F* alpha=00003D0.05}
for up to $m_{+}=6$ and up to $q=6$.

\section{Estimation of LATE and Identified Sub-Populations\label{Section: Estimation}}

We discuss estimation of the LATE parameter $\beta$ in \eqref{Eq. Structural eq. IV model}
in both the cases of a known and unknown sub-population $\mathbf{S}_{0,T}$,
as well as estimation of $\mathbf{S}_{0,T}$ itself in the latter
case. When $\mathbf{S}_{0,T}$ is known, estimation of $\beta$ is
an application of IV estimation for which $Z_{t}\mathbf{1}\{t\in\mathbf{S}_{0,T}\}$
is treated as the vector of instruments. Let this estimator be denoted
as $\widehat{\beta}(\mathbf{S}_{0,T})$.

On the other hand, when the sub-population $\mathbf{S}_{0,T}$ is
unknown, we must estimate it first. Although $\mathbf{S}_{0,T}$ can
be estimated consistently in the special case of a binary instrument
under the conditions of Proposition \ref{Proposition: Compliers FS indivdually}
and Theorem \ref{Theorem: Compliers}, it can also be estimated more
generally. We discuss two methods. The first is more computationally
straightforward but the second is more efficient because it uses the
information in both structural and reduced-form equations \eqref{Eq. Structural eq. IV model}-\eqref{Eq. Reduced Form Eq. IV model}.
We follow the structural change literature and assume that $\pi_{0}$
and $m_{0}$ are known, i.e., the practitioner has previously used
the tests from Section \ref{Section: ID Failure Test} to determine
$\pi_{0}$ and $m_{0}$.

We begin with the first estimator. Consider the $T\times T$ matrix
$C_{T}$ that selects the $\pi T$ rows of a matrix corresponding
to the indices in $\mathbf{S}_{T}$ while setting the remaining $(1-\pi)T$
rows to zero. For example, for a $T\times k$ matrix $A$, if $\mathbf{S}_{T}=\left\{ 1,\ldots,\,0.25T,\,0.75T+1,\ldots,\,T\right\} $,
\[
C_{T}A=\left[A^{\left(1,:\right)\prime}:\cdots:A^{\left(0.25T,:\right)\prime}:0_{k\times1}:\cdots:0_{k\times1}:A^{\left(0.75T+1,:\right)\prime}:\cdots:A^{\left(T,:\right)\prime}\right]'.
\]
Let $\overline{A}(C_{T})=M_{X}C_{T}A$ so that for a given $\mathbf{S}_{T}$,
the OLS estimators of $\theta$ and $\gamma_{2}$ in \eqref{Eq. Reduced Form Eq. IV model}
can be expressed as $\widehat{\theta}_{OLS}(\mathbf{S}_{T})=(\overline{Z}(C_{T})^{\prime}\overline{Z}(C_{T}))^{-1}\overline{Z}(C_{T})^{\prime}D$
and $\widehat{\gamma}_{2,OLS}(\mathbf{S}_{T})=(X^{\prime}M_{C_{T}Z}X)^{-1}X^{\prime}M_{C_{T}Z}D$.
Our first estimator of $\mathbf{S}_{0,T}$ minimizes the sum of squared
residuals of the reduced-form: 
\[
\widehat{\mathbf{S}}_{T,OLS}=\underset{\mathbf{S}_{T}\in\Xi_{\epsilon,\pi_{0},m_{0},T}}{\mathrm{argmin}}\left(D-C_{T}Z\widehat{\theta}_{OLS}(\mathbf{S}_{T})-X\widehat{\gamma}_{2,OLS}(\mathbf{S}_{T})\right)^{\prime}\left(D-C_{T}Z\widehat{\theta}_{OLS}(\mathbf{S}_{T})-X\widehat{\gamma}_{2,OLS}(\mathbf{S}_{T})\right).
\]
Correspondingly, we estimate $\beta$ with $\widehat{\beta}(\widehat{\mathbf{S}}_{T,OLS})$.
In the supplement, we present the consistency results for $\widehat{\mathbf{S}}_{T,OLS}$
and $\widehat{\beta}(\widehat{\mathbf{S}}_{T,OLS})$, and we consider
a second estimator of $\mathbf{S}_{0,T}$, namely a GLS criterion
that minimizes an efficiently weighted combination of the sum of squared
residuals of both the structural \eqref{Eq. Structural eq. IV model}
and  reduced-form equation \eqref{Eq. Reduced Form Eq. IV model}.

In model \eqref{Eq. Structural eq. IV model} the LATE parameter $\beta$
is constant, so $\pi$-LATE is the full population LATE and $\widehat{\beta}(\widehat{\mathbf{S}}_{T,OLS})$
is consistent for the LATE parameter $\beta$. This can be a precise
estimate even when a first-stage $F$ test detects full sample weak
identification because it uses the most-strongly identified subsample.
When the model \eqref{Eq. Structural eq. IV model} is misspecified,
so that LATEs may be nonlinear and time-varying, $\widehat{\beta}(\widehat{\mathbf{S}}_{T,OLS})$
is still consistent for a weighted average of the LATEs in the $\mathbf{S}_{0,T}$
subsample if the $\mathbf{S}_{0,T}$ subsample exhibits strong identification.

The estimator $\widehat{\mathbf{S}}_{T,OLS}$ and the test statistic
$F_{T}^{*}$ solve an optimization problem over many partitions. This
is computationally more complex than problems in the structural breaks
literature, as it involves optimizing both over sample partitions
and identification strength. We address this challenge by proposing
an efficient algorithm based on dynamic programming, extending the
approach of \citet{bai/perron:03} to our setting.\footnote{While \citet{antonie/boldea:2018} consider the case of a single break,
and \citet{magnusson/mavroeidis:2014} study a related context, neither
provide a computational solution\textemdash referring to the problem
as ``computationally demanding.''}

\section{\label{Section: Inference}Identification-Robust Inference}

We consider tests on $\beta$ in \eqref{Eq. Structural eq. IV model}
that are robust to weak identification in both the cases for which
the sub-population $\mathbf{S}_{0,T}$ is known and unknown. The hypothesis
testing problem is $H_{0}:\,\beta=\beta_{0}$ versus $H_{1}:\,\beta\neq\beta_{0}.$
Here we present results for the case of unknown sub-population $\mathbf{S}_{0,T}$
and weak instruments. We defer to the supplement the formal treatment
of the case of known $\mathbf{S}_{0,T}$ and strong instruments. Rewrite
\eqref{Eq. Structural eq. IV model}-\eqref{Eq. Reduced Form Eq. IV model}
as 
\begin{align}
y & =\overline{Z}(C_{0,T})\theta a_{\beta}'+X\eta+v,\:\:\:y=\left[Y:D\right],\:v=\left[v_{1}:e\right],\:a_{\beta}=\left(\beta,\,1\right)',\:\eta=\left[\gamma:\phi\right],\label{Eq. (2.5) AMS}
\end{align}
where $v_{1}=u+\beta e$, $\gamma=\gamma_{1}+\phi\beta$ and $\phi=\gamma_{2}+(X^{\prime}X)^{-1}X^{\prime}C_{0,T}Z\theta$.
When $\mathbf{S}_{0,T}$ is known, it is straightforward to use existing
tests in the identification-robust linear IVs literature to test $H_{0}$
{[}cf. \citet{anderson/rubin:1949}, \citet{andrews/moreira/stock:2006},
\citet{kleibergen:2002} and \citet{moreira:2003}{]}. However, Proposition
\ref{Lemma 1 AMS} in the supplement shows that $Z^{\prime}M_{X}y$
is not a sufficient statistic for $(\beta,\theta')'$ but $\overline{Z}(C_{0,T})^{\prime}y$
is, implying that existing tests suffer a loss in efficiency because
they treat $Z$ rather than $C_{0,T}Z$ as the matrix of IVs. Efficient
tests are therefore functions of $\overline{Z}(C_{0,T})^{\prime}y$.
\citet{magnusson/mavroeidis:2014} consider a model similar to \eqref{Eq. (2.5) AMS}.
Our model specifies that $\theta$ is nonzero in the sub-population
$\mathbf{S}_{0,T}$ and is zero in $\mathbf{S}_{0,T}^{c}$ where $\mathbf{S}_{0,T}^{c}$
is the complement of $\mathbf{S}_{0,T}$. \citet{magnusson/mavroeidis:2014}
allow the first-stage coefficient $\theta_{t}$ to be generally time-varying
for some of their tests. Their tests are based on the full sample
of observations whereas our tests are based on a lower-dimensional
statistic since we do not use the sub-population $\mathbf{S}_{0,T}^{c}$.
This allows us to obtain gains in efficiency.

When $\mathbf{S}_{0,T}$ is known we can apply the results of \citet{andrews/moreira/stock:2006}
to form identification-robust tests of $H_{0}$ vs $H_{1}$ that are
functions of $\overline{Z}(C_{0,T})^{\prime}y$ and are robust to
both heteroskedasticity and autocorrelation (HAR) in the reduced-form
errors $\{v_{t}\}$. Suppose $\widehat{\Sigma}_{N_{1}}(\mathbf{S}_{0,T})$,
$\widehat{\Sigma}_{N_{1},N_{2}}(\mathbf{S}_{0,T})$ and $\widehat{\Sigma}_{N_{2}}(\mathbf{S}_{0,T})$
are consistent estimators of $\Sigma_{N_{1}}(\mathbf{S}_{0})$, $\Sigma_{N_{1},N_{2}}(\mathbf{S}_{0})$
and $\Sigma_{N_{2}}(\mathbf{S}_{0})$ under $H_{0}$, where these
latter quantities are defined by 
\begin{gather}
\Sigma_{v\overline{Z}}\left(\mathbf{S}_{0}\right)=\begin{bmatrix}\Sigma_{N_{1}}\left(\mathbf{S}_{0}\right) & \Sigma{}_{N_{1}N_{2}}\left(\mathbf{S}_{0}\right)'\\
\Sigma_{N_{1}N_{2}}\left(\mathbf{S}_{0}\right) & \Sigma_{N_{2}}^{*}\left(\mathbf{S}_{0}\right)
\end{bmatrix},\label{Eq. LRV matrix}\\
\Sigma_{N_{2}}\left(\mathbf{S}_{0}\right)=\Sigma_{N_{2}}^{*}\left(\mathbf{S}_{0}\right)-\Sigma_{N_{1}N_{2}}\left(\mathbf{S}_{0}\right)\Sigma_{N_{1}}^{-1}\left(\mathbf{S}_{0}\right)\Sigma_{N_{1}N_{2}}\left(\mathbf{S}_{0}\right)'\nonumber 
\end{gather}
for $\Sigma_{v\overline{Z}}\left(\mathbf{S}_{0}\right)=\Sigma_{v\overline{Z}}\left(\mathbf{S}_{0},\mathbf{S}_{0}\right)$,
with 
\begin{gather*}
\Sigma_{v\overline{Z}}\left(\mathbf{S},\mathbf{S}'\right)=\lim_{T\rightarrow\infty}\mathrm{Cov}\left(T^{-1/2}\sum_{t=1}^{T}\begin{bmatrix}v'_{t}b_{0}\overline{Z}_{t}\left(C_{T}\right)\\
v'_{t}\Sigma_{v}^{-1}a_{0,\beta}\overline{Z}_{t}\left(C_{T}\right)
\end{bmatrix},T^{-1/2}\sum_{t=1}^{T}\begin{bmatrix}v'_{t}b_{0}\overline{Z}_{t}\left(C_{T}'\right)\\
v'_{t}\Sigma_{v}^{-1}a_{0,\beta}\overline{Z}_{t}\left(C_{T}'\right)
\end{bmatrix}\right)
\end{gather*}
for $\mathbf{S}=\lim_{T\rightarrow\infty}T^{-1}\mathbf{S}_{T}$, $\mathbf{S}'=\lim_{T\rightarrow\infty}T^{-1}\mathbf{S}_{T}'$
$b_{0}=(1,-\beta_{0})^{\prime}$ and $a_{0,\beta}=(\beta_{0},1)^{\prime}$,
where $v_{t}$ and $\overline{Z}_{t}\left(C_{T}\right)$ are the transposes
of the $t$th rows of $v$ and $\overline{Z}(C_{T})$.\footnote{See the supplement for details on how to construct these estimators
and for consistency results.} Let $\widehat{\Sigma}_{v}\left(\mathbf{S}_{0,T}\right)=\left(T-q-p\right)^{-1}\widehat{v}\left(\mathbf{S}_{0,T}\right)'\widehat{v}\left(\mathbf{S}_{0,T}\right)$
with $\widehat{v}\left(\mathbf{S}_{0,T}\right)=y-P_{\overline{Z}\left(C_{0,T}\right)}y-P_{X}y$.
Define 
\begin{align}
{N}_{1,T}\left(\mathbf{S}_{0,T}\right) & =\widehat{\Sigma}_{N_{1}}^{-1/2}\left(\mathbf{S}_{0,T}\right)T^{-1/2}\overline{Z}\left(C_{0,T}\right)'yb_{0}\qquad\mathrm{and}\label{Eq. (9.7) AMS}\\
{N}_{2,T}\left(\mathbf{S}_{0,T}\right) & =\widehat{\Sigma}_{N_{2}}^{-1/2}\left(\mathbf{S}_{0,T}\right)\left(T^{-1/2}\overline{Z}\left(C_{0,T}\right)'y\widehat{\Sigma}_{v}^{-1}\left(\mathbf{S}_{0,T}\right)a_{0,\beta}-\widehat{\Sigma}_{N_{1}N_{2}}\left(\mathbf{S}_{0,T}\right)\widehat{\Sigma}_{N_{1}}^{-1/2}\left(\mathbf{S}_{0,T}\right){N}_{1,T}\left(\mathbf{S}_{0,T}\right)\right).\nonumber 
\end{align}
Consider the following HAR versions of the Anderson-Rubin (AR), Lagrange
multiplier (LM) and likelihood ratio statistics based on the sufficient
statistic $\overline{Z}(C_{0,T})'y$: 
\begin{align}
AR_{T}(\mathbf{S}_{0,T}) & =M_{1,T}(\mathbf{S}_{0,T}),\qquad\qquad LM_{T}(\mathbf{S}_{0,T})=\frac{M_{1,2,T}(\mathbf{S}_{0,T})^{2}}{M_{2,T}(\mathbf{S}_{0,T})},\label{Eq. (3.4) AMS}\\
LR_{T}(\mathbf{S}_{0,T}) & =\frac{1}{2}\left(M_{1,T}(\mathbf{S}_{0,T})-M_{2,T}(\mathbf{S}_{0,T})+\sqrt{\left(M_{1,T}(\mathbf{S}_{0,T})-M_{2,T}(\mathbf{S}_{0,T})\right)^{2}+4M_{1,2,T}(\mathbf{S}_{0,T})^{2}}\right),\nonumber 
\end{align}
where $M_{1,T}(\mathbf{S}_{0,T})={N}_{1,T}\left(\mathbf{S}_{0,T}\right)^{\prime}{N}_{1,T}\left(\mathbf{S}_{0,T}\right)$,
$M_{1,2,T}(\mathbf{S}_{0,T})={N}_{1,T}\left(\mathbf{S}_{0,T}\right)^{\prime}{N}_{2,T}\left(\mathbf{S}_{0,T}\right)$
and $M_{2,T}($ $\mathbf{S}_{0,T})={N}_{2,T}\left(\mathbf{S}_{0,T}\right)^{\prime}{N}_{2,T}\left(\mathbf{S}_{0,T}\right)$.
The conditional likelihood ratio (CLR) test of level $\alpha$ rejects
$H_{0}$ when $LR_{T}(\mathbf{S}_{0,T})>\kappa_{\alpha}(N_{2,T}(\mathbf{S}_{0,T}))$,
where the critical value function $\kappa_{\alpha}(\cdot)$ is defined
such that $\kappa_{\alpha}(n_{2})$ is the $1-\alpha$ quantile of
the large-sample conditional distribution of $LR_{T}(\mathbf{S}_{0,T})$
under $H_{0}$, given $N_{2,T}(\mathbf{S}_{0,T})=n_{2}$: 
\[
\frac{1}{2}\left(\mathcal{Z}_{q}'\mathcal{Z}_{q}-n_{2}'n_{2}+\sqrt{\left(\mathcal{Z}_{q}'\mathcal{Z}_{q}-n_{2}'n_{2}\right)^{2}+4(\mathcal{Z}_{q}'n_{2})^{2}}\right),
\]
where $\mathcal{Z}_{q}\sim\mathscr{N}(0,I_{q})$. The critical value
function $\kappa_{\alpha}(\cdot)$ is approximated in \citet{moreira:2003}.
The LM and AR tests reject $H_{0}$ when $LM_{T}>\chi_{1}^{2}(1-\alpha)$
and $AR_{T}>\chi_{q}^{2}(1-\alpha)$, where $\chi_{q}^{2}(1-\alpha)$
denotes the $1-\alpha$ quantile of a chi-squared distribution with
$q$ degrees of freedom.

When $\mathbf{S}_{0,T}$ is known the results of \textcolor{MyBlue}{Andrews et al.}
\citeyearpar{andrews/moreira/stock:2006} imply that the CLR, LM
and AR tests have limiting null rejection probabilities equal to $\alpha$
under weak IV asymptotics, $\theta=c/T^{1/2}$ for some nonstochastic
$c\in\mathbb{R}^{q}$, under a weakening of Assumptions \ref{Assumption 1 AMS}-\ref{Assumption Uniform Consistent Covariance Matrix}
below for which these assumptions need only hold pointwise in $\mathbf{S}_{T}$.
These tests are asymptotically similar and therefore have asymptotically
correct size in the presence of weak IVs.

For the case of an unknown sub-population, the identification-robust
tests in the extant literature no longer apply because the set of
instruments $C_{0,T}Z$ is unknown and must be estimated. In this
section, we show how to form HAR CLR, LM and AR tests with correct
asymptotic null rejection probabilities under both weak and strong
IV asymptotics. To estimate the true sub-population $\mathbf{S}_{0,T}$
when constructing these tests let 
\begin{equation}
\widehat{\mathbf{S}}_{T}=\arg\max_{\mathbf{S}_{T}\in\mathcal{S}}M_{2,T}(\mathbf{S}_{T}),\qquad\qquad\mathrm{where}\qquad\qquad\mathcal{S}=\underset{1\leq m\leq m_{+}}{\cup}\underset{\pi\in(\epsilon,\,1]}{\cup}\Xi_{\epsilon,\pi,m,T}.\label{Eq. S_hat MLE}
\end{equation}
Proposition \ref{Lemma 1 AMS unknown subpopulation} in the supplement
shows that the process $\{\overline{Z}(C_{T})'y\}_{\mathbf{S}_{T}\in\mathcal{S}}$
is sufficient for $(\beta,\theta')'$ in a canonical Gaussian setting
analogous to that in \textcolor{MyBlue}{Andrews et al.} \citeyearpar{andrews/moreira/stock:2006}
so that there is no loss in efficiency from using the unknown sub-population
AR, LM and LR statistics, $LR_{T}(\widehat{\mathbf{S}}_{T})$, $LM_{T}(\widehat{\mathbf{S}}_{T})$
and $AR_{T}(\widehat{\mathbf{S}}_{T})$, which are only functions
of the process $\{\overline{Z}(C_{T})'y\}_{\mathbf{S}_{T}\in\mathcal{S}}$.

We establish the asymptotic validity of the HAR CLR, LM and AR tests
in the unknown sub-population setting under a weak set of high-level
sufficient conditions on the IVs, exogenous variables and errors.
Define $w\left(\mathbf{S}_{T}\right)=\left[C_{T}Z:X\right]$.
\begin{assumption}
\label{Assumption 1 AMS}Let $\mathbf{S}=T^{-1}\mathbf{S}_{T}$, $\mathbf{S}'=T^{-1}\mathbf{S}'_{T}$,
and $\mathcal{S}_{\infty}=\{\mathbf{S}:\,\mathbf{S}=\lim_{T\rightarrow\infty}\mathbf{S}_{T}\,\mathrm{for\,\mathrm{some}}$
$\mathrm{sequence\,\mathbf{S}}_{T}\in\mathcal{S}.$ There exists a
matrix-valued function $Q:\,\mathcal{S}_{\infty}\times\mathcal{S}_{\infty}\rightarrow\mathbb{R}^{\left(q+p\right)\times\left(q+p\right)}$
such that $\sup_{\mathbf{S}_{T},\,\mathbf{S}'_{T}\in\mathcal{S}}\left\Vert T^{-1}w\left(\mathbf{S}_{T}\right)'w\left(\mathbf{S}'_{T}\right)-Q\left(\mathbf{S},\,\mathbf{S}'\right)\right\Vert \overset{\mathbb{P}}{\rightarrow}0.$
Moreover, the diagonal limits are uniformly positive definite:$\inf_{\mathbf{S}\in\mathcal{S}_{\infty}}\lambda_{\min}\left(Q\left(\mathbf{S},\,\mathbf{S}'\right)\right)>0.$ 
\end{assumption}
\begin{assumption}
\label{Assumption 2 AMS} $T^{-1}v'v\overset{\mathbb{P}}{\rightarrow}\Sigma_{v}$
for some $2\times2$ p.d. matrix $\Sigma_{v}$.
\end{assumption}
\begin{assumption}
\label{Assumption 3 AMS} For $\mathbf{S}_{T},\mathbf{S}_{T}'\in\mathcal{S}$
and $\mathbf{S}=\lim_{T\rightarrow\infty}T^{-1}\mathbf{S}_{T}$, $\mathbf{S}'=\lim_{T\rightarrow\infty}T^{-1}\mathbf{S}_{T}'$,
$T^{-1/2}\mathrm{vec}(w\left(\mathbf{S}_{T}\right)'v)\Rightarrow\mathscr{G}\left(\mathbf{S}\right)$,
where $\mathscr{G}(\cdot)$ is a mean-zero Gaussian process indexed
by $\mathbf{S}\subseteq(0,1]$ with $2\left(q+p\right)\times2\left(q+p\right)$
covariance function $\Psi\left(\mathbf{S},\,\mathbf{S}'\right)=\lim_{T\rightarrow\infty}T^{-1}\mathrm{Cov}(\mathrm{vec}(w\left(\mathbf{S}_{T}\right)'v),\mathrm{vec}(w\left(\mathbf{S}'_{T}\right)'v))$.
\end{assumption}
In Assumption \ref{Assumption 3 AMS}, $\mathrm{vec}\left(\cdot\right)$
denotes the vec operator. The quantities $Q\left(\cdot\right)$, $\Sigma_{v}$,
and $\Psi\left(\cdot\right)$ are assumed to be unknown. Assumptions
\ref{Assumption 1 AMS}-\ref{Assumption 2 AMS} hold under suitable
conditions by a (uniform) law of large numbers. Assumption \ref{Assumption 3 AMS}
holds under suitable conditions by a functional central limit theorem.
Assumptions \ref{Assumption 1 AMS}-\ref{Assumption 3 AMS} are consistent
with non-normal, heteroskedastic, autocorrelated errors and IVs and
regressors that may be random or non-random.\footnote{In the supplement we provide primitive sufficient conditions for Assumptions
\ref{Assumption 1 AMS}-\ref{Assumption 3 AMS}.}

We assume that we can consistently estimate $\Sigma_{v\overline{Z}}\left(\mathbf{S}\right)\equiv\Sigma_{v\overline{Z}}\left(\mathbf{S},\mathbf{S}\right)$
uniformly in $\mathbf{S}_{T}$.
\begin{assumption}
\label{Assumption Uniform Consistent Covariance Matrix}We have an
estimator $\widehat{\Sigma}_{v\overline{Z}}(\mathbf{S}_{T})$ such
that $\widehat{\Sigma}_{v\overline{Z}}(\mathbf{S}_{T})\overset{\mathbb{P}}{\rightarrow}\Sigma_{v\overline{Z}}(\mathbf{S})$
uniformly in $\mathbf{S}_{T}\in\mathcal{S}$ for $\mathbf{S}=\lim_{T\rightarrow\infty}T^{-1}\mathbf{S}_{T}$.
\end{assumption}
Note that this assumption immediately implies the uniform consistency
of $\widehat{\Sigma}_{N_{2}}(\mathbf{S}_{T})=\widehat{\Sigma}_{N_{2}}^{*}(\mathbf{S}_{T})-\widehat{\Sigma}_{N_{1}N_{2}}(\mathbf{S}_{T})\widehat{\Sigma}_{N_{1}}^{-1}(\mathbf{S}_{T})\widehat{\Sigma}_{N_{1}N_{2}}(\mathbf{S}_{T})'$
as well. Consistent estimators of $\Sigma_{v\overline{Z}}$ are HAC
and DK-HAC estimators.\footnote{In the supplement we provide weak sufficient conditions, even allowing
for certain forms of nonstationarity, that ensure this assumption
holds.} 

Finally, we impose a second-order stationarity condition for $v'_{t}b_{0}\overline{Z}_{t}\left(C_{T}\right)$
and $v'_{t}\Sigma_{v}^{-1}a_{0,\beta}\overline{Z}_{t}\left(C_{T}\right)$.
 
\begin{assumption}
\label{Assumption 2nd Moment Stability}Let $\pi(\mathbf{S})$ equal
the Lebesgue measure of $\mathbf{S}\subseteq(0,1]$. Assume that $\Sigma_{v\overline{Z}}\left(\mathbf{S},\,\mathbf{S}'\right)=\pi\left(\mathbf{S}\cap\mathbf{S}'\right)\Sigma_{v\overline{Z}}$
where $\mathbf{S},\,\mathbf{S}'\subseteq(0,1]$ and $\Sigma_{v\overline{Z}}$
is p.d.
\end{assumption}
Assumption \ref{Assumption 2nd Moment Stability} is implied by a
uniform law of large numbers and functional central limit theorem
for partial sum processes under second-order stationarity. Under weak
IV asymptotics, $T^{-1}\widehat{\mathbf{S}}_{T}$ is not consistent
for $\mathbf{S}_{0}$. Assumption \ref{Assumption 2nd Moment Stability}
is needed in order to show that $N_{1,T}(\cdot)$ and $N_{2,T}(\cdot)$
are asymptotically independent processes. Under strong IV asymptotics
we can dispense with Assumption \ref{Assumption 2nd Moment Stability}
because $T^{-1}\widehat{\mathbf{S}}_{T}\overset{\mathbb{P}}{\rightarrow}\mathbf{S}_{0}$
and the limit of the processes $N_{1,T}(\cdot)$ and $N_{2,T}(\cdot)$
have zero covariance when evaluated at a fixed $\mathbf{S}_{0}$.

Define the LR, LM and AR statistics in this context according to \eqref{Eq. (3.4) AMS},
replacing $\mathbf{S}_{0,T}$ with $\widehat{\mathbf{S}}_{T}$. We
now establish the correct asymptotic null rejection probabilities
of the sub-population-estimated plug-in HAR CLR, LM and AR tests under
weak identification. 
\begin{thm}
\label{Theorem 7 AMS HAR}Let Assumptions \ref{Assumption 1 AMS}-\ref{Assumption 2nd Moment Stability}
hold and suppose $\theta=c/T^{1/2}$ for some nonstochastic $c\in\mathbb{R}^{q}$.
We have: (i) ${AR}_{T}(\widehat{\mathbf{S}}_{T})\overset{d}{\rightarrow}\chi_{q}^{2}$
under $H_{0};$ (ii) ${LM}_{T}(\widehat{\mathbf{S}}_{T})\overset{d}{\rightarrow}\chi_{1}^{2}$
under $H_{0};$ (iii) $\mathbb{P}_{\beta_{0}}(LR_{T}(\widehat{\mathbf{S}}_{T})>\kappa_{\alpha}(N_{2,T}(\widehat{\mathbf{S}}_{T})))\rightarrow\alpha$
where $\mathbb{P}_{\beta_{0}}(\cdot)$ is the probability computed
under $H_{0}$. 
\end{thm}
The key to establishing these asymptotic validity results is to show
that each of the above statements hold conditional on the realization
of $N_{2,T}(\cdot)$. This can be readily established from the facts
that the stochastic processes $N_{1,T}(\cdot)$ and $N_{2,T}(\cdot)$
are asymptotically independent by construction, $\widehat{\mathbf{S}}_{T}$
is a function of $N_{2,T}(\cdot)$ and $N_{1,T}(\mathbf{S}_{T})\Rightarrow\mathscr{N}\left(0,\,I_{q}\right)$
under $H_{0}$.\footnote{In addition to identification-robust tests of $H_{0}$ vs $H_{1}$,
since the causal interpretation of $\beta$ depends upon the sub-population
$\mathbf{S}_{0,T}$, practitioners may wish to simultaneously report
the result of these tests along with a corresponding estimate of the
sub-population. More specifically, failure to reject $H_{0}$ should
be interpreted as failure to reject that the estimand is equal to
$\beta_{0}$, where the estimand is interpreted as a weighted average
of the LATEs for the estimated sub-population $\widehat{\mathbf{S}}_{T}$.
Given that the tests of $H_{0}$ remain asymptotically valid conditional
on the realization of $N_{2,T}(\cdot)$ and the fact that $\widehat{\mathbf{S}}_{T}$
is a function of $N_{2,T}(\cdot)$, the tests remain asymptotically
valid when interpreted conditional on the value of $\widehat{\mathbf{S}}_{T}$.} 

\section{\label{Section: Empirical Evidence}Empirical Evidence on LATE of
Monetary Policy}

We illustrate our methods by revisiting the identification of monetary
policy effects in the framework of \citet{nakamura/steinsson:2018},
introduced in Section \ref{Section: Application: Money Neutrality}.
They use a bivariate model \eqref{Eq. (1) NS and RS, Outcome variable}
to estimate the causal effect of $\widetilde{D}_{t}$ on $\widetilde{Y}_{t}$,
employing both event-study and heteroskedasticity-based identification
approaches. The dependent variable $\widetilde{Y}_{t}$ is either
the nominal or real 2-Year instantaneous Treasury forward rate and
the policy variable is either the daily change in nominal 2-Year Treasury
yields or the 30-minute change in a ``policy news'' series constructed
as the first principal component of the unanticipated 30-minute changes
in five selected interest rates. Heteroskedasticity-based identification
assumes the variance of the monetary shock rises on FOMC announcement
days, while the variance of other shocks remains constant {[}cf. eq.
\eqref{Eq. Vol_P >Vol_C}{]}. FOMC dates define the policy sample
$\mathbf{P}$, and analogous non-FOMC dates define the control sample
$\mathbf{C}.$ Nakamura and Steinsson's instrument for $\widetilde{D}_{t}^{2}$
is defined as $Z_{t}=\mathbf{1}\left\{ t\in\mathbf{P}\right\} $,
corresponding to the model in Section \ref{Section: Application: Money Neutrality}.
We focus on the same period: January 1, 2004, to March 19, 2014.

\citet{lewis:2020} recently analyzes this problem by developing a
first-stage $F$-test for weak identification. He finds that weak
identification is not rejected when $\widetilde{D}_{t}$ is the 1-day
change in nominal 2-Year Treasury yields, but is strongly rejected
when $\widetilde{D}_{t}$ is the 30-minute policy news series. This
supports \citeauthor{nakamura/steinsson:2018}'s \citeyearpar{nakamura/steinsson:2018}
observation that the daily policy variable may suffer from weaker
identification. Unlike \citet{nakamura/steinsson:2018}, \citet{lewis:2020}
estimates the model using GMM and does not impose the assumption that
the non-monetary policy shock $\eta_{t}$ has equal variance across
the treatment and control samples.

Section \ref{Subsection: Pre Test for Full Population Identification Failure NS18}
reports results of our test for full sample identification failure.
Section \ref{Subsection: Estimation in Strongly Identified Sub-sample NS18}
presents causal effect estimates based on the most strongly-identified
subsample. Section \ref{Subsection: Identifcation Robust Inference NS18}
provides identification-robust inference results, and Section \ref{Subsection: Identification and Estimation of Compliers}
estimates compliers at the individual level and tests the exclusion
restriction.

\subsection{\label{Subsection: Pre Test for Full Population Identification Failure NS18}Testing
for Identification Failure }

We present the results of our test for identification failure over
all sub-populations from Section \ref{Section: ID Failure Test} in
Table \ref{Table: F tests} considering values of $\pi_{L}$ from
0.6 to 1. For the 30-minute policy news variable, the $F_{T}^{*}$
statistic is very large and identification failure is rejected. This
supports the finding in \citet{lewis:2020} and intuition in \citet{nakamura/steinsson:2018}
that the 30-minute policy news variable leads to stronger identification
in the full sample. In contrast, for the 1-day change in nominal Treasury
yields, identification failure cannot be strongly rejected in the
full sample: the $F_{T}^{*}$ statistic at $\pi_{L}=1$ (i.e., full
sample) is only slightly larger than the 1\% critical value. The $F_{T}^{*}$
statistic increases substantially as $\pi_{L}$ decreases and it is
very far from the critical values. This is clear evidence that identification
is much stronger over subsamples. At $\pi_{L}=0.9$ it reaches 33.87,
clearly rejecting identification failure in the $\pi$-subsample (with
$\pi=0.9$ or $0.95$) over which the supremum of $F_{T}\left(\mathbf{S}_{T}\right)$
is computed. The $F_{T}^{*}$ statistic increases monotonically with
smaller $\pi_{L}$ due to the increasing number of partitions considered.
For example, at $\pi_{L}=0.8$, $F_{T}^{*}$ is 54.78\textemdash nearly
seven times the full sample value. Overall, the results indicate that
strong identification may hold when using a 1-day window, but only
within subsamples comprising at most 90\% of the data. The weak identification
reported by \citet{lewis:2020} using a 1-day window around FOMC announcements
likely does not stem solely from volatility returning to normal after
announcements. Rather, a small subsample (10\textendash 20\% of the
data) exhibits weak or failed identification, contributing to the
weaker identification exhibited in the full sample.

\begin{table}[H]
\caption{\label{Table: F tests}Tests for Identification Failure over all Sub-Populations}

\smallskip{}

\begin{centering}
{\footnotesize{}%
\begin{tabular}{cccccc}
\hline 
\multicolumn{6}{c}{{\small$F_{T}^{*}$ statistic and critical values}}\tabularnewline
 & \multicolumn{5}{c}{{\small$F_{T}^{*}$}}\tabularnewline
{\small$\widetilde{D}_{t}\backslash\pi_{L}$} & {\small 0.6} & {\small 0.7} & {\small 0.8} & {\small 0.9} & {\small 1}\tabularnewline
\hline 
\hline 
\begin{cellvarwidth}[t]
\centering
{\small 30-minute}\\
{\small ``policy news''}
\end{cellvarwidth} & {\small$10^{4}\times95.36$} & {\small$10^{4}\times56.50$} & {\small$10^{4}\times32.45$} & {\small$10^{4}\times18.75$} & {\small$10^{4}\times7.42$}\tabularnewline
\begin{cellvarwidth}[t]
\centering
{\small 1-day nominal}\\
{\small{} Treasury yields}
\end{cellvarwidth} & {\small 155.69} & {\small 88.22} & {\small 54.78} & {\small 33.88} & {\small 8.09}\tabularnewline
{\small 1\% critical values } & {\small 17.89} & {\small 15.58} & {\small 13.11} & {\small 10.66} & {\small 6.46}\tabularnewline
{\small 5\% critical values } & {\small 12.74} & {\small 10.70} & {\small 8.88} & {\small 7.02} & {\small 3.85}\tabularnewline
\hline 
\end{tabular}}{\footnotesize\par}
\par\end{centering}
{\scriptsize{}%
\noindent\begin{minipage}[t]{1\columnwidth}%
{\scriptsize$F_{T}^{*}$ statistics for first-stage identification
failure. $\widetilde{D}_{t}$ is either the 30-minute policy news
series or 1-day change in nominal Treasury yields. $\pi_{L}$ is the
minimum fraction of the sample over which the supremum of the $F\left(\mathbf{S}_{T}\right)$
is computed. Maximum number of breaks is set to $m_{+}=5$. }%
\end{minipage}}{\scriptsize\par}
\end{table}

\subsection{\label{Subsection: Estimation in Strongly Identified Sub-sample NS18}Estimation
in Strongly-Identified Subsample}

We turn to estimation of $\pi_{0}$ and $\mathbf{S}_{0,T}$ using
the methods from Section \ref{Section: Estimation}, and then to estimating
the LATE of monetary policy based on the strongly-identified subsample,
$\widehat{\beta}(\mathbf{\widehat{S}}_{T,OLS})$, or simply, $\widehat{\pi}$-sample,
where $\widehat{\pi}=|\widehat{\mathbf{S}}_{T,OLS}|/T$.\footnote{Recall that when the instrument is binary, the Wald and IV estimands
coincide. Hence, the results and intuition developed in Sections \ref{Section: Statistical Framework for Identification of Causal Effects}-\ref{Section: Application: Money Neutrality}
directly apply to the TSLS estimand analyzed here.} We focus on $\widehat{\beta}(\widehat{\mathbf{S}}_{T,OLS})$; results
using $\widehat{\beta}(\mathbf{\widehat{S}}_{T,FGLS})$ are similar.
Figure \ref{Figure: pi Sample} plots the 1-day changes in 2-Year
yields for the control and policy samples and highlights the regimes
included in the strongly-identified subsample $\mathbf{\widehat{S}}_{T,OLS}$.
The estimate $\widehat{\pi}=0.8$ implies that in 80\% of the sample,
the first-stage is strong and identification holds. In the control
sample, the excluded periods include the first seven months of 2005
and the regime surrounding the financial crisis (2007-2009). As shown
in the figure, volatility during the crisis period is much higher
than in the rest of the control group and higher than the average
volatility in the treatment group. This subsample appears to drive
the apparent full sample weak identification. Since our method searches
for maximum identification strength, it correctly excludes this period
when computing $\pi$-LATE.\footnote{The other excluded period (January to July 2005) does not display
obviously high volatility but shows some persistence, with a short-duration
cluster below the mean toward the end.} The interpretation is that in both excluded regimes\textemdash especially
during the financial crisis\textemdash market uncertainty was elevated
even on non-FOMC days, violating the identification assumption.
\noindent\begin{flushleft}
\begin{center}
\begin{figure}[h]
\begin{raggedright}
\includegraphics[clip,width=17cm,totalheight=8cm]{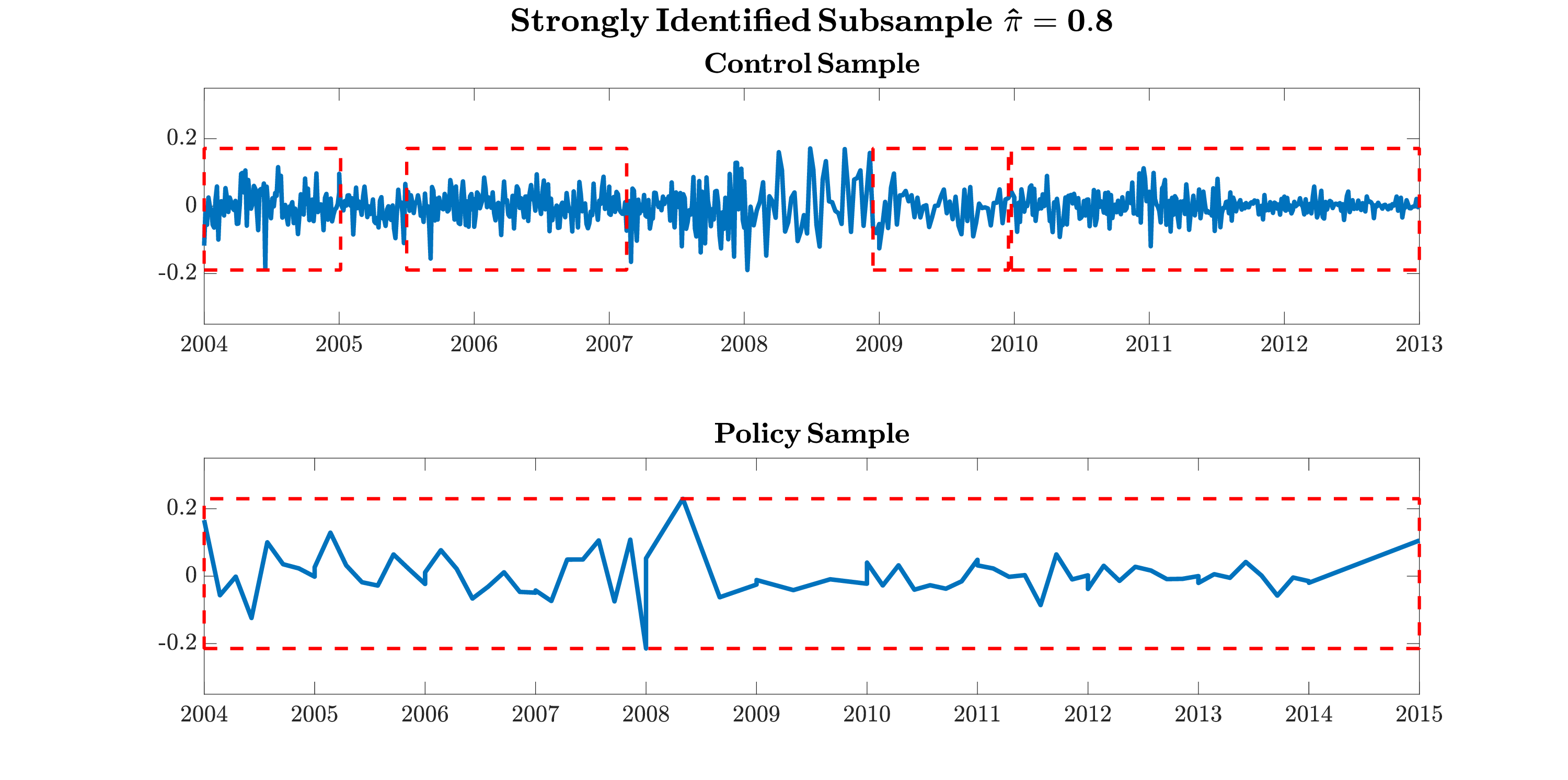}
\par\end{raggedright}
\raggedright{}\caption{\label{Figure: pi Sample}{\scriptsize Plot of $\widetilde{D}_{t}$
(2-Year Treasury yields) in the control sample (top panel) and policy
sample (bottom panel). The red rectangles indicate subsamples included
in the strongly-identified subsample $\mathbf{\widehat{S}}_{T,OLS}$
where $\widehat{\pi}=0.8$.}}
\end{figure}
\end{center}
\par\end{flushleft}

We now estimate the causal effect of monetary policy using the $\widehat{\pi}$-sample,
where by construction the LATE is most strongly-identified. We compare
these results with full sample estimates obtained using two-stage
least squares (TSLS) and GMM, following \citet{nakamura/steinsson:2018}
and \citet{lewis:2020}, respectively. Table \ref{Table: Estimation NS18}
presents the results. Starting with the full sample estimates: when
the policy variable is the 30-minute policy news series, TSLS and
GMM yield very similar point estimates for both nominal and real forward
rates, and both are statistically significant using standard and robust
confidence intervals.\footnote{The robust confidence intervals for the GMM estimates are based on
the subset $K$-test in \citet{lewis:2020}.}

As noted by \citet{lewis:2020}, the assumption that non-monetary
shocks have equal variance across treatment and control groups does
not bias the TSLS estimates, as they closely match the GMM ones. One
explanation is that the GMM estimate of $a$ (capturing reverse causality
from forward rates to policy news) is both near zero and statistically
significant (not reported). Since potential bias from this assumption
is proportional to $a(\sigma_{\eta,P}^{2}-\sigma_{\eta,C}^{2}),$
and $a$ is close to zero, the resulting bias is negligible even if
the variances $\sigma_{\eta,P}^{2}$ and $\sigma_{\eta,C}^{2}$ differ.

Turning to the case where the policy variable is the 1-day change
in 2-Year Treasury yields, the TSLS and GMM estimates differ markedly
from each other and from those based on the 30-minute policy news
series. Notably, the GMM estimate of $\beta$ is negative for nominal
forwards and positive for real forwards, but in neither case is it
statistically significant\textemdash whether using standard or robust
confidence intervals.

\begin{table}[H]
\caption{\label{Table: Estimation NS18}Estimation of $\beta$}

\smallskip{}

\begin{centering}
{\footnotesize}{\small{}%
\begin{tabular}{ccccc}
\hline 
 & \multicolumn{2}{c}{{\small 30-minute Policy News}} & \multicolumn{2}{c}{{\small 1-day 2-Year Yield}}\tabularnewline
{\small dep. var.} & {\small Nominal} & {\small Real} & {\small Nominal} & {\small Real}\tabularnewline
\hline 
 & \multicolumn{4}{c}{{\small Full Sample}}\tabularnewline
 & \multicolumn{4}{c}{{\small TSLS}}\tabularnewline
{\small$\beta$} & {\small 1.10{*}{*}} & {\small 0.96{*}{*}{*}} & {\small 1.14{*}{*}{*}} & {\small 0.97{*}{*}{*}}\tabularnewline
{\small standard CI} & {\small{[}0.17, 2.02{]}} & {\small{[}0.41, 1.51{]}} & {\small{[}0.83, 1.45{]}} & {\small{[}0.40, 1.565{]}}\tabularnewline
 & \multicolumn{4}{c}{{\small GMM}}\tabularnewline
{\small$\beta$} & {\small 1.07{*}{*}} & {\small 0.94{*}{*}{*}} & {\small -0.27} & {\small 1.31}\tabularnewline
{\small standard CI} & {\small{[}0.17, 1.98{]}} & {\small{[}0.36, 1.51{]}} & {\small{[}-4.90, 4.36{]}} & {\small{[}-3.74, 6.35{]}}\tabularnewline
{\small robust CI} & {\small{[}0.27, 3.25{]}} & {\small{[}0.44, 2.38{]}} & {\small{[}-77.27, 0.94{]}} & {\small{[}-253.70, 1.92{]}}\tabularnewline
 & \multicolumn{4}{c}{{\small$\pi$-sample based on $\widehat{\mathbf{S}}_{T,OLS}$ with
$\widehat{\pi}=0.8$}}\tabularnewline
 & \multicolumn{4}{c}{{\small TSLS}}\tabularnewline
{\small$\beta$} & {\small 1.11{*}{*}} & {\small 0.97{*}{*}{*}} & {\small 1.13{*}{*}{*}} & {\small 0.92{*}{*}{*}}\tabularnewline
{\small standard CI} & {\small{[}0.19, 2.02{]}} & {\small{[}0.42, 1.51{]}} & {\small{[}0.92, 1.30{]}} & {\small{[}0.56, 1.28{]}}\tabularnewline
 & \multicolumn{4}{c}{{\small GMM}}\tabularnewline
{\small$\beta$} & {\small 1.07{*}{*}} & {\small 0.94{*}{*}{*}} & {\small 0.65{*}} & {\small 0.86{*}{*}}\tabularnewline
{\small standard CI} & {\small{[}0.17, 1.96{]}} & {\small{[}0.38, 1.50{]}} & {\small{[}-0.02, 1.31{]}} & {\small{[}0.29, 1.43{]}}\tabularnewline
\hline 
\end{tabular}}{\small\par}
\par\end{centering}
\centering{}{\scriptsize{}%
\begin{minipage}[t]{0.8\columnwidth}%
{\scriptsize TSLS and GMM estimates of $\beta$. The GMM estimates
allow for changes also in the variance of $\eta_{t}$ across regimes.
The dependent variable is the 1-day change in either nominal or real
2-Year instantaneous Treasury forward rate. The policy variable is
either the 30-minute changes in the ``policy news'' variable or
1-day changes in the 2-Year nominal Treasury yield. The standard 95\%
confidence interval is based on the standard normal critical values.
For the GMM estimates, the robust 95\% confidence interval is based
on the subset $K$-test in \citet{lewis:2020}. Asterisks indicate
statistical significance at the 10\%, 5\%, or 1\% level based on standard
intervals.}%
\end{minipage}}{\scriptsize}{\scriptsize\par}
\end{table}

As discussed by \citet{lewis:2020}, these estimates are difficult
to interpret in economically meaningful terms. He also shows that
the GMM estimates of $a$ are nonzero and proposed a second dimension
of policy news to account for the findings. However, the opposing
signs of $\beta$ across nominal and real forwards complicate this
interpretation. Ultimately, he concludes that these results are inconsistent
with \citeauthor{nakamura/steinsson:2018}'s \citeyearpar{nakamura/steinsson:2018}
``background noise'' view of the non-monetary shock $\eta_{t}$
which assumes that its volatility remains unchanged between FOMC and
non-FOMC days.

We contribute to this discussion by presenting TSLS and GMM estimates
based on the most strongly-identified $\widehat{\pi}$-sample. We
focus first on standard confidence intervals and defer weak identification-robust
inference to Table \ref{Table: Inference NS18}. The bottom panel
of Table \ref{Table: Estimation NS18} shows that, for the 30-minute
policy news variable, the TSLS and GMM estimates, including their
statistical significance, are virtually unchanged. As expected\textemdash given
the apparent strong identification in the full sample\textemdash results
are broadly similar when using the $\widehat{\pi}$-sample.\footnote{The confidence intervals in the $\widehat{\pi}$-sample are even slightly
tighter.}

Finally, we turn to the $\widehat{\pi}$-sample estimates using the
1-day window for the policy. The GMM estimates differ sharply from
those in the full sample: for both nominal and real forwards, they
now have the same sign and are statistically significant. This suggests
that the opposite signs reported by \citet{lewis:2020} likely stemmed
from weak identification, rendering those estimates unreliable.\footnote{While the TSLS estimates are nearly unchanged from the full sample,
this should not be taken as evidence of their reliability. Under weak
IVs, their similarity to the $\widehat{\pi}$-sample results may simply
be coincidental.} Notably, the GMM estimates are now similar in magnitude to those
based on the 30-minute policy variable, supporting a more meaningful
interpretation.\footnote{We also verified that the GMM estimate of $a$ is 0.70 for nominal
forwards and -0.91 for real forwards. It is intuitive that the estimate
of $a$ is close to zero when using a 30-minute window but significantly
different from zero with a 1-day window. In the narrow 30-minute window
around an FOMC announcement, reverse causality from $\widetilde{Y}_{t}$
to $\widetilde{D}_{t}$ is limited, as monetary news is more pronounced
than other shocks\textemdash though some endogeneity may still arise
from omitted factors affecting both. In contrast, over a full day,
asset price movements can influence short-term interest rates, making
reverse causality more likely.}

Overall, this analysis highlights the advantage of using the most
strongly-identified $\widehat{\pi}$-sample. Given weak identification
in the full sample when using 1-day Treasury yields as the policy
variable, the corresponding estimates should be discarded. In contrast,
evidence from the $\widehat{\pi}$-sample shows that TSLS and GMM
produce similar, positive estimates for $\beta,$ consistent with
monetary policy affecting real forward rates, as predicted by New
Keynesian models, and supporting the existence of a forward guidance
channel.\footnote{However, the results do not yet support a second meaningful dimension
of news, as proposed by{\small{} }\citet{lewis:2020}, since the
sign of the GMM estimate of $a$ is unstable across nominal and real
forwards. Regarding \citeauthor{nakamura/steinsson:2018}'s \citeyearpar{nakamura/steinsson:2018}
``background noise'' interpretation of non-monetary shocks, we find
no clear evidence against it: in the $\widehat{\pi}$-sample, identification
appears strong, and TSLS and GMM estimates consistently share the
same sign and similar magnitudes.}

\subsection{\label{Subsection: Identifcation Robust Inference NS18}Weak Identification-Robust
Inference}

We apply the weak identification-robust tests proposed in Section
\ref{Section: Inference} and compare them to existing full sample
tests $LM_{T}$ and $LR_{T}$.\footnote{We do not report the $AR_{T}$ test since for $q=1$ it is equivalent
to the $LM_{T}$ test.} We test the  null hypothesis $H_{0}:\,\beta=0$ against $H_{1}:\,\beta\neq0$.
Results are shown in Table \ref{Table: Inference NS18}. When the
policy variable is the 30-minute policy news, identification is strong
in the full sample. Accordingly, both the proposed and existing tests
yield similar results: all tests reject at the 5\% level for both
nominal and real forwards (not reported for brevity). \citet{nakamura/steinsson:2018}
showed that the effect of policy news peaks at the 2-Year maturity
and declines with longer maturities. Consistent with this, we find
weaker statistical significance for the 5-Year. In line with theoretical
predictions, the long-run impact of monetary policy shocks on real
interest rates (i.e., the 10 Year forwards) approaches zero (not reported).

\begin{table}
\caption{\label{Table: Inference NS18}Identification-Robust Inference on $\beta$}

\smallskip{}

\begin{centering}
{\small{}}{\small{}%
\begin{tabular}{ccccccccccccc}
\hline 
 & \multicolumn{6}{c}{{\small 30-minute Policy News}} & \multicolumn{6}{c}{{\small 1-day change in 2-Year Yields}}\tabularnewline
{\small 2-Year Forwards} & \multicolumn{3}{c}{{\small Nominal}} & \multicolumn{3}{c}{{\small Real}} & \multicolumn{3}{c}{{\small Nominal}} & \multicolumn{3}{c}{{\small Real}}\tabularnewline
{\small$\alpha$} & {\small 0.10} & {\small 0.05} & {\small 0.01} & {\small 0.10} & {\small 0.05} & {\small 0.01} & {\small 0.10} & {\small 0.05} & {\small 0.01} & {\small 0.10} & {\small 0.05} & {\small 0.01}\tabularnewline
\hline 
{\small$LM_{T}$} & {\small$\checkmark$} & {\small$\checkmark$} & {\small$\times$} & {\small$\checkmark$} & {\small$\checkmark$} & {\small$\checkmark$} & {\small$\times$} & {\small$\times$} & {\small$\times$} & {\small$\checkmark$} & {\small$\checkmark$} & {\small$\times$}\tabularnewline
{\small$CLR_{T}$} & {\small$\checkmark$} & {\small$\checkmark$} & {\small$\times$} & {\small$\checkmark$} & {\small$\checkmark$} & {\small$\checkmark$} & {\small$\times$} & {\small$\times$} & {\small$\times$} & {\small$\checkmark$} & {\small$\checkmark$} & {\small$\times$}\tabularnewline
{\small$LM_{T}(\widehat{\mathbf{S}}_{T})$} & {\small$\checkmark$} & {\small$\checkmark$} & {\small$\times$} & {\small$\checkmark$} & {\small$\checkmark$} & {\small$\checkmark$} & {\small$\checkmark$} & {\small$\checkmark$} & {\small$\times$} & {\small$\checkmark$} & {\small$\checkmark$} & {\small$\checkmark$}\tabularnewline
{\small$CLR_{T}(\widehat{\mathbf{S}}_{T})$} & {\small$\checkmark$} & {\small$\checkmark$} & {\small$\times$} & {\small$\checkmark$} & {\small$\checkmark$} & {\small$\times$} & {\small$\checkmark$} & {\small$\checkmark$} & {\small$\checkmark$} & {\small$\checkmark$} & {\small$\checkmark$} & {\small$\checkmark$}\tabularnewline
\hline 
\end{tabular}}{\small}{\small\par}
\par\end{centering}
{\scriptsize{}}{\scriptsize{}%
\noindent\begin{minipage}[t]{1\columnwidth}%
{\scriptsize Weak identification-robust tests on $\beta$. The dependent
variable $\widetilde{Y}_{t}$ is the 2-Year forward rates. $\widetilde{D}_{t}$
is either the 30-minute policy news series or the 1-day nominal Treasury
yields. Significance levels are $\alpha=0.10,\,0.05,\,0.01$. A $\checkmark$
indicates rejection $H_{0};$ a $\times$ non-rejection.}%
\end{minipage}}{\scriptsize\par}
\end{table}

Let us instead consider the 1-day change in 2-Year yields as the policy
variable. The existing $LM_{T}$ and $LR_{T}$ tests do not reject
the null at any standard significance level for nominal forwards,
and at the 1\% level for real forwards. In contrast, our proposed
tests based on $\widehat{\mathbf{S}}_{T}$ show much stronger rejections,
aligning with the results using the 30-minute policy news series,
which indicate a positive causal effect on 2-Year forwards.

\subsection{Identification and Estimation of Compliers, and Exclusion Restriction }

\label{Subsection: Identification and Estimation of Compliers}We
now identify compliers individually by applying Theorem \ref{Theorem: Compliers}.
Under heteroskedasticity-based identification, the sample rolling
window averages in Theorem \ref{Theorem: Compliers} correspond to
rolling window variances, i.e., $\overline{D}_{P,t,n_{1}}$ and $\overline{D}_{C,t,n_{0}}$
are equal to $\overline{\sigma}_{P,t,n_{1}}^{2}$ and $\overline{\sigma}_{C,t,n_{0}}^{2}$
in this context, where $\overline{\sigma}_{P,t,n_{1}}^{2}$ and $\overline{\sigma}_{C,t,n_{0}}^{2}$
are defined analogously but using $\tilde{D}_{t}^{2}$ for $D_{t}$.\footnote{More specifically, we have $\overline{\sigma}_{C,t_{0}-1,n_{0}}^{2}=\frac{1}{n_{0}}\sum_{s\in N_{0}\left(t_{0}\right)}\tilde{D}_{s}^{2},$
and $\overline{\sigma}_{P,t_{0},n_{1}}^{2}=\frac{1}{n_{1}}\sum_{s\in N_{1}\left(t_{0}\right)}\tilde{D}_{s}^{2}$.\textcolor{blue}{{}}} We use two-sided rolling windows with $n_{0}=101$ and $n_{1}=15$.\footnote{We also used one-sided backward rolling windows, and the results are
essentially unchanged. } For each $t_{0}$ we test the null hypothesis $H_{0}:\,\mathbb{E}(D_{t_{0}}^{2}\left(1\right))-\mathbb{E}(D_{t_{0}}^{2}\left(0\right))=0$
($t_{0}$ is a non-complier) versus the one tailed alternative $H_{1}:\,\mathbb{E}(D_{t_{0}}^{2}\left(1\right))-\mathbb{E}(D_{t_{0}}^{2}\left(0\right))>0$
($t_{0}$ is a complier). We use the $t$-statistic 
\begin{align*}
t_{t_{0}} & =\begin{cases}
\frac{\sqrt{n_{0}}\left(\overline{\sigma}_{P,t_{0},n_{1}}^{2}-\overline{\sigma}_{C,t_{0}-1,n_{0}}^{2}\right)}{\sqrt{J_{\mathrm{HAC},t_{0}-1}\left(1+n_{0}/n_{1}\right)}} & t_{0}\in\mathbf{P}\\
\frac{\sqrt{n_{0}}\left(\overline{\sigma}_{P,s^{*}\left(t_{0}\right),n_{1}}^{2}-\overline{\sigma}_{C,t_{0},n_{0}}^{2}\right)}{\sqrt{J_{\mathrm{HAC},t_{0}}\left(1+n_{0}/n_{1}\right)}} & t_{0}\in\mathbf{C},
\end{cases}
\end{align*}
where $J_{\mathrm{HAC},t_{0}}$ is the Newey-West estimator with $\left\lfloor n_{0}^{1/3}\right\rfloor $
lags applied to $\tilde{D}_{s}^{2}-n_{0}^{-1}\sum_{k\in N_{0}\left(t_{0}+1\right)}\tilde{D}_{k}^{2}$.

\begin{singlespace}
\noindent\begin{flushleft}
\begin{center}
\begin{figure}[h]
\begin{raggedright}
\includegraphics[clip,width=17cm,totalheight=8cm]{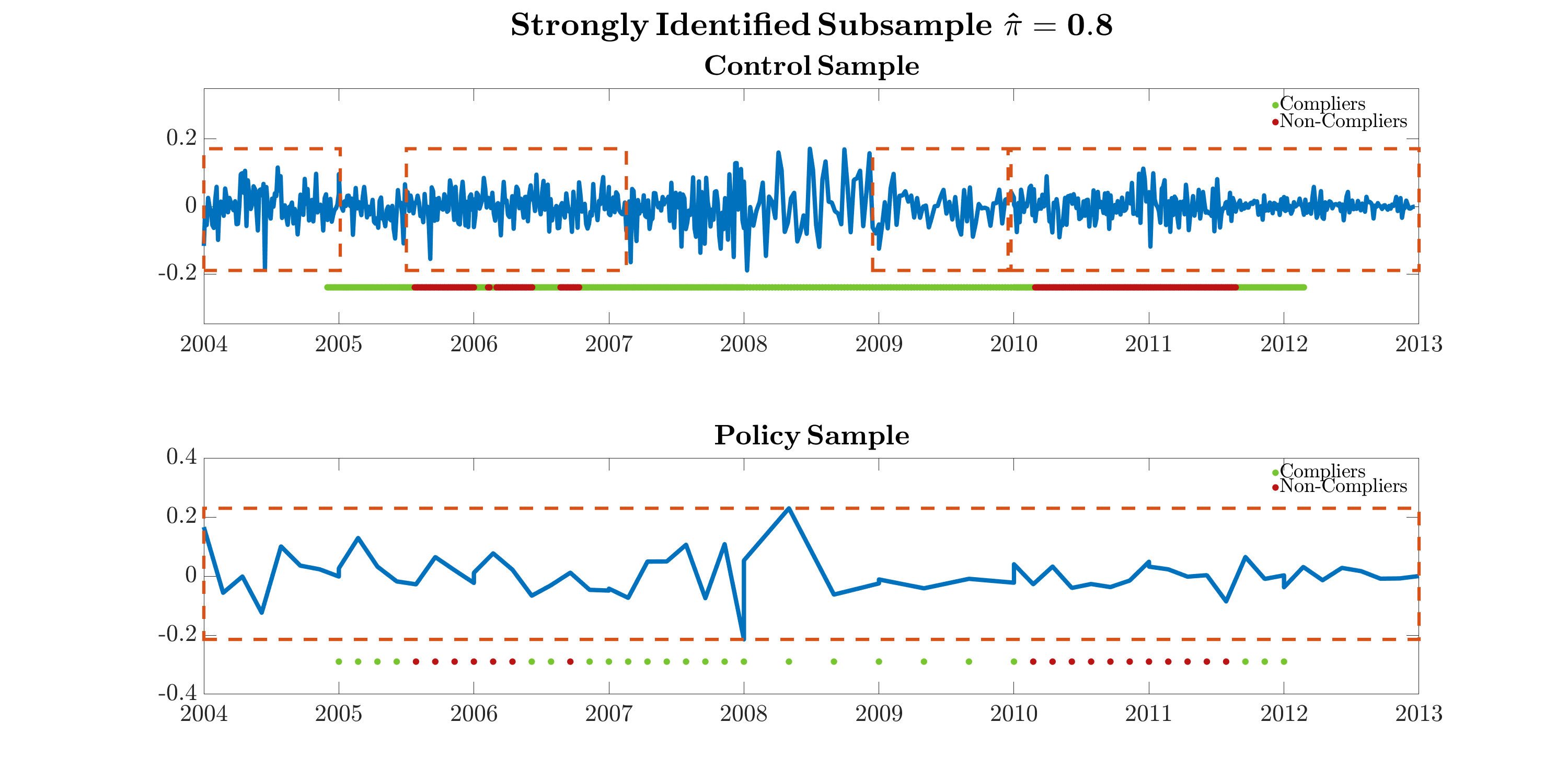}
\par\end{raggedright}
\raggedright{}\caption{\label{Figure Compliers}{\scriptsize Plot of $\widetilde{D}_{t}$
(2-Years Treasury yields) in the control sample (top panel) and policy
sample (bottom panel). The orange rectangles indicate subsamples included
in the strongly-identified set $\mathbf{\widehat{S}}_{T,OLS}$ where
$\widehat{\pi}=0.8$. Green filled circles indicate compliers; red
filled circles indicate non-compliers. Time points without colored
markers correspond to cases where rolling sample variances could not
be computed due to proximity to the start or end of the sample.}}
\end{figure}
\end{center}
\par\end{flushleft}
\end{singlespace}

The results at the 0.10 significance level are shown in Figure \ref{Figure Compliers}.
Approximately 62\% of observations are classified as compliers. The
non-compliers are mostly concentrated in the period from 2010 to mid-2011,
which corresponds to the early phase of the zero lower bound (ZLB)
period following the 2008\textendash 09 recession. During this time,
the Fed relied primarily on qualitative forward guidance\textemdash e.g.,
stating that economic conditions were "likely to warrant exceptionally
low levels of the federal funds rate for some time." In August 2011,
the Fed shifted to more explicit, calendar-based guidance, stating
that such conditions were "likely to warrant exceptionally low levels
of the federal funds rate at least through mid-2013." Thus, the non-complier
period aligns with the phase of the ZLB when forward guidance was
less aggressive as the policy announcements by then only imply a near
zero-rate horizon for the following three to four quarters, significantly
shorter than what the ZLB constraint would actually have implied.
Non-compliers also appear intermittently between mid-2005 and mid-2006.
One possible explanation is that this period coincided with the latter
stages of the Fed's highly predictable tightening cycle, during which
policy rates were increased by 25 basis points at successive meetings
and forward guidance emphasized a ``measured pace'' of tightening.
Because markets largely anticipated these policy actions, FOMC announcements
may have conveyed relatively little new information, resulting in
weaker local compliance. However, these episodes are less persistent
than those observed during the early ZLB period.\footnote{The set of compliers does not coincide with the set of observations
in the strongly-identified $\widehat{\pi}$-sample. However, this
does not necessarily imply a violation of monotonicity (cf. Proposition
\ref{Proposition: Compliers FS indivdually}). First, the complier
status is determined via a $t$-test whereas the strongly-identified
$\widehat{\pi}$-sample is determined via estimation. Second, the
complier status is determined by the rolling window variances at each
$t$, whereas inclusion in the $\widehat{\pi}$-sample depends on
how these variances contribute to the average volatility in the control
sample relative to that in the policy sample. In other words, inclusion
in the $\widehat{\pi}$-sample is a global criterion, influenced by
volatility elsewhere in the sample, while complier status reflects
whether the first stage holds individually.}

Finally, we use Theorem \ref{Theorem: Compliers} and Proposition
\ref{Proposition: exclusion-restriction} to test the exclusion restriction
(cf. Assumption \ref{Assumption: Exclusion }). Here the exclusion
restriction means that the volatility of the non-monetary policy shocks
remain constant across FOMC and non-FOMC days. We consider the whole
set of non-compliers $\mathcal{NC}.$ We test the null hypothesis
that the exclusion restriction holds by using the following $t$-statistic:
\begin{align*}
t_{\mathrm{exclusion}}=\frac{\sqrt{|\mathcal{NC}_{\mathbf{C}}|}\left(\overline{Y}_{\mathcal{NC}_{\mathbf{P}}^{s}}-\overline{Y}_{\mathcal{NC}_{\mathbf{C}}^{s}}\right)}{\sqrt{J_{\mathrm{HAC},\mathcal{NC}^{s}\left(1+|\mathcal{NC}_{\mathbf{C}}|/|\mathcal{NC}_{\mathbf{P}}|\right)}}}
\end{align*}
where $\overline{Y}_{\mathcal{NC}_{\mathbf{P}}^ {}}=\frac{1}{|\mathcal{NC}_{\mathbf{P}}^ {}|}\sum_{t\in\mathcal{NC}_{\mathbf{P}}^ {}}Y_{t}$,
$\overline{Y}_{\mathcal{NC}_{\mathbf{C}}^ {}}=\frac{1}{|\mathcal{NC}_{\mathbf{C}}^ {}|}\sum_{t\in\mathcal{NC}_{\mathbf{C}}^ {}}Y_{t}$,
$Y_{t}=\tilde{D}_{t}\tilde{Y}_{t}$ and $J_{\mathrm{HAC},\mathcal{NC}^{s}}$
is the Newey-West estimator applied to $\overline{Y}_{\mathcal{NC}_{\mathbf{P}}^ {}}-\overline{Y}_{\mathcal{NC}_{\mathbf{C}}^ {}}$.
For the real (nominal) 2-Year forward rate, we find $t_{\mathrm{exclusion}}=0.01$
($t_{\mathrm{exclusion}}=0.60$), and thus fail to reject the exclusion
restriction. This supports the ``background noise'' interpretation
of \citet{nakamura/steinsson:2018} as non-monetary shocks appear
to have constant variance across FOMC and non-FOMC days.

\section{\label{Section Conclusions}Conclusions}

This paper discusses identification, estimation and inference on dynamic
LATE. We show that compliers can be identified individually and the
exclusion restriction can be tested using a $t$-test. While weak
identification is common in the full sample in practice, strong identification
often appears to hold in a sizable subsample. We propose a method
to isolate this strongly-identified subsample, enabling consistent
estimation and inference.

\paragraph*{Supplemental Materials:}

The online supplement {[}cf. \textcolor{MyBlue}{Casini et al.} \citeyearpar{casini/mccloskey/pala/rolla_Dynamic_Late_Supp}{]}
includes Monte Carlo simulations, proofs of the results of Sections
\ref{Section: Statistical Framework for Identification of Causal Effects}-\ref{Section: ID Failure Test}
and \ref{Section: Inference}. The non-online supplement {[}cf. \textcolor{MyBlue}{Casini et al.}
\citeyearpar{casini/mccloskey/pala/rolla_Dynamic_Late_Supp_Not_Online}{]}
contains the theoretical results and corresponding proofs for the
estimators in Section \ref{Section: Estimation} and additional results.

\begin{singlespace}

\bibliographystyle{econometrica}
\bibliography{References}
\addcontentsline{toc}{section}{References}

\end{singlespace}

\newpage{}

\newpage{}

\clearpage 
\pagenumbering{arabic}
\renewcommand*{\thepage}{A-\arabic{page}}
\appendix

\pagebreak{}

\section*{}
\addcontentsline{toc}{part}{Supplemental Material}

\begin{center}
\title{\textbf{\Large{Online Supplement to ``Dynamic Local Average Treatment Effects in Time Series"}}} 
\maketitle
\end{center}
\medskip{} 
\medskip{} 
\medskip{} 
\thispagestyle{empty}

\begin{center}
 \author{\textsc{\textcolor{MyBlue}{Alessandro Casini}}
\quad \quad \quad \quad \quad  \,
\textsc{\textcolor{MyBlue}{Adam McCloskey}}}
\\
\small{University of Rome Tor Vergata} 
\quad \, 
\small{University of Colorado at Boulder}
\\
\medskip{}
\medskip{}
\author{\textsc{\textcolor{MyBlue}{Luca Rolla}}
\quad \quad \quad \quad \quad \quad \quad  \quad \,
\textsc{\textcolor{MyBlue}{Raimondo Pala}}}
\\
\small{University of Rome Tor Vergata}
\quad \, 
\small{University of Rome Tor Vergata}
\\
\medskip{}
\medskip{} 
\medskip{} 
\medskip{} 
\date{\small{\today}} 
\medskip{} 
\medskip{} 
\medskip{} 
\end{center}
\begin{abstract}
{\footnotesize This supplemental material is for online publication.
It contains additional results on identification-robust inference,
Monte Carlo simulations and proofs of the results.}{\footnotesize\par}
\end{abstract}
\setcounter{page}{0}
\setcounter{section}{0}
\renewcommand*{\theHsection}{\the\value{section}}

\newpage{}

\noindent 

\allowdisplaybreaks


\renewcommand{\thepage}{S-\arabic{page}}   
\renewcommand{\thesection}{S.\Alph{section}}   
\renewcommand{\theequation}{S.\Alph{section}.\arabic{equation}}




\section{Critical Values of $F_{T}^{*}$}

\begin{table}[H]
\caption{\label{Table CVs of F* alpha=00003D0.05}Critical values of $F_{T}^{*}$}

\centering{}%
\begin{tabular}{ccccccc}
\hline 
 & \multicolumn{6}{c}{\begin{cellvarwidth}[t]
\centering
\begin{lyxlist}{}
\item [{$\alpha=0.10$}]~
\end{lyxlist}
\end{cellvarwidth}}\tabularnewline
{\small$\pi_{L}\backslash q$} & {\small 1} & {\small 2} & {\small 3} & {\small 4} & {\small 5} & {\small 10}\tabularnewline
\hline 
\hline 
{\small 0.50} & 11.56 & {\small 7.68} & {\small 6.04} & {\small 5.11} & {\small 4.51} & {\small 3.15}\tabularnewline
{\small 0.60} & 10.22 & {\small 6.96} & {\small 5.53} & {\small 4.72} & {\small 4.19} & {\small 3.00}\tabularnewline
{\small 0.70} & {\small 8.54} & {\small 6.10} & {\small 4.92} & {\small 4.33} & {\small 3.87} & {\small 2.82}\tabularnewline
{\small 0.80} & {\small 6.89} & {\small 5.26} & {\small 4.32} & {\small 3.77} & {\small 3.51} & {\small 2.59}\tabularnewline
{\small 0.90} & {\small 5.46} & {\small 4.16} & {\small 3.47} & {\small 3.15} & {\small 2.92} & {\small 2.29}\tabularnewline
{\small 1.00} & {\small 2.68} & {\small 2.32} & {\small 2.07} & {\small 1.94} & {\small 1.85} & {\small 1.61}\tabularnewline
 & \multicolumn{6}{c}{{\small$\alpha=0.05$}}\tabularnewline
{\small$\pi_{L}\backslash q$} & {\small 1} & {\small 2} & {\small 3} & {\small 4} & {\small 5} & {\small 10}\tabularnewline
{\small 0.50} & {\small 14.03} & {\small 9.02} & {\small 6.94} & {\small 5.78} & {\small 5.03} & {\small 3.44}\tabularnewline
{\small 0.60} & {\small 12.74} & {\small 8.21} & {\small 6.38} & {\small 5.34} & {\small 4.70} & {\small 3.29}\tabularnewline
{\small 0.70} & {\small 10.70} & {\small 7.27} & {\small 5.75} & {\small 4.97} & {\small 4.38} & {\small 3.10}\tabularnewline
{\small 0.80} & {\small 8.88} & {\small 6.36} & {\small 5.10} & {\small 4.39} & {\small 3.96} & {\small 2.86}\tabularnewline
{\small 0.90} & {\small 7.02} & {\small 5.19} & {\small 4.15} & {\small 3.71} & {\small 3.41} & {\small 2.56}\tabularnewline
{\small 1.00} & {\small 3.85} & {\small 3.05} & {\small 2.58} & {\small 2.37} & {\small 2.23} & {\small 1.84}\tabularnewline
 & \multicolumn{6}{c}{{\small$\alpha=0.01$}}\tabularnewline
{\small$\pi_{L}\backslash q$} & {\small 1} & {\small 2} & {\small 3} & {\small 4} & {\small 5} & {\small 10}\tabularnewline
{\small 0.50} & {\small 19.83} & {\small 11.65} & {\small 8.88} & {\small 7.16} & {\small 6.22} & {\small 4.01}\tabularnewline
{\small 0.60} & {\small 17.89} & {\small 10.88} & {\small 8.21} & {\small 6.74} & {\small 5.75} & {\small 3.89}\tabularnewline
{\small 0.70} & {\small 15.58} & {\small 9.78} & {\small 7.62} & {\small 6.37} & {\small 5.50} & {\small 3.67}\tabularnewline
{\small 0.80} & {\small 13.11} & {\small 8.72} & {\small 6.69} & {\small 5.61} & {\small 4.93} & {\small 3.45}\tabularnewline
{\small 0.90} & {\small 10.66} & {\small 7.28} & {\small 5.67} & {\small 4.80} & {\small 4.38} & {\small 3.10}\tabularnewline
{\small 1.00} & {\small 6.46} & {\small 4.53} & {\small 3.72} & {\small 3.30} & {\small 3.00} & {\small 2.35}\tabularnewline
\end{tabular}
\end{table}

\section{\label{Section: Additional Resuls Identification-Robust Inference }Additional
Results on Identification-Robust Inference}

We present the sufficiency results referenced in Section \ref{Section: Inference}
as well as other results. 

\subsection{\label{Subsection: Known Subpopulation}Known Sub-Population}

When $\mathbf{S}_{0,T}$ is known, it is straightforward to use existing
tests in the identification-robust linear IVs literature to test $H_{0}$
with known optimality properties under certain conditions. However,
one must be careful in defining the appropriate statistics when applying
existing tests in this setting in order to maintain efficiency.

With fixed regressors $X$ and $Z$ and reduced-form errors $v$ that
are i.i.d. across rows with each row being bivariate normally distributed
with a mean of zero and a known nonsingular covariance matrix $\Sigma_{v}$,
careful application of the results of \citeReferencesSupp{andrews/moreira/stock:2006}
imply that $\overline{Z}(C_{0,T})^{\prime}y$ is a sufficient statistic
for $(\beta,\theta^{\prime})^{\prime}$. This implies that there can
be no loss in efficiency from focusing on tests that are functions
of only $\overline{Z}(C_{0,T})^{\prime}y$. On the other hand, the
following proposition implies a loss in efficiency from tests that
are functions of only $Z^{\prime}M_{X}y$, which a casual user may
be tempted to use when constructing identification-robust tests for
$\beta$.
\begin{assumption}
\label{Assumption: S.B.X-(Non-redundancy)} $\mathrm{rank}\left(\left(I-P_{M_{X}Z}\right)M_{X}C_{0,T}Z\right)>0$.
\end{assumption}
Assumption \ref{Assumption: S.B.X-(Non-redundancy)} requires that
the residualized instrument variation in $\mathbf{S}_{0,T}$ is not
fully spanned by the full-sample residualized instrument. This rules
out degenerate designs in which $M_{X}C_{0,T}Z$ and $M_{X}Z$ have
the same relevant span.
\begin{prop}
\label{Lemma 1 AMS}For the model in \eqref{Eq. (2.5) AMS} with fixed
regressors $X$ and $Z$ and reduced-form errors $v$ that are i.i.d.
across rows with each row being bivariate normal with a zero mean
and known p.d. covariance matrix $\Sigma_{v}$. If $\pi_{0}<1$ and
Assumption \ref{Assumption: S.B.X-(Non-redundancy)} holds, then $Z^{\prime}M_{X}y$
is not sufficient for $\left(\beta,\,\theta'\right)'$.
\end{prop}
This result implies that existing tests (i.e.,. CLR, LM and AR tests)
and extensions thereof are not efficient when $Z$ is treated as the
matrix of IVs rather than $C_{0,T}Z$.

The results of \citeReferencesSupp{andrews/moreira/stock:2006} imply
that the CLR, LM and AR tests have limiting null rejection probabilities
equal to $\alpha$ under weak instrument asymptotics. In addition,
results in \citeReferencesSupp{andrews/moreira/stock:2006} imply
asymptotic near-optimality properties of the CLR test under a stronger
set of assumptions that may not hold in the presence of serial correlation
in $\{v_{t}\}$.

\subsection{\label{Subsection: Unknown Subpopulation}Unknown Sub-Population}

In Section \ref{Section: Inference} we propose CLR, LM and AR statistics
where we plug-in the estimate for the unknown sub-population:~$CLR_{T}(\widehat{\mathbf{S}}_{T})$,
$LM_{T}(\widehat{\mathbf{S}}_{T})$ and $AR_{T}(\widehat{\mathbf{S}}_{T})$.
To motivate their use in the unknown sub-population setting, we establish
the analog of the sufficiency result of \citeReferencesSupp{andrews/moreira/stock:2006}
in this setting. 
\begin{prop}
\label{Lemma 1 AMS unknown subpopulation}For the model in \eqref{Eq. (2.5) AMS}
with fixed regressors $X$ and $Z$ and reduced form errors $v$ that
are i.i.d.~across rows with each row being bivariate normal with
a zero mean, known p.d. covariance matrix $\Sigma_{v}$ and an unknown
sub-population $\mathbf{S}_{0,T}\in\mathcal{S}$, the Gaussian process
$\{\overline{Z}(C_{T})'y\}_{\mathbf{S}_{T}\in\mathcal{S}}$ is a sufficient
statistic for $(\beta,\theta')'$.
\end{prop}
Since the AR, LM and CLR statistics are only functions of $\{\overline{Z}(C_{T})'y\}_{\mathbf{S}_{T}\in\mathcal{S}}$,
this result implies that these tests entail no loss in efficiency
relative to tests using the entire data.

We finally show that under weak IV asymptotics, $\widehat{\mathbf{S}}_{T}$
is the maximum likelihood estimator of $\mathbf{S}_{T}$ under $H_{0}$,
so that the sub-population estimate we use when constructing and interpreting
the identification-robust tests is efficient.
\begin{prop}
\label{MLE for subpopulation} For the model in \eqref{Eq. (2.5) AMS}
with fixed regressors $X$ and $Z$ and reduced form errors $v$ that
are i.i.d.~across rows with each row being bivariate normal with
a zero mean, p.d. covariance matrix $\Sigma_{v}$ and unknown sub-population
$\mathbf{S}_{T}\in\mathcal{S}$, if Assumptions \ref{Assumption 1 AMS}
and \ref{Assumption 3 AMS}-\ref{Assumption Uniform Consistent Covariance Matrix}
hold and $\theta=c/T^{1/2}$ for some nonstochastic $c\in\mathbb{R}^{q}$,
$\widehat{\mathbf{S}}_{T}$ is asymptotically equivalent to the maximum
likelihood estimator of $\mathbf{S}_{T}$ under $H_{0}$.
\end{prop}
The result in Proposition \ref{MLE for subpopulation} continues to
hold under strong IVs, i.e., $\theta\neq0$ is fixed. See \citeReferencesSupp{casini/mccloskey/pala/rolla_Dynamic_Late_Supp_Not_Online}.

\citeReferencesSupp{magnusson/mavroeidis:2014} propose tests for
$\beta$ that are robust to changes in $\theta$ whether through persistent
time variation or breaks. For the latter case, they assumed the number
of breaks is known, whereas our approach does not require this prior
knowledge. For the weak IVs case, \citeReferencesSupp{magnusson/mavroeidis:2014}
consider a single break and build split-sample tests based on the
sufficient statistic $\{Z\left(\tau\right)'y\}_{\tau\in\left[0,\,1\right]}$
where $Z\left(\tau\right)=[[\left\{ Z'_{t}\right\} _{t=1}^{\left\lfloor \tau T\right\rfloor }\,\mathbf{0']}'\,\vdots\,[\mathbf{0'}\,\left\{ Z'_{t}\right\} _{t=\left\lfloor \tau T\right\rfloor +1}^{T}]']$.
This high-dimensional statistic effectively uses the full data sequence
evaluated at all potential split points. In contrast, our framework
focuses on subsamples where $\theta$ is nonzero, allowing us to construct
a lower-dimensional sufficient statistic. In other words, their statistic
is not minimal sufficient {[}cf. \citeReferencesSupp{lehmann/romano:05}{]},
whereas ours is\textemdash making our tests more efficient in this
setting.

\section{\label{Section: Monte-Carlo-Simulations}Monte Carlo Simulations}

\subsection{Finite-Sample Size and Power of $F_{T}^{*}$}

We study the finite-sample rejection frequencies of the $F_{T}^{*}$
test using a simulation experiment calibrated to real data from the
analysis in \citeReferencesSupp{nakamura/steinsson:2018} introduced
in Section \ref{Section: Application: Money Neutrality}. We use the
same sequences of policy dates and control dates, $\mathbf{P}$ and
$\mathbf{C}$, to construct the instrument as 
\begin{align*}
Z_{t} & =\begin{cases}
\frac{T}{T_{P}}, & t\in\mathbf{P}\\
-\frac{T}{T_{C}}, & t\in\mathbf{C}.
\end{cases}
\end{align*}
Under heteroskedasticity-based identification the first-stage equation
can be equivalently written as $D_{t}=\theta_{}Z_{t}D_{t}+e_{t}$
for some $\theta$ and $e_{t}$ {[}cf. \citeReferencesSupp{rigobon/sack:2003}
and \citeReferencesSupp{lewis:2020}{]}. Hence, we generate $D_{t}$
according to the following data-generating process (DGP): 
\begin{align}
D_{t} & =\begin{cases}
\frac{e_{t}}{1-\theta_{1}Z_{t}}, & t\leq\left\lfloor T/4\right\rfloor \\
\frac{e_{t}}{1-\theta_{2}Z_{t}}, & \left\lfloor T/4\right\rfloor +1\leq t\leq\left\lfloor T/4\right\rfloor +\left\lfloor \left(1-\pi_{0}\right)T\right\rfloor \\
\frac{e_{t}}{1-\theta_{3}Z_{t}}, & \left\lfloor T/4\right\rfloor +\left\lfloor \left(1-\pi_{0}\right)T\right\rfloor +1\leq t\leq T,
\end{cases}\label{Eq. Calibrated DGP}
\end{align}
where $\pi_{0}=0.4,\,0.6,\,0.8$ and $T=400.$ We set $T_{P}$ and
$T_{C}$ equal to the number of policy and control dates that occur
in the first $T$ observations in \citeauthor{nakamura/steinsson:2018}'s
\citeyearpar{nakamura/steinsson:2018} sample. We specify $e_{t}=\rho_{e}e_{t-1}+v_{e,t}$,
where $\rho_{e}\in\{0,\,0.25,\,0.5,\,0.75\}$, $v_{e,t}\sim\mathrm{i.i.d.}\,\mathscr{N}\left(0,\,\sigma_{v}^{2}\right)$
and $\sigma_{v}^{2}$ is set equal to the sample variance of the policy
variable (2-years nominal Treasury yields). We set $\theta_{1}=\theta_{2}=\theta_{3}=0$
under the null hypothesis. Under the alternative, $\theta_{1}=\theta_{3}>0$
and $\theta_{2}=0.$

We also consider the following DGP:\footnote{This is also used to compare the performance of the estimators of
$\beta$ discussed in Section \ref{Section: Estimation} and of the
identification-robust tests discussed in Section \ref{Section: Inference}.} 
\begin{align}
Y_{t} & =\beta D_{t}+\gamma_{1}X_{t}+u_{t},\label{Eq. (DGP Y sims)}
\end{align}
where $X_{t}\sim\mathrm{i.i.d.\,\mathscr{N}\left(1,\,1\right)}$ for
all $t$ and 
\begin{align}
D_{t} & =\begin{cases}
\theta_{1}Z_{t}+\gamma_{2}X_{t}+e_{t}, & t\leq\left\lfloor T/4\right\rfloor \\
\theta_{2}Z_{t}+\gamma_{2}X_{t}+e_{t}, & \left\lfloor T/4\right\rfloor +1\leq t\leq\left\lfloor T/4\right\rfloor +\left\lfloor \left(1-\pi_{0}\right)T\right\rfloor \\
\theta_{3}Z_{t}+\gamma_{2}X_{t}+e_{t}, & \left\lfloor T/4\right\rfloor +\left\lfloor \left(1-\pi_{0}\right)T\right\rfloor +1\leq t\leq T,
\end{cases}\label{Eq. (DGP D sims)}
\end{align}
$Z_{t}\sim\mathrm{i.i.d.\,\mathscr{N}\left(1,\,1\right),}$ and $u_{t}$
and $e_{t}$ are i.i.d. jointly normal with mean zero and covariance
\begin{align}
\Sigma_{ue} & =\begin{bmatrix}1 & \rho\\
\rho & 1
\end{bmatrix},\label{Eq. cov matrix Sigma_eu}
\end{align}
with $\rho\in\left\{ 0.25,\,0.50,\,0.75\right\} $ and $\gamma_{1}=\gamma_{2}\in\left\{ 0,\,1\right\} .$
Under the null hypothesis we set $\theta_{1}=\theta_{2}=\theta_{3}=0$.
Under the alternative hypothesis we set $\theta_{1}=\theta_{3}=dT^{-1/2}$
with $d\in\left\{ 4,\,10,\,16\right\} $ and $\theta_{2}=0$. We also
consider two additional specifications for $\theta_{2}$. In the first,
$\theta_{2}=-0.5dT^{-1/2}$, so $\theta_{2}$ has the opposite sign
to $\theta_{1}$ and $\theta_{3}$. In the second, $\theta_{2}=-0.5$.
In both cases the instrument is relevant throughout the sample (no
identification failure). However, in the second regime the first-stage
effect tends to offset instrument relevance in the other regimes because
of the sign reversal. We set $\pi_{0}\in\{0.6,\,0.8\}$ and $T=200$.
We also consider a variant of the DGP with serially correlated data.
We assume $u_{t}=\rho_{u}u_{t-1}+v_{u,t}$ and $e_{t}=\rho_{e}e_{t-1}+v_{e,t}$
with $v_{u,t}$ and $v_{e,t}$ being jointly normal with mean zero
and covariance $\Sigma_{ue}$ as in \eqref{Eq. cov matrix Sigma_eu}
with $\rho\in\{0.25,\,0.50,\,0.75\}$.

Throughout the simulation study, $F_{T}^{*}$ is implemented with
$\pi_{L}=0.6,\,\epsilon=0.05$, $m_{+}=5$. For both $F_{T}^{*}$
and the full sample $F_{T}$ we use the Newey-West estimator with
bandwidth equal to the popular rule $\left\lfloor T^{-1/3}\right\rfloor $
for $\widehat{J}(\cdot)$.\footnote{We also consider data-dependent bandwidths. The results are similar
and not reported.} The significance level is 5\% and the number of simulations is 5,000.
Figure \ref{Figure_1_Sect_4} plots the rejection rates of $F_{T}^{*}$
and the full sample $F_{T}$ for the calibrated DGP in \eqref{Eq. Calibrated DGP}
with $\pi_{0}=0.6$. Both $F_{T}$ and $F_{T}^{*}$ yield accurate
rejection rates, providing evidence for the reliability of our empirical
results in Section \ref{Section: Empirical Evidence}. $F_{T}^{*}$
is much more powerful than $F_{T}$ by about across all values of
$\rho_{e}$. 
\noindent\begin{flushleft}
\begin{center}
\begin{figure}[h]
\begin{raggedright}
\includegraphics[clip,width=17cm,totalheight=7cm]{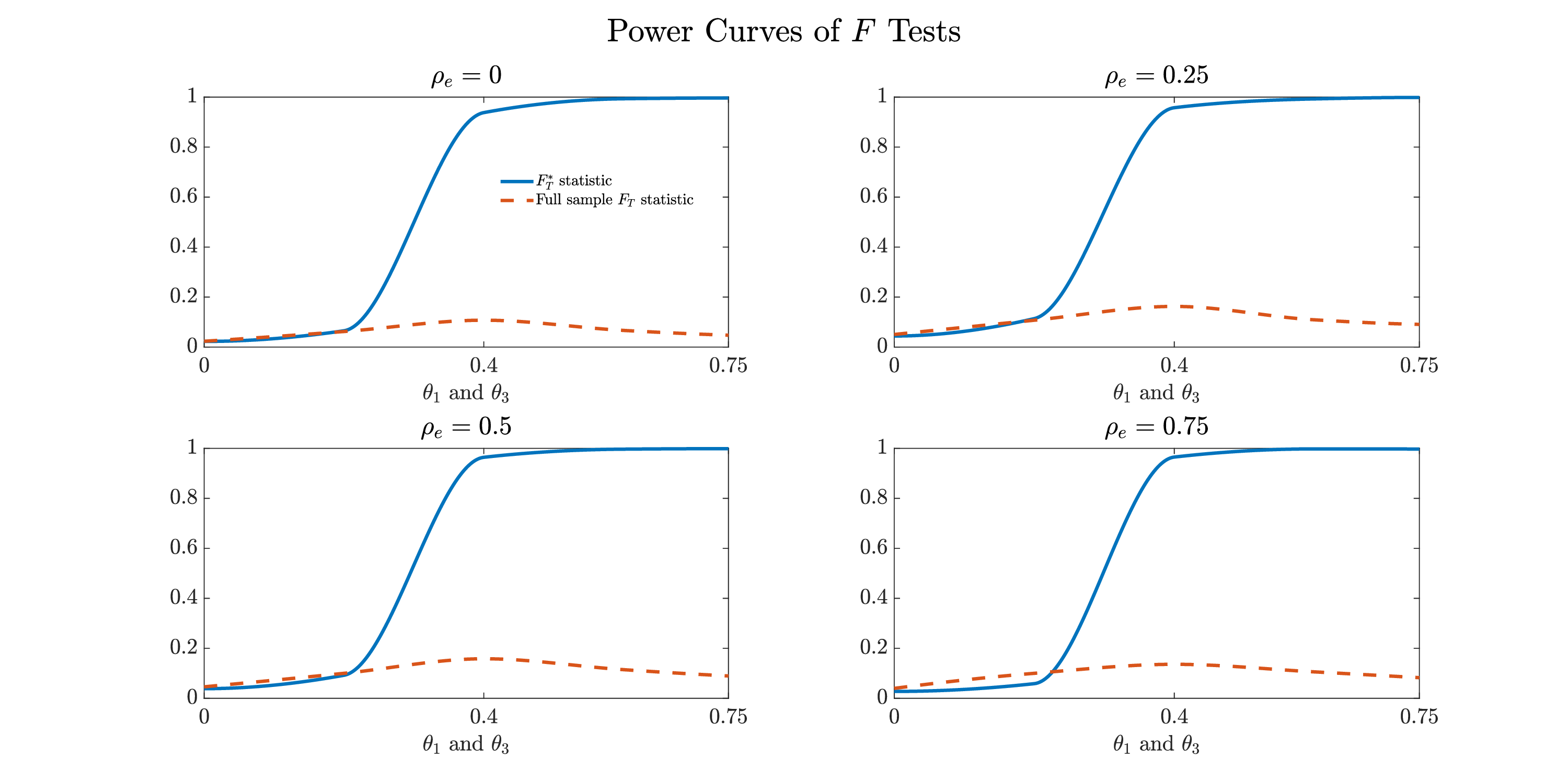}
\par\end{raggedright}
\raggedright{}\caption{\label{Figure_1_Sect_4}{\scriptsize Power curves of $F_{T}^{*}$ and
$F_{T}$ for the DGP calibrated to \citeauthor{nakamura/steinsson:2018}'s
\citeyearpar{nakamura/steinsson:2018} sample with $T=400$.}}
\end{figure}
\end{center}
\par\end{flushleft}

Moving on to the DGP specified in \eqref{Eq. (DGP Y sims)}-\eqref{Eq. (DGP D sims)},
Table \ref{Table: Size Power F tests} presents the size and size-adjusted
power of the $F$ tests for the case $\theta_{2}=-0.5$ under the
alternative. Both statistics show accurate null rejection rates. The
size-adjusted power of $F_{T}^{*}$ is much higher than that of  the
full sample $F_{T}$ for all values of $d$. Power gains are larger
when $\pi_{0}=0.6$ than when 0.8 since in the former case the strongly-identified
subsample is smaller, making the full sample $F_{T}$ rely more heavily
on subsamples that suffer from identification failure.

\begin{table}[H]
\caption{\label{Table: Size Power F tests}Size and size-adjusted power of
$F$ tests under alternative hypothesis with $\theta_{2}=-0.5$}

\smallskip{}

\begin{centering}
{\footnotesize}{\small{}%
\begin{tabular}{cccccccc}
\hline 
{\footnotesize$\rho=0.25$, $\pi_{0}=0.6$} & {\footnotesize$\theta_{1}=\theta_{2}=\theta_{3}=0$ (null)} & {\footnotesize$d=4$} & {\footnotesize$d=8$} & {\footnotesize$d=12$} & {\footnotesize$d=16$} & {\footnotesize$d=20$} & {\footnotesize$d=24$}\tabularnewline
\hline 
\hline 
{\footnotesize full sample $F_{T}$} & {\footnotesize 0.061} & {\footnotesize 0.023} & {\footnotesize 0.123} & {\footnotesize 0.478} & {\footnotesize 0.797} & {\footnotesize 0.933} & {\footnotesize 0.981}\tabularnewline
{\footnotesize$F_{T}^{*}$} & {\footnotesize 0.058} & {\footnotesize 0.128} & {\footnotesize 0.426} & {\footnotesize 0.838} & {\footnotesize 0.971} & {\footnotesize 0.996} & {\footnotesize 1.000}\tabularnewline
{\footnotesize$\rho=0.75$, $\pi_{0}=0.6$} & {\footnotesize$\theta_{1}=\theta_{2}=\theta_{3}=0$ (null)} & {\footnotesize$d=4$} & {\footnotesize$d=8$} & {\footnotesize$d=12$} & {\footnotesize$d=16$} & {\footnotesize$d=20$} & {\footnotesize$d=24$}\tabularnewline
{\footnotesize full sample $F_{T}$} & {\footnotesize 0.063} & {\footnotesize 0.034} & {\footnotesize 0.107} & {\footnotesize 0.394} & {\footnotesize 0.712} & {\footnotesize 0.879} & {\footnotesize 0.957}\tabularnewline
{\footnotesize$F_{T}^{*}$} & {\footnotesize 0.035} & {\footnotesize 0.142} & {\footnotesize 0.389} & {\footnotesize 0.774} & {\footnotesize 0.943} & {\footnotesize 0.989} & {\footnotesize 0.998}\tabularnewline
{\footnotesize$\rho=0.25$, $\pi_{0}=0.8$} & {\footnotesize$\theta_{1}=\theta_{2}=\theta_{3}=0$ (null)} & {\footnotesize$d=2$} & {\footnotesize$d=6$} & {\footnotesize$d=10$} & {\footnotesize$d=14$} & {\footnotesize$d=18$} & {\footnotesize$d=22$}\tabularnewline
{\footnotesize full sample $F_{T}$} & {\footnotesize 0.061 } & {\footnotesize 0.038} & {\footnotesize 0.642} & {\footnotesize 0.969} & {\footnotesize 0.998} & {\footnotesize 1.000} & {\footnotesize 1.000}\tabularnewline
{\footnotesize$F_{T}^{*}$} & {\footnotesize 0.058} & {\footnotesize 0.087} & {\footnotesize 0.853} & {\footnotesize 0.992} & {\footnotesize 1.000} & {\footnotesize 1.000} & {\footnotesize 1.000}\tabularnewline
{\footnotesize$\rho=0.75$, $\pi_{0}=0.8$} & {\footnotesize$\theta_{1}=\theta_{2}=\theta_{3}=0$ (null)} & {\footnotesize$d=2$} & {\footnotesize$d=6$} & {\footnotesize$d=10$} & {\footnotesize$d=14$} & {\footnotesize$d=18$} & {\footnotesize$d=22$}\tabularnewline
{\footnotesize full sample $F_{T}$} & {\footnotesize 0.063} & {\footnotesize 0.049} & {\footnotesize 0.468} & {\footnotesize 0.907} & {\footnotesize 0.993} & {\footnotesize 0.999} & {\footnotesize 1.000}\tabularnewline
{\footnotesize$F_{T}^{*}$} & {\footnotesize 0.035} & {\footnotesize 0.093} & {\footnotesize 0.696} & {\footnotesize 0.980} & {\footnotesize 0.999} & {\footnotesize 1.000} & {\footnotesize 1.000}\tabularnewline
\end{tabular}}{\small\par}
\par\end{centering}
\centering{}{\scriptsize{}%
\begin{minipage}[t]{0.8\columnwidth}%
\end{minipage}}{\scriptsize\par}
\end{table}

Figure \ref{Figure_2_Sect_4} plots the size-adjusted power of the
$F$ tests for the specification $\theta_{2}=-0.5dT^{-1/2}$.\footnote{Note that the size of the tests is that reported in Table \ref{Table: Size Power F tests}
since the DGPs for the two specifications are the same under the null.} In this specification, under the alternative the instrument is relevant
throughout the sample. However, because $\theta_{2}$ has opposite
sign to $\theta_{1}$ and $\theta_{3}$ the contribution of the second
regime tends to offset those of the first and third. Consistent with
this, the plots show that $F_{T}^{*}$ is substantially more powerful
than the full sample $F_{T}$. The resulting power gains exceed those
obtained when $\theta_{2}$ shares the same sign as $\theta_{1}$
and $\theta_{3}$. 
\noindent\begin{flushleft}
\begin{center}
\begin{figure}[h]
\begin{raggedright}
\includegraphics[clip,width=17cm,totalheight=7cm]{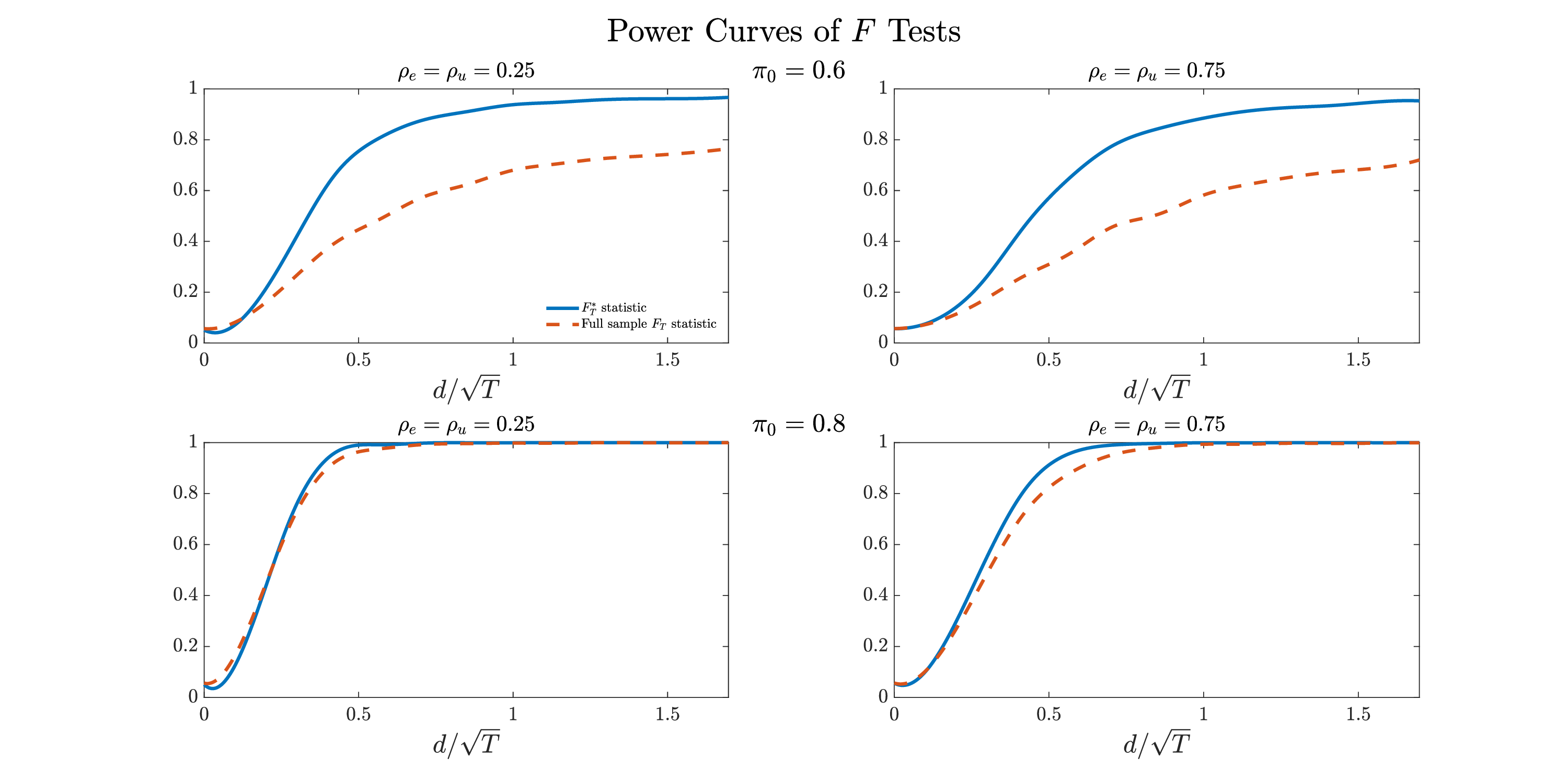}
\par\end{raggedright}
\raggedright{}\caption{\label{Figure_2_Sect_4}{\scriptsize Power curves of $F_{T}^{*}$ and
full sample $F_{T}$ for $\theta_{2}=-0.5dT^{-1/2}$ under the alternative
hypothesis with $T=200$.}}
\end{figure}
\end{center}
\par\end{flushleft}

\subsection{Finite-sample Size and Power of Weak Identification-Robust Tests}

We study the finite-sample rejection frequencies of the proposed tests
and compare their performance to the tests analyzed by \citeReferencesSupp{andrews/moreira/stock:2006}
and \citeReferencesSupp{magnusson/mavroeidis:2014}. The analysis
is based on the same DGP described in \eqref{Eq. (DGP Y sims)}-\eqref{Eq. (DGP D sims)}
with parameters $\rho\in\{0.25,\,0.5,\,0.75\}$, $\gamma_{1}=\gamma_{2}\in\left\{ 0,\,1\right\} $,
$\theta_{1}=\theta_{3}=dT^{-1/2}$ with $d\in\{4,\,8,\,10,\,16,\,24\}$
and $\theta_{2}=0.$ We consider values of $\pi_{0}\in\{0.4,\,0.6,\,0.8\}$
and $T=\{200,\,400\}$. Under the null hypothesis we set $\beta=0$. 

We compare the performance of our proposed test statistics $AR_{T}(\widehat{\mathbf{S}}_{T}),\,LM_{T}(\widehat{\mathbf{S}}_{T})$
and $CLR_{T}(\widehat{\mathbf{S}}_{T})$ with their full sample counterparts
and with the test statistics $\mathrm{Split}$-S, $\mathrm{Split}$-CLR,
qLL-S, ave-S and exp-S analyzed by \citeReferencesSupp{andrews/moreira/stock:2006}
and with the tests statistics $\mathrm{Split}$-S, $\mathrm{Split}$-CLR,
qLL-S, ave-S and exp-S proposed by \citeReferencesSupp{magnusson/mavroeidis:2014}.
For all tests, we consider heteroskedasticity and autocorrelation-robust
versions using the Newey-West estimator with bandwidth equal to the
popular rule $1/\left\lfloor T^{1/3}\right\rfloor $.\footnote{We also experiment with data-dependent bandwidths, though results
are similar and therefore omitted.} For qLL-S, the tuning parameters $c$ and $\widetilde{c}$ are set
to 10, following the recommendation in \citeReferencesSupp{magnusson/mavroeidis:2014}.
 The significance level is fixed at 5\%, and the number of Monte
Carlo replications is set to 10,000 throughout the analysis.

Tables \ref{Table: Size Identification-Robust Tests T400}-\ref{Table: Size Identification-Robust Tests M1 M2 T=00003D200}
report the null rejections frequencies of the various test statistics.
In Table \ref{Table: Size Identification-Robust Tests T400} the sample
size is set to $T=400$, $\rho=0.25$ and the errors are assumed to
be i.i.d.\footnote{When the number of instruments is one, the $AR$ tests are not reported,
as they are numerically equivalent to the $LM$ tests. For results
with serially correlated errors, see the supplement. The conclusions
are similar to i.i.d. errors.} The results show that qLL-S and exp-S tests systematically yield
rejection rates below the nominal significance level, while the Split-CLR
is systematically oversized. $LM_{T}$ and $LM_{T}(\widehat{\mathbf{S}}_{T})$
show accurate null rejection frequencies for all values of $d$, whereas
$CLR_{T}$ and $CLR_{T}(\widehat{\mathbf{S}}_{T})$ tend to produce
slightly oversized rejection rates when the instrument is weak (i.e.,
$d=4$ and $d=8$). We note that $LM_{T}(\widehat{\mathbf{S}}_{T})$
and $CLR_{T}(\widehat{\mathbf{S}}_{T})$ have often more accurate
rejection rates than their full sample counterparts. These findings
are consistent across values of $\pi_{0}$. In Table \ref{Table: Size Identification-Robust Tests M1 M2 T=00003D200}
we consider a smaller sample size of $T=200$ and $\rho\in\{0.25,\,0.50,\,0.75\}$.
The qualitative patterns remain similar. As $\rho$ increases $LM_{T}(\widehat{\mathbf{S}}_{T})$
and $CLR_{T}(\widehat{\mathbf{S}}_{T})$ become slightly more oversized
than their full sample counterparts for $d=4,$ although the opposite
occurs for larger $d$. Thus, the proposed tests demonstrate good
size control, even in small samples.

In the supplement, we examine the impact of serial correlation on
the null rejection rates. Under strong serial dependence ($\rho_{e}=\rho_{u}=0.75$)
all tests exhibit rejection rates that exceed the nominal significance
level. Specifically, $LM_{T}(\widehat{\mathbf{S}}_{T})$ and $CLR_{T}(\widehat{\mathbf{S}}_{T})$
are a bit more oversized than $LM_{T}$ and $CLR_{T}$ but similar
to qLL-S. Under weak serial dependence $\rho_{e}=\rho_{u}=0.25$,
the proposed tests $LM_{T}(\widehat{\mathbf{S}}_{T})$ and $CLR_{T}(\widehat{\mathbf{S}}_{T})$
are only slightly more oversized than their full sample counterparts,
$LM_{T}$ and $CLR_{T}$. 

Finally, we turn to the comparison of size-adjusted power, as reported
in Figures \ref{Figure_1_Sect_6}-\ref{Figure_3_Sect_6}. Neither
test achieves unit power when the instrument is weak. The results
indicate that the proposed tests consistently achieve the highest
size-adjusted power across all specifications considered. The power
gains are substantial, averaging around approximately 20\textendash 30\%
and reaching 40-50\% in the most favorable cases. The latter coincide
with the specifications $\theta_{2}=-0.5dT^{-1/2}$ and $\theta_{2}=-0.5$.
Consistent with the theoretical predictions, the gains are more pronounced
for smaller values of $\pi_{0}$, provided that $\pi_{0}$ is not
too small. Overall, these finite-sample results support our theoretical
relative efficiency results. 

\begin{table}[H]
\caption{\label{Table: Size Identification-Robust Tests T400}Finite-Sample
Null Rejection Frequencies of Tests}

\smallskip{}

\begin{centering}
{\footnotesize}{\small{}%
\begin{tabular}{ccccccccc}
\hline 
{\footnotesize$T=400$} & \multicolumn{4}{c}{{\footnotesize$\gamma_{1}=\gamma_{2}=0$ and $\rho_{e}=\rho_{u}=0$}} & \multicolumn{4}{c}{{\footnotesize$\gamma_{1}=\gamma_{2}=1$ and $\rho_{e}=\rho_{u}=0$}}\tabularnewline
{\footnotesize$\rho=0.25$, $\pi_{0}=0.6$} & {\footnotesize$d=4$} & {\footnotesize$d=8$} & {\footnotesize$d=12$} & {\footnotesize$d=16$} & {\footnotesize$d=4$} & {\footnotesize$d=8$} & {\footnotesize$d=12$} & {\footnotesize$d=16$}\tabularnewline
\hline 
\hline 
{\footnotesize$LM_{T}$} & {\footnotesize 0.057} & {\footnotesize 0.057} & {\footnotesize 0.053} & {\footnotesize 0.056} & {\footnotesize 0.059} & {\footnotesize 0.060} & {\footnotesize 0.058} & {\footnotesize 0.056}\tabularnewline
{\footnotesize$CLR_{T}$} & {\footnotesize 0.078} & {\footnotesize 0.065} & {\footnotesize 0.060} & {\footnotesize 0.062} & {\footnotesize 0.083 } & {\footnotesize 0.074} & {\footnotesize 0.067} & {\footnotesize 0.065}\tabularnewline
{\footnotesize$LM_{T}(\widehat{\mathbf{S}}_{T})$} & {\footnotesize 0.061} & {\footnotesize 0.052} & {\footnotesize 0.049} & {\footnotesize 0.050} & {\footnotesize 0.065} & {\footnotesize 0.055} & {\footnotesize 0.054} & {\footnotesize 0.048}\tabularnewline
{\footnotesize$CLR_{T}(\widehat{\mathbf{S}}_{T})$} & {\footnotesize 0.072} & {\footnotesize 0.054} & {\footnotesize 0.051} & {\footnotesize 0.050} & {\footnotesize 0.079} & {\footnotesize 0.074} & {\footnotesize 0.055} & {\footnotesize 0.049}\tabularnewline
{\footnotesize$\mathrm{split-S}$} & {\footnotesize 0.039} & {\footnotesize 0.038} & {\footnotesize 0.036} & {\footnotesize 0.036} & {\footnotesize 0.042} & {\footnotesize 0.042} & {\footnotesize 0.038} & {\footnotesize 0.034}\tabularnewline
{\footnotesize$\mathrm{split-CLR}$} & {\footnotesize 0.115} & {\footnotesize 0.122} & {\footnotesize 0.117} & {\footnotesize 0.115} & {\footnotesize 0.110} & {\footnotesize 0.130} & {\footnotesize 0.122} & {\footnotesize 0.123}\tabularnewline
{\footnotesize$\mathrm{qqL-S}$} & {\footnotesize 0.027} & {\footnotesize 0.031} & {\footnotesize 0.028} & {\footnotesize 0.056} & {\footnotesize 0.031} & {\footnotesize 0.029} & {\footnotesize 0.031} & {\footnotesize 0.026}\tabularnewline
{\footnotesize$\mathrm{ave-S}$} & {\footnotesize 0.044} & {\footnotesize 0.039} & {\footnotesize 0.043} & {\footnotesize 0.038} & {\footnotesize 0.043} & {\footnotesize 0.043} & {\footnotesize 0.042} & {\footnotesize 0.038}\tabularnewline
{\footnotesize$\mathrm{exp-S}$} & {\footnotesize 0.019} & {\footnotesize 0.020} & {\footnotesize 0.017} & {\footnotesize 0.019} & {\footnotesize 0.017} & {\footnotesize 0.022} & {\footnotesize 0.021} & {\footnotesize 0.019}\tabularnewline
{\footnotesize$\rho=0.25$, $\pi_{0}=0.4$} & {\footnotesize$d=4$} & {\footnotesize$d=8$} & {\footnotesize$d=12$} & {\footnotesize$d=16$} & {\footnotesize$d=4$} & {\footnotesize$d=8$} & {\footnotesize$d=12$} & {\footnotesize$d=16$}\tabularnewline
{\footnotesize$LM_{T}$} & {\footnotesize 0.058} & {\footnotesize 0.058} & {\footnotesize 0.058} & {\footnotesize 0.058} & {\footnotesize 0.059} & {\footnotesize 0.058} & {\footnotesize 0.056} & {\footnotesize 0.060}\tabularnewline
{\footnotesize$CLR_{T}$} & {\footnotesize 0.087} & {\footnotesize 0.079} & {\footnotesize 0.073} & {\footnotesize 0.072} & {\footnotesize 0.083} & {\footnotesize 0.081} & {\footnotesize 0.072} & {\footnotesize 0.073}\tabularnewline
{\footnotesize$LM_{T}(\widehat{\mathbf{S}}_{T})$} & {\footnotesize 0.061} & {\footnotesize 0.060} & {\footnotesize 0.065} & {\footnotesize 0.064} & {\footnotesize 0.065} & {\footnotesize 0.065} & {\footnotesize 0.062} & {\footnotesize 0.066}\tabularnewline
{\footnotesize$CLR_{T}(\widehat{\mathbf{S}}_{T})$} & {\footnotesize 0.081} & {\footnotesize 0.071} & {\footnotesize 0.073} & {\footnotesize 0.071} & {\footnotesize 0.079} & {\footnotesize 0.076} & {\footnotesize 0.070} & {\footnotesize 0.072}\tabularnewline
{\footnotesize$\mathrm{split-S}$} & {\footnotesize 0.036} & {\footnotesize 0.036} & {\footnotesize 0.038} & {\footnotesize 0.037} & {\footnotesize 0.042} & {\footnotesize 0.039} & {\footnotesize 0.039} & {\footnotesize 0.038}\tabularnewline
{\footnotesize$\mathrm{split-CLR}$} & {\footnotesize 0.104} & {\footnotesize 0.119} & {\footnotesize 0.120} & {\footnotesize 0.121} & {\footnotesize 0.110} & {\footnotesize 0.119} & {\footnotesize 0.119} & {\footnotesize 0.122}\tabularnewline
{\footnotesize$\mathrm{qqL-S}$} & {\footnotesize 0.027} & {\footnotesize 0.028} & {\footnotesize 0.029} & {\footnotesize 0.031} & {\footnotesize 0.028} & {\footnotesize 0.030} & {\footnotesize 0.032} & {\footnotesize 0.032}\tabularnewline
{\footnotesize$\mathrm{ave-S}$} & {\footnotesize 0.040} & {\footnotesize 0.043} & {\footnotesize 0.047} & {\footnotesize 0.043} & {\footnotesize 0.044} & {\footnotesize 0.043} & {\footnotesize 0.037} & {\footnotesize 0.044}\tabularnewline
{\footnotesize$\mathrm{exp-S}$} & {\footnotesize 0.015} & {\footnotesize 0.017} & {\footnotesize 0.017} & {\footnotesize 0.018} & {\footnotesize 0.017} & {\footnotesize 0.020} & {\footnotesize 0.020} & {\footnotesize 0.019}\tabularnewline
{\footnotesize$\rho=0.25$, $\pi_{0}=0.8$} & {\footnotesize$d=4$} & {\footnotesize$d=8$} & {\footnotesize$d=12$} & {\footnotesize$d=16$} & {\footnotesize$d=4$} & {\footnotesize$d=8$} & {\footnotesize$d=12$} & {\footnotesize$d=16$}\tabularnewline
{\footnotesize$LM_{T}$} & {\footnotesize 0.055} & {\footnotesize 0.056} & {\footnotesize 0.057} & {\footnotesize 0.058} & {\footnotesize 0.056} & {\footnotesize 0.060} & {\footnotesize 0.058} & {\footnotesize 0.060}\tabularnewline
{\footnotesize$CLR_{T}$} & {\footnotesize 0.069} & {\footnotesize 0.061} & {\footnotesize 0.060} & {\footnotesize 0.060} & {\footnotesize 0.076} & {\footnotesize 0.067} & {\footnotesize 0.062} & {\footnotesize 0.062}\tabularnewline
{\footnotesize$LM_{T}(\widehat{\mathbf{S}}_{T})$} & {\footnotesize 0.060} & {\footnotesize 0.061} & {\footnotesize 0.053} & {\footnotesize 0.049} & {\footnotesize 0.061} & {\footnotesize 0.059} & {\footnotesize 0.056} & {\footnotesize 0.052}\tabularnewline
{\footnotesize$CLR_{T}(\widehat{\mathbf{S}}_{T})$} & {\footnotesize 0.068} & {\footnotesize 0.063} & {\footnotesize 0.054} & {\footnotesize 0.049} & {\footnotesize 0.071} & {\footnotesize 0.062} & {\footnotesize 0.058} & {\footnotesize 0.053}\tabularnewline
{\footnotesize$\mathrm{split-S}$} & {\footnotesize 0.038} & {\footnotesize 0.041} & {\footnotesize 0.047} & {\footnotesize 0.045} & {\footnotesize 0.043} & {\footnotesize 0.047} & {\footnotesize 0.044} & {\footnotesize 0.041}\tabularnewline
{\footnotesize$\mathrm{split-CLR}$} & {\footnotesize 0.117} & {\footnotesize 0.128} & {\footnotesize 0135} & {\footnotesize 0.135} & {\footnotesize 0.122} & {\footnotesize 0.133} & {\footnotesize 0.133} & {\footnotesize 0.128}\tabularnewline
{\footnotesize$\mathrm{qqL-S}$} & {\footnotesize 0.028} & {\footnotesize 0.032} & {\footnotesize 0.029} & {\footnotesize 0.030} & {\footnotesize 0.025} & {\footnotesize 0.030} & {\footnotesize 0.032} & {\footnotesize 0.031}\tabularnewline
{\footnotesize$\mathrm{ave-S}$} & {\footnotesize 0.045} & {\footnotesize 0.044} & {\footnotesize 0.047} & {\footnotesize 0.040} & {\footnotesize 0.042} & {\footnotesize 0.046} & {\footnotesize 0.044} & {\footnotesize 0.041}\tabularnewline
{\footnotesize$\mathrm{exp-S}$} & {\footnotesize 0.028} & {\footnotesize 0.019} & {\footnotesize 0.017} & {\footnotesize 0.020} & {\footnotesize 0.018} & {\footnotesize 0.022} & {\footnotesize 0.019} & {\footnotesize 0.018}\tabularnewline
\end{tabular}}{\small\par}
\par\end{centering}
\centering{}{\scriptsize{}%
\begin{minipage}[t]{0.8\columnwidth}%
{\scriptsize Model M1 and M2. The null hypothesis is $H_{0}:\,\beta=0$. }%
\end{minipage}}{\scriptsize\par}
\end{table}

\begin{table}[H]
\caption{\label{Table: Size Identification-Robust Tests M1 M2 T=00003D200}Finite-Sample
Null Rejection Frequencies of Tests}

\smallskip{}

\centering{}{\footnotesize}{\small{}%
\begin{tabular}{cccccccccc}
\hline 
 & \multicolumn{9}{c}{{\footnotesize$\gamma_{1}=\gamma_{2}=0$ and $\rho_{e}=\rho_{u}=0$}}\tabularnewline
 & \multicolumn{3}{c}{{\footnotesize$\rho=0.25$}} & \multicolumn{3}{c}{{\footnotesize$\rho=0.50$}} & \multicolumn{3}{c}{{\footnotesize$\rho=0.75$}}\tabularnewline
{\footnotesize$T=200$, $\pi_{0}=0.6$} & {\footnotesize$d=4$} & {\footnotesize$d=10$} & {\footnotesize$d=16$} & {\footnotesize$d=4$} & {\footnotesize$d=10$} & {\footnotesize$d=16$} & {\footnotesize$d=4$} & {\footnotesize$d=10$} & {\footnotesize$d=16$}\tabularnewline
\hline 
\hline 
{\footnotesize$LM_{T}$} & {\footnotesize 0.061} & {\footnotesize 0.061} & {\footnotesize 0.061} & {\footnotesize 0.062} & {\footnotesize 0.062} & {\footnotesize 0.059} & {\footnotesize 0.062} & {\footnotesize 0.062} & {\footnotesize 0.062}\tabularnewline
{\footnotesize$CLR_{T}$} & {\footnotesize 0.084} & {\footnotesize 0.073} & {\footnotesize 0.070} & {\footnotesize 0.085} & {\footnotesize 0.073} & {\footnotesize 0.063} & {\footnotesize 0.080} & {\footnotesize 0.071} & {\footnotesize 0.070}\tabularnewline
{\footnotesize$LM_{T}(\widehat{\mathbf{S}}_{T})$} & {\footnotesize 0.068} & {\footnotesize 0.054} & {\footnotesize 0.053} & {\footnotesize 0.075} & {\footnotesize 0.057} & {\footnotesize 0.059} & {\footnotesize 0.082} & {\footnotesize 0.057} & {\footnotesize 0.054}\tabularnewline
{\footnotesize$CLR_{T}(\widehat{\mathbf{S}}_{T})$} & {\footnotesize 0.081} & {\footnotesize 0.057} & {\footnotesize 0.054} & {\footnotesize 0.086} & {\footnotesize 0.059} & {\footnotesize 0.059} & {\footnotesize 0.090} & {\footnotesize 0.058} & {\footnotesize 0.054}\tabularnewline
{\footnotesize$\mathrm{split-S}$} & {\footnotesize 0.034} & {\footnotesize 0.035} & {\footnotesize 0.034} & {\footnotesize 0.035} & {\footnotesize 0.034} & {\footnotesize 0.035} & {\footnotesize 0.036} & {\footnotesize 0.035} & {\footnotesize 0.034}\tabularnewline
{\footnotesize$\mathrm{split-CLR}$} & {\footnotesize 0.105} & {\footnotesize 0.113} & {\footnotesize 0.111} & {\footnotesize 0.106} & {\footnotesize 0.115} & {\footnotesize 0.115} & {\footnotesize 0.110} & {\footnotesize 0.115} & {\footnotesize 0.115}\tabularnewline
{\footnotesize$\mathrm{qqL-S}$} & {\footnotesize 0.015} & {\footnotesize 0.017} & {\footnotesize 0.017} & {\footnotesize 0.019} & {\footnotesize 0.019} & {\footnotesize 0.014} & {\footnotesize 0.017} & {\footnotesize 0.018} & {\footnotesize 0.018}\tabularnewline
{\footnotesize$\mathrm{ave-S}$} & {\footnotesize 0.035} & {\footnotesize 0.038} & {\footnotesize 0.035} & {\footnotesize 0.038} & {\footnotesize 0.036} & {\footnotesize 0.039} & {\footnotesize 0.036} & {\footnotesize 0.039} & {\footnotesize 0.042}\tabularnewline
{\footnotesize$\mathrm{exp-S}$} & {\footnotesize 0.012} & {\footnotesize 0.012} & {\footnotesize 0.012} & {\footnotesize 0.012} & {\footnotesize 0.012} & {\footnotesize 0.014} & {\footnotesize 0.012} & {\footnotesize 0.012} & {\footnotesize 0.012}\tabularnewline
 & \multicolumn{9}{c}{{\footnotesize$\gamma_{1}=\gamma_{2}=1$ and $\rho_{e}=\rho_{u}=0$}}\tabularnewline
 & \multicolumn{3}{c}{{\footnotesize$\rho=0.25$}} & \multicolumn{3}{c}{{\footnotesize$\rho=0.50$}} & \multicolumn{3}{c}{{\footnotesize$\rho=0.75$}}\tabularnewline
{\footnotesize$T=200$, $\pi_{0}=0.6$} & {\footnotesize$d=4$} & {\footnotesize$d=10$} & {\footnotesize$d=16$} & {\footnotesize$d=4$} & {\footnotesize$d=10$} & {\footnotesize$d=16$} & {\footnotesize$d=4$} & {\footnotesize$d=10$} & {\footnotesize$d=16$}\tabularnewline
{\footnotesize$LM_{T}$} & {\footnotesize 0.066} & {\footnotesize 0.066} & {\footnotesize 0.066} & {\footnotesize 0.064} & {\footnotesize 0.064} & {\footnotesize 0.064} & {\footnotesize 0.069} & {\footnotesize 0.069} & {\footnotesize 0.069}\tabularnewline
{\footnotesize$CLR_{T}$} & {\footnotesize 0.089} & {\footnotesize 0.077} & {\footnotesize 0.075} & {\footnotesize 0.089} & {\footnotesize 0.075} & {\footnotesize 0.073} & {\footnotesize 0.089} & {\footnotesize 0.080} & {\footnotesize 0.078}\tabularnewline
{\footnotesize$LM_{T}(\widehat{\mathbf{S}}_{T})$} & {\footnotesize 0.070} & {\footnotesize 0.060} & {\footnotesize 0.056} & {\footnotesize 0.082} & {\footnotesize 0.061} & {\footnotesize 0.056} & {\footnotesize 0.095} & {\footnotesize 0.058} & {\footnotesize 0.056}\tabularnewline
{\footnotesize$CLR_{T}(\widehat{\mathbf{S}}_{T})$} & {\footnotesize 0.085} & {\footnotesize 0.063} & {\footnotesize 0.057} & {\footnotesize 0.094} & {\footnotesize 0.064} & {\footnotesize 0.056} & {\footnotesize 0.105} & {\footnotesize 0.060} & {\footnotesize 0.057}\tabularnewline
{\footnotesize$\mathrm{split-S}$} & {\footnotesize 0.040} & {\footnotesize 0.040} & {\footnotesize 0.039} & {\footnotesize 0.041} & {\footnotesize 0.040} & {\footnotesize 0.038} & {\footnotesize 0.040} & {\footnotesize 0.037} & {\footnotesize 0.038}\tabularnewline
{\footnotesize$\mathrm{split-CLR}$} & {\footnotesize 0.110} & {\footnotesize 0.123} & {\footnotesize 0.121} & {\footnotesize 0.112} & {\footnotesize 0.112} & {\footnotesize 0.119} & {\footnotesize 0.118} & {\footnotesize 0.119} & {\footnotesize 0.120}\tabularnewline
{\footnotesize$\mathrm{qqL-S}$} & {\footnotesize 0.020} & {\footnotesize 0.021} & {\footnotesize 0.021} & {\footnotesize 0.024} & {\footnotesize 0.019} & {\footnotesize 0.024} & {\footnotesize 0.025} & {\footnotesize 0.023} & {\footnotesize 0.024}\tabularnewline
{\footnotesize$\mathrm{ave-S}$} & {\footnotesize 0.041} & {\footnotesize 0.040} & {\footnotesize 0.039} & {\footnotesize 0.042} & {\footnotesize 0.036} & {\footnotesize 0.037} & {\footnotesize 0.040} & {\footnotesize 0.041} & {\footnotesize 0.038}\tabularnewline
{\footnotesize$\mathrm{exp-S}$} & {\footnotesize 0.014} & {\footnotesize 0.015} & {\footnotesize 0.015} & {\footnotesize 0.015} & {\footnotesize 0.016} & {\footnotesize 0.015} & {\footnotesize 0.014} & {\footnotesize 0.016} & {\footnotesize 0.014}\tabularnewline
\end{tabular}}{\small\par}
\end{table}

\noindent\begin{flushleft}
\begin{center}
\begin{figure}[h]
\begin{raggedright}
\includegraphics[clip,width=17cm,totalheight=9.5cm]{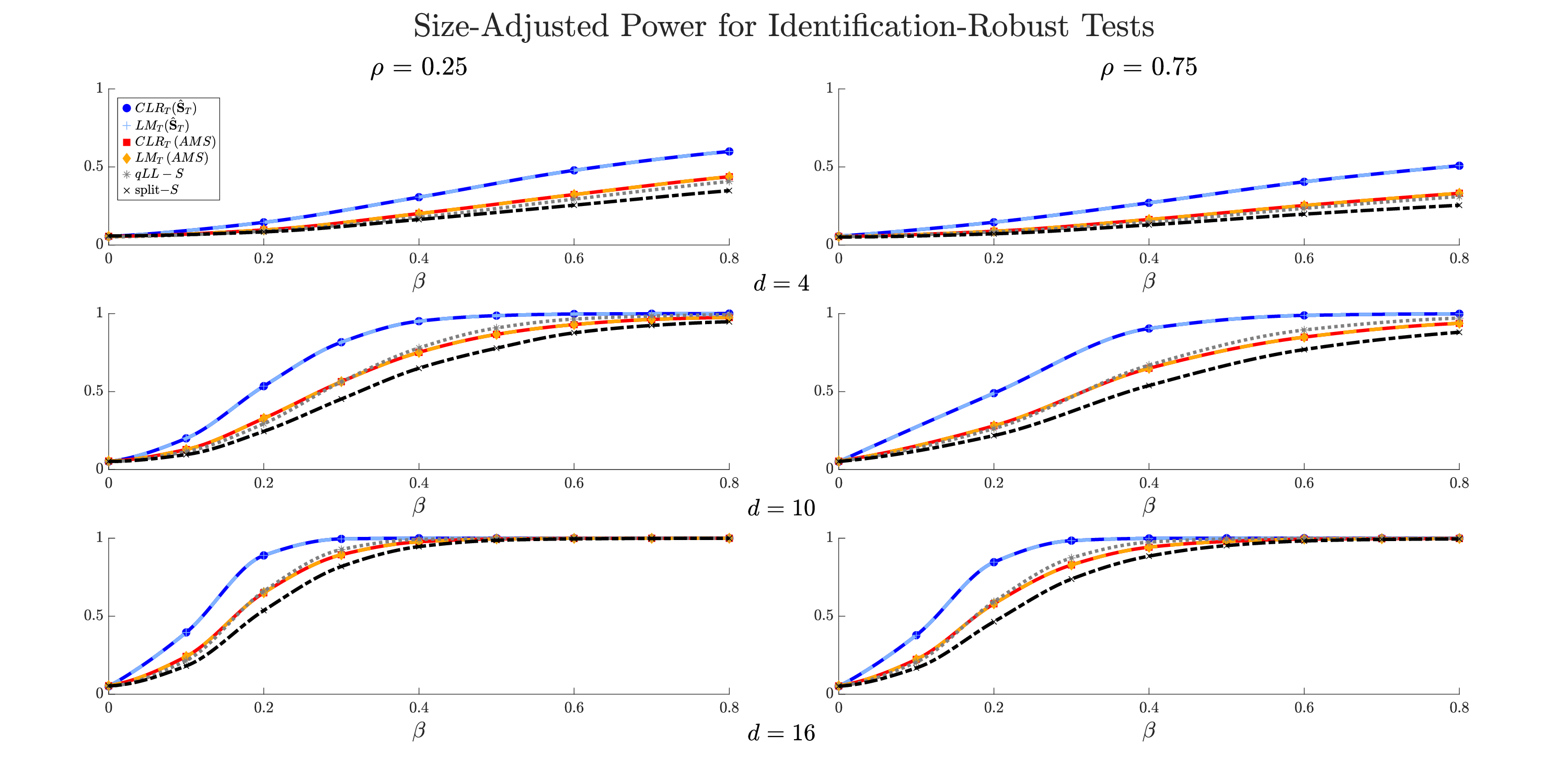}
\par\end{raggedright}
\raggedright{}\caption{\label{Figure_1_Sect_6}{\scriptsize Size-adjusted power of identification
robust tests for $\theta_{2}=0$ with $T=200$ and $\pi_{0}=0.6$. }}
\end{figure}
\end{center}
\par\end{flushleft}

\noindent\begin{flushleft}
\begin{center}
\begin{figure}[H]
\begin{raggedright}
\includegraphics[clip,width=17cm,totalheight=9.5cm]{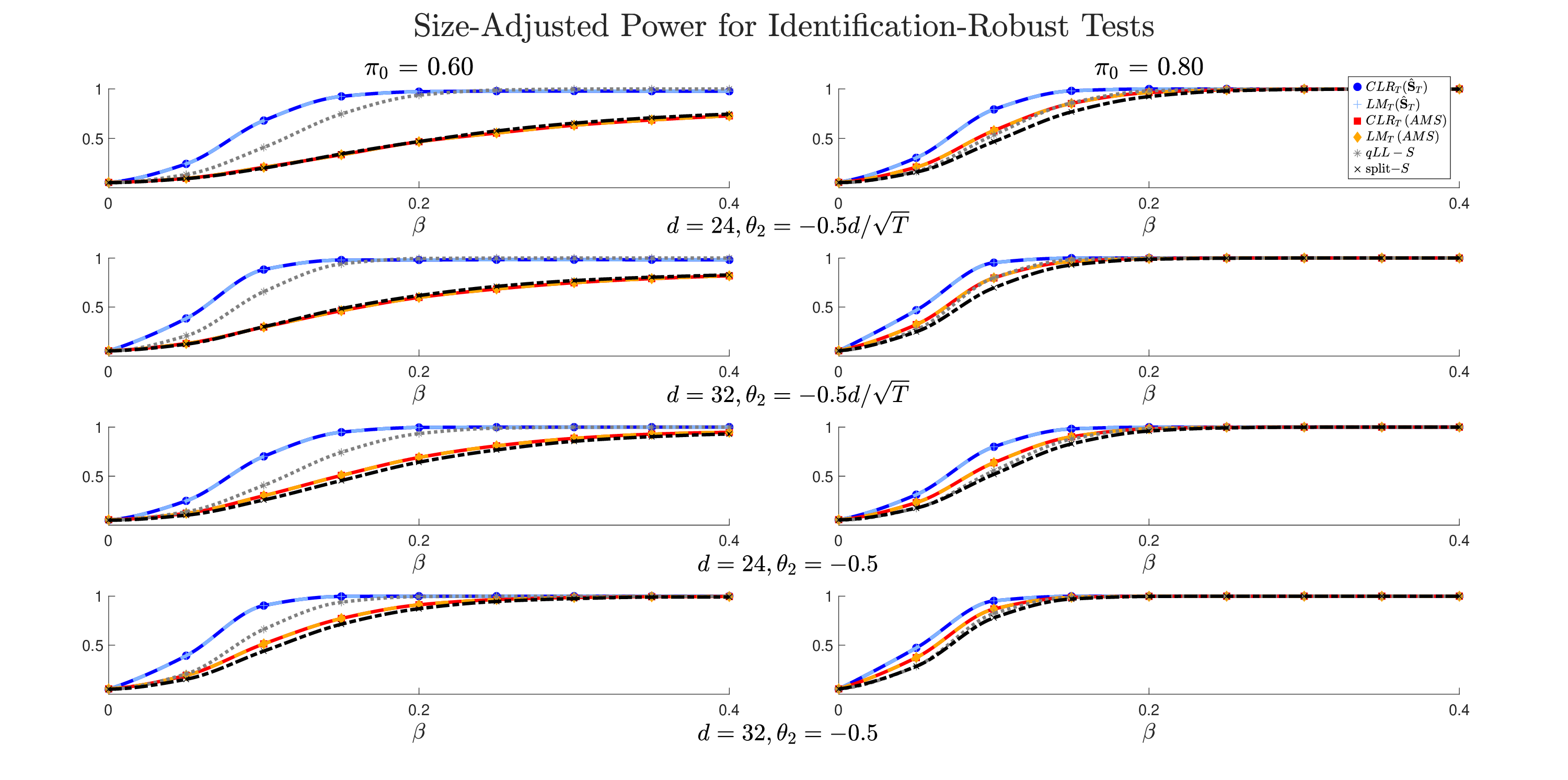}
\par\end{raggedright}
\raggedright{}\caption{\label{Figure_3_Sect_6}{\scriptsize Size-adjusted power of identification
robust tests for $\theta_{2}=-0.5dT^{-1/2}$ and $\theta_{2}=-0.5$
with $T=200$ and $\rho=0.25$.}}
\end{figure}
\end{center}
\par\end{flushleft}

\section{\label{Section Mathematical-Appendix}Mathematical Proofs}

\subsection{Proofs of the Results of Sections \ref{Section: Statistical Framework for Identification of Causal Effects}-\ref{Section: Application: Money Neutrality}}

\subsubsection{Preliminary Lemmas}
\begin{lem}
\label{Lemma: Lem 5 I=000026A}Let Assumption \ref{Assumption: Monotonicity}
hold, $g_{t}\left(\cdot\right)$ be locally absolutely continuous
on $\mathbf{D}\subseteq\mathbb{R}$ and $\mathbb{E}[\int_{\mathbf{D}}|\partial g_{t}(d)/\partial d|\mathrm{d}d|\widetilde{V}_{t}]<\infty$.
For $t\in\mathbf{S}_{0,T},$ $\widetilde{v}$ in the support of $\widetilde{V}_{t}$,
and $z,\,z'\in\mathbf{Z},$ we have
\begin{align*}
\mathbb{E} & \left(g_{t}\left(D_{t}\left(z'\right)\right)-g_{t}\left(D_{t}\left(z\right)\right)|\,\widetilde{V}_{t}=\widetilde{v}\right)\\
 & =\int_{\mathbf{D}}\mathbb{E}\left(\frac{\partial}{\partial d}g_{t}\left(d\right)|\,D_{t}\left(z\right)\leq d\leq D_{t}\left(z'\right),\,\widetilde{V}_{t}=\widetilde{v}\right)\mathbb{P}\left(D_{t}\left(z\right)\leq d\leq D_{t}\left(z'\right)|\,\widetilde{V}_{t}=\widetilde{v}\right)\mathrm{d}d.
\end{align*}
\end{lem}
\noindent\textit{Proof of Lemma }\ref{Lemma: Lem 5 I=000026A}. Suppose
that Assumption \ref{Assumption: Monotonicity} holds with $D_{t}\left(z'\right)\geq D_{t}\left(z\right)$.
We have
\begin{align*}
 & \mathbb{E}\left(g_{t}\left(D_{t}\left(z'\right)\right)-g_{t}\left(D_{t}\left(z\right)\right)|\,\widetilde{V}_{t}=\widetilde{v}\right)\\
 & =\mathbb{E}\left(\int_{D_{t}\left(z\right)}^{D_{t}\left(z'\right)}\frac{\partial g_{t}}{\partial d}\left(d\right)\mathrm{d}d|\,\widetilde{V}_{t}=\widetilde{v}\right)\\
 & =\mathbb{E}\left(\int_{\mathbf{D}}\frac{\partial g_{t}}{\partial d}\left(d\right)\mathbf{1}\left\{ D_{t}\left(z\right)\leq d\leq D_{t}\left(z'\right)\right\} \mathrm{d}d|\,\widetilde{V}_{t}=\widetilde{v}\right)\\
 & =\int_{\mathbf{D}}\mathbb{E}\left(\frac{\partial g_{t}}{\partial d}\left(d\right)\mathbf{1}\left\{ D_{t}\left(z\right)\leq d\leq D_{t}\left(z'\right)\right\} |\,\widetilde{V}_{t}=\widetilde{v}\right)\mathrm{d}d\\
 & =\int_{\mathbf{D}}\mathbb{E}\left(\frac{\partial g_{t}}{\partial d}\left(d\right)|\,D_{t}\left(z\right)\leq d\leq D_{t}\left(z'\right),\,\widetilde{V}_{t}=\widetilde{v}\right)\mathbb{P}\left(D_{t}\left(z\right)\leq d\leq D_{t}\left(z'\right)|\,\widetilde{V}_{t}=\widetilde{v}\right)\mathrm{d}d,
\end{align*}
where the first equality follows from local absolute continuity and
the fundamental theorem of calculus and the third equality follows
from Fubini's theorem and integrability. $\square$

\subsubsection{Proof of Proposition \ref{Proposition: Continuous pi-LATE}}

Consider first the denominator of $\beta_{\pi,t,h}\left(\widetilde{v}\right)$.
We appeal to Assumption \ref{Assumption: Independence} and apply
Lemma \ref{Lemma: Lem 5 I=000026A} with $g_{t}\left(d\right)=d$
to obtain for $t\in\mathbf{S}_{0,T}$, 
\begin{align*}
\mathbb{E}\left(D_{t}|\,Z_{t}=z',\,\widetilde{V}_{t}=\widetilde{v}\right)-\mathbb{E}\left(D_{t}|\,Z_{t}=z,\,\widetilde{V}_{t}=\widetilde{v}\right) & =\int_{\mathbf{D}}\mathbb{P}\left(D_{t}\left(z\right)\leq d\leq D_{t}\left(z'\right)|\,\widetilde{V}_{t}=\widetilde{v}\right)\mathrm{d}d.
\end{align*}
Noting that $Y_{t,h}$ is a function of $D_{t}=d$, we appeal to Assumptions
\ref{Assumption: Independence}-\ref{Assumption: Exclusion } and
\ref{Assumption KP 2024}(i) and apply Lemma \ref{Lemma: Lem 5 I=000026A}
with $g_{t}\left(d\right)=Y_{t,h}^{*}\left(d\right)$ to the numerator
of $\beta_{\pi,t,h}\left(\widetilde{v}\right)$ to obtain,
\begin{align*}
\mathbb{E} & \left(Y_{t+h}|\,Z_{t}=z',\,\widetilde{V}_{t}=\widetilde{v}\right)-\mathbb{E}\left(Y_{t+h}|\,Z_{t}=z,\,\widetilde{V}_{t}=\widetilde{v}\right)\\
 & =\int_{\mathbf{D}}\mathbb{E}\left(\frac{\partial Y_{t,h}^{*}}{\partial d}\left(d\right)|\,D_{t}\left(z\right)\leq d\leq D_{t}\left(z'\right),\,\widetilde{V}_{t}=\widetilde{v}\right)\times\mathbb{P}\left(D_{t}\left(z\right)\leq d\leq D_{t}\left(z'\right)|\,\widetilde{V}_{t}=\widetilde{v}\right)\mathrm{d}d.
\end{align*}
Using the derived expressions for the numerator and denominator of
$\beta_{\pi,t,h}\left(\widetilde{v}\right)$ yields the expression
given in the proposition, which is well-defined by Assumption \ref{Assumption: First-Stage}.
Note that by definition $w_{t}\left(d|\,\widetilde{v}\right)\geq0$
and $\int_{\mathbf{D}}w_{t}\left(d|\,\widetilde{v}\right)\mathrm{d}d=1.$
$\square$

\subsubsection{Proof of Theorem \ref{Theorem: Compliers}}

\begin{lem}
\label{Lemma: D1>D0 E(D1)>E(D0)}Let Assumption \ref{Assumption: Pure behavior}
hold. For each $t$, $D_{t}\left(1\right)>D_{t}\left(0\right)$ with
probability one if and only if $\mathbb{E}\left(D_{t}\left(1\right)\right)>\mathbb{E}\left(D_{t}\left(0\right)\right)$. 
\end{lem}
\noindent\textit{Proof of Lemma }\ref{Lemma: D1>D0 E(D1)>E(D0)}.
 First, note that $\mathbb{P}\left(D_{t}\left(1\right)>D_{t}\left(0\right)\right)=1$
immediately implies $\mathbb{E}\left(D_{t}\left(1\right)\right)>\mathbb{E}\left(D_{t}\left(0\right)\right)$.
To see the reverse implication, suppose $\mathbb{E}\left(D_{t}\left(1\right)\right)>\mathbb{E}\left(D_{t}\left(0\right)\right)$.
Then it must be the case that $\mathbb{P}\left(D_{t}(1)>D_{t}(0)\right)>0$.
But then Assumption \ref{Assumption: Pure behavior} immediately implies
$\mathbb{P}\left(D_{t}(0)<D_{t}(1)\right)=1$. $\square$

\medskip{}

Consider first the policy sample. Given $t\in\mathbf{P}$ we have
$Z_{t}=1$ and $D_{t}=D_{t}\left(1\right).$ By Lemma \ref{Lemma: D1>D0 E(D1)>E(D0)}
$t\in\mathbf{P}$ is a complier if and only if $\mathbb{E}\left(D_{t}\left(1\right)\right)>\mathbb{E}\left(D_{t}\left(0\right)\right)$.
By Assumption \ref{Assumption: Local LLN}(ii) we have $\mathbb{E}\left(D_{t}\left(0\right)\right)-\mathbb{E}\left(D_{t-1}\left(0\right)\right)=0$.
The latter implies $t\in\mathbf{P}$ is a complier if and only if
\begin{align*}
\mathbb{E}\left(D_{t}\left(1\right)\right)-\mathbb{E}\left(D_{t-1}\left(0\right)\right) & =\mathbb{E}\left(D_{t}\left(1\right)\right)-\mathbb{E}\left(D_{t}\left(0\right)\right)\\
 & >\mathbb{E}\left(D_{t}\left(0\right)\right)-\mathbb{E}\left(D_{t}\left(0\right)\right)=0.
\end{align*}
By Assumption \ref{Assumption: Local LLN}(i), $\overline{D}_{C,t-1,n_{0}}\overset{\mathbb{P}}{\rightarrow}\mathbb{E}\left(D_{t-1}\left(0\right)\right)$
as $n_{0}\rightarrow\infty$. By Assumption \ref{Assumption Local LLN Treatment Sample}(i),
$\overline{D}_{P,t,n_{1}}\overset{\mathbb{P}}{\rightarrow}\mathbb{E}\left(D_{t}\left(1\right)\right)$
as $n_{1}\rightarrow\infty$. Thus, $\overline{D}_{P,t,n_{1}}-\overline{D}_{C,t-1,n_{0}}\overset{\mathbb{P}}{\rightarrow}c$
as $n_{0},n_{1}\rightarrow\infty$, where $c>0$ if and only if $t\in\mathbf{P}$
is a complier.

Now consider the control sample. Since $t\in\mathbf{C}$ we have
$Z_{t}=0$ and $D_{t}=D_{t}\left(0\right).$ Using Assumption \ref{Assumption Local LLN Treatment Sample}(ii)
and Lemma \ref{Lemma: D1>D0 E(D1)>E(D0)} we have $t\in\mathbf{C}$
is a complier if and only if 
\begin{align*}
\mathbb{E}\left(D_{s^{*}\left(t\right)}\left(1\right)\right)-\mathbb{E}\left(D_{t}\left(0\right)\right) & =\mathbb{E}\left(D_{t}\left(1\right)\right)-\mathbb{E}\left(D_{t}\left(0\right)\right)\\
 & >\mathbb{E}\left(D_{t}\left(0\right)\right)-\mathbb{E}\left(D_{t}\left(0\right)\right)=0.
\end{align*}
By Assumption \ref{Assumption: Local LLN}(i), $\overline{D}_{C,t,n_{0}}\overset{\mathbb{P}}{\rightarrow}\mathbb{E}\left(D_{t}\left(0\right)\right)$
as $n_{0}\rightarrow\infty$. By Assumption \ref{Assumption Local LLN Treatment Sample}(i),
$\overline{D}_{P,s^{*}\left(t\right),n_{1}}\overset{\mathbb{P}}{\rightarrow}\mathbb{E}(D_{s^{*}\left(t\right)}\left(1\right))$
as $n_{1}\rightarrow\infty$. Thus, $\overline{D}_{P,s^{*}\left(t\right),n_{1}}-\overline{D}_{C,t,n_{0}}\overset{\mathbb{P}}{\rightarrow}\widetilde{c}$
as $n_{0},n_{1}\rightarrow\infty$, where $\widetilde{c}>0$ if and
only if $t\in\mathbf{C}$ is a complier. $\square$

\subsubsection{Proof of Proposition \ref{Proposition: Compliers FS indivdually}}

Under Assumption \ref{Assumption: Monotonicity} with $D_{t}(1)\geq D_{t}(0)$,
non-compliers are characterized by $\mathbb{P}(D_{t}(1)=D_{t}(0))=1$
using Assumption \ref{Assumption: Pure behavior} so that any non-complier
cannot belong to $\mathbf{S}_{0,T}$. On the other hand, if $t$ is
a complier, $\mathbb{E}[D_{t}(1)]\neq\mathbb{E}[D_{t}(0)]$ since
$\mathbb{P}(D_{t}(1)>D_{t}(0))=1$ by the definition of a complier
so that $t\in\mathbf{S}_{0,T}$. $\square$

\subsubsection{Proof Proposition \ref{Proposition: exclusion-restriction}}

For $t\in\mathcal{NC}_{\mathbf{C}}^{s}$, $Z_{t}=0$ so that $Y_{t}=Y_{t}^{*}(D_{t}(0),0)$
and Assumption \ref{Assumption: Non-Complier LLN}(ii) implies 
\[
|\mathcal{NC}_{\mathbf{C}}^{s}|^{-1}\sum_{t\in\mathcal{NC}_{\mathbf{C}}^{s}}Y_{t}\overset{\mathbb{P}}{\rightarrow}\mathbb{E}[Y_{t}|t\in\mathcal{NC}_{\mathbf{C}}^{s}]=\mathbb{E}[Y_{t}^{*}(D_{t}(0),0)|t\in\mathcal{NC}_{\mathbf{C}}^{s}]=\mathbb{E}[Y_{t}^{*}(D_{t},0)|t\in\mathcal{NC}_{\mathbf{C}}^{s}],
\]
as $|\mathcal{NC}_{\mathbf{C}}^{s}|\rightarrow\infty$ since $t$
is a non-complier. Similarly, Assumption \ref{Assumption: Non-Complier LLN}(ii)
implies 
\[
|\mathcal{NC}_{\mathbf{P}}^{s}|^{-1}\sum_{t\in\mathcal{NC}_{\mathbf{P}}^{s}}Y_{t}\overset{\mathbb{P}}{\rightarrow}\mathbb{E}[Y_{t}^{*}(D_{t},1)|t\in\mathcal{NC}_{\mathbf{P}}^{s}]
\]
as $|\mathcal{NC}_{\mathbf{P}}^{s}|\rightarrow\infty$. But Assumption
\ref{Assumption: Non-Complier LLN}(i) implies $\mathbb{E}[Y_{t}^{*}(D_{t},0)|t\in\mathcal{NC}_{\mathbf{C}}^{s}]=\mathbb{E}[Y_{t}^{*}(D_{t},0)|t\in\mathcal{NC}_{\mathbf{P}}^{s}]$,
which is in turn equal to $\mathbb{E}[Y_{t}^{*}(D_{t},1)|t\in\mathcal{NC}_{\mathbf{P}}^{s}]$
under Assumption \ref{Assumption: Exclusion }. $\square$

\subsection{Proofs of the Results of Sections \ref{Section: ID Failure Test},
\ref{Section: Inference} and \ref{Section: Additional Resuls Identification-Robust Inference }}

\subsubsection{Proof of Theorem \ref{Theorem: Asymptotic Distribution Sup Fstar test}}

Suppose that $\mathbf{S}_{T}\in\Xi_{\epsilon,\pi,m,T}$. Note that
$\widetilde{Z}\left(S_{T}\right)=M_{S_{T}X}S_{T}Z$ and 
\begin{align*}
\widetilde{D}\left(S_{T}\right) & =M_{S_{T}X}S_{T}D=M_{S_{T}X}S_{T}\left(X\gamma_{2}+e\right)=M_{S_{T}X}S_{T}e
\end{align*}
under $H_{\theta,0}$. Thus, 
\begin{align*}
\widetilde{D}\left(S_{T}\right)'\widetilde{Z}\left(S_{T}\right) & =(S_{T}e)'M_{S_{T}X}M_{S_{T}X}S_{T}Z=(S_{T}e)'M_{S_{T}X}S_{T}Z=(S_{T}e)'\widetilde{Z}(S_{T}).
\end{align*}
By Assumptions \ref{Assumption: w ULLN}-\ref{Assumption: we FCLT}
we have 
\begin{align*}
T^{-1/2}\widetilde{Z}\left(S_{T}\right)'\widetilde{D}\left(S_{T}\right) & =T^{-1/2}\sum_{t\in\mathbf{S}_{T}}\widetilde{Z}_{t}e_{t}\Rightarrow B\left(\lambda_{L,1},\,\lambda_{R,m}\right),
\end{align*}
where $B\left(\lambda_{L,1},\,\lambda_{R,m}\right)$ is $q$-dimensional
Brownian motion with covariance $J(S)=\underset{T\rightarrow\infty}{\lim}\left(\pi T\right)^{-1}$
$\mathrm{Var}(\left(S_{T}Z\right)'M_{S_{T}X}S_{T}e)$ and $S=\underset{T\rightarrow\infty}{\lim}S_{T}$.
Since $\widehat{J}(S_{T})\overset{\mathbb{P}}{\rightarrow}J\left(S\right)$
uniformly by Assumption \ref{Assumption: hat-J uniform consistency},
we have 
\begin{align*}
F_{T}\left(\mathbf{S}_{T}\right) & \Rightarrow\frac{B\left(\lambda_{L,1},\,\lambda_{R,m}\right)'J(S)^{-1}B\left(\lambda_{L,1},\,\lambda_{R,m}\right)}{q\pi}=\frac{1}{q\pi}\left\Vert \sum_{i=1}^{m}\left(W_{q}\left(\lambda_{R,i}\right)-W_{q}\left(\lambda_{L,i}\right)\right)\right\Vert ^{2}.
\end{align*}
The result then follows from the continuous mapping theorem. $\square$

\subsubsection{Proof of Theorem \ref{Theorem 7 AMS HAR}}
\begin{lem}
\label{Lemma 5 AMS Weak IV}Let Assumptions \ref{Assumption 1 AMS}-\ref{Assumption 3 AMS}
hold and suppose $\theta=c/T^{1/2}$ for some nonstochastic $c\in\mathbb{R}^{q}$.
We have $\widehat{\Sigma}_{v}\left(\mathbf{S}_{T}\right)\overset{\mathbb{P}}{\rightarrow}\Sigma_{v}$
uniformly in $\mathbf{S}_{T}\in\mathcal{S}$. 
\end{lem}
\noindent\textit{Proof of Lemma \ref{Lemma 5 AMS Weak IV}. }Recall
that $C_{T}$ is the selection matrix that corresponds to $\mathbf{S}_{T}.$
Using $y=\overline{Z}\left(C_{0,T}\right)\theta a'_{\beta}+X\eta+v$,
$P_{X}\overline{Z}\left(C_{T}\right)=0$ and $P_{\overline{Z}\left(C_{T}\right)}X=0$,
we have 
\begin{align}
\widehat{v}\left(\mathbf{S}_{T}\right) & =y-P_{\overline{Z}\left(C_{T}\right)}y-P_{X}y=y-P_{\overline{Z}\left(C_{T}\right)}y-X\eta-P_{X}v\label{Eq. (1) Proof Lemma 5 AMS Weak IV}\\
 & =\overline{Z}\left(C_{0,T}\right)\theta a'_{\beta}+v-P_{\overline{Z}\left(C_{T}\right)}y-P_{X}v\nonumber \\
 & =\overline{Z}\left(C_{0,T}\right)\theta a'_{\beta}+v-P_{\overline{Z}\left(C_{T}\right)}\overline{Z}\left(C_{0,T}\right)\theta a'_{\beta}-P_{\overline{Z}\left(C_{T}\right)}v-P_{X}v\nonumber \\
 & =M_{\overline{Z}\left(C_{T}\right)}\overline{Z}\left(C_{0,T}\right)\theta a'_{\beta}+v-P_{\overline{Z}\left(C_{T}\right)}v-P_{X}v.\nonumber 
\end{align}
Then, using $\overline{Z}\left(C_{T}\right)'X=0$, we have 
\begin{align}
\widehat{v}\left(\mathbf{S}_{T}\right)' & \widehat{v}\left(\mathbf{S}_{T}\right)\label{Eq. (14.29-1) AMS}\\
 & =\left(M_{\overline{Z}\left(C_{T}\right)}\overline{Z}\left(C_{0,T}\right)\theta a'_{\beta}+v-P_{\overline{Z}\left(C_{T}\right)}v-P_{X}v\right)'\nonumber \\
 & \quad\times\left(M_{\overline{Z}\left(C_{T}\right)}\overline{Z}\left(C_{0,T}\right)\theta a'_{\beta}+v-P_{\overline{Z}\left(C_{T}\right)}v-P_{X}v\right)\nonumber \\
 & =a_{\beta}\theta'\overline{Z}\left(C_{0,T}\right)'M_{\overline{Z}\left(C_{T}\right)}\overline{Z}\left(C_{0,T}\right)\theta a'_{\beta}+a_{\beta}\theta'\overline{Z}\left(C_{0,T}\right)'M_{\overline{Z}\left(C_{T}\right)}v\nonumber \\
 & \quad+v'M_{\overline{Z}\left(C_{T}\right)}\overline{Z}\left(C_{0,T}\right)\theta a'_{\beta}+v'v-v'P_{\overline{Z}\left(C_{T}\right)}v-v'P_{X}v.\nonumber 
\end{align}
Using Assumptions \ref{Assumption 1 AMS}, \ref{Assumption 3 AMS}
and $\theta=cT^{-1/2}$ we have 
\begin{align}
T^{-1} & v'M_{\overline{Z}\left(C_{T}\right)}\overline{Z}\left(C_{0,T}\right)\theta a'_{\beta}\label{Eq. (14.29-2) AMS}\\
 & =T^{-1}v'\overline{Z}\left(C_{0,T}\right)\frac{c}{\sqrt{T}}a'_{\beta}-T^{-1}v'\overline{Z}\left(C_{T}\right)\left(\overline{Z}\left(C_{T}\right)'\overline{Z}\left(C_{T}\right)\right)^{-1}\overline{Z}\left(C_{T}\right)'\overline{Z}\left(C_{0,T}\right)\frac{c}{\sqrt{T}}a'_{\beta}\nonumber \\
 & =T^{-1/2}O_{\mathbb{P}}\left(1\right)\frac{c}{\sqrt{T}}a'_{\beta}-T^{-1/2}\left(T^{-1/2}v'\overline{Z}\left(C_{T}\right)\right)\left(T^{-1}\overline{Z}\left(C_{T}\right)'\overline{Z}\left(C_{T}\right)\right)^{-1}T^{-1}\overline{Z}\left(C_{T}\right)'\overline{Z}\left(C_{0,T}\right)\frac{c}{\sqrt{T}}a'_{\beta}\nonumber \\
 & =O_{\mathbb{P}}\left(T^{-1}\right),\qquad\qquad\qquad\qquad\mathrm{and}\nonumber 
\end{align}
\begin{align}
 & T^{-1}a{}_{\beta}\theta'\overline{Z}\left(C_{0,T}\right)'M_{\overline{Z}\left(C_{T}\right)}\overline{Z}\left(C_{0,T}\right)\theta a'_{\beta}\label{Eq. (14.29-3) AMS}\\
 & =T^{-1}a_{\beta}\theta'\overline{Z}\left(C_{0,T}\right)'\overline{Z}\left(C_{0,T}\right)\theta a'_{\beta}-T^{-1}a_{\beta}\theta'\overline{Z}\left(C_{0,T}\right)'\overline{Z}\left(C_{T}\right)\left(\overline{Z}\left(C_{T}\right)'\overline{Z}\left(C_{T}\right)\right)^{-1}\overline{Z}\left(C_{T}\right)'\overline{Z}\left(C_{0,T}\right)\theta a'_{\beta}\nonumber \\
 & =a_{\beta}\frac{c'}{\sqrt{T}}T^{-1}\overline{Z}\left(C_{0,T}\right)'\overline{Z}\left(C_{0,T}\right)\frac{c}{\sqrt{T}}a'_{\beta}\nonumber \\
 & \quad-a_{\beta}\frac{c'}{\sqrt{T}}T^{-1}\overline{Z}\left(C_{0,T}\right)'\overline{Z}\left(C_{T}\right)\left(T^{-1}\overline{Z}\left(C_{T}\right)'\overline{Z}\left(C_{T}\right)\right)^{-1}T^{-1}\overline{Z}\left(C_{T}\right)'\overline{Z}\left(C_{0,T}\right)\frac{c}{\sqrt{T}}a'_{\beta}=O_{\mathbb{P}}\left(T^{-1}\right)\nonumber 
\end{align}
so that 
\begin{align}
T^{-1}\widehat{v}'\left(\mathbf{S}_{T}\right)\widehat{v}\left(\mathbf{S}_{T}\right) & -\Sigma_{v}=\left(T^{-1}v'v-\Sigma_{v}\right)-T^{-1}v'P_{\overline{Z}\left(C_{T}\right)}v-T^{-1}v'P_{X}v+O_{\mathbb{P}}\left(T^{-1}\right).\label{Eq. (14.29) AMS}
\end{align}
By Assumption \ref{Assumption 2 AMS}, the first term on the right-hand
side of \eqref{Eq. (14.29) AMS} converges in probability to zero.
The second term satisfies 
\begin{align*}
T^{-1}v'P_{\overline{Z}\left(C_{T}\right)}v & =T^{-1}\left\Vert P_{\overline{Z}\left(C_{T}\right)}v\right\Vert ^{2}=T^{-1}O_{\mathbb{P}}\left(1\right)=o_{\mathbb{P}}\left(1\right),
\end{align*}
under bounded second moments, fixed $q$, and weak dependence conditions.
All convergence results above hold uniformly in $\mathbf{S}_{T}$.
The argument for the third term of \eqref{Eq. (14.29) AMS} is analogous.
$\square$

\medskip{}

Lemma \ref{Lemma 5 AMS Weak IV} shows that $\widehat{\Sigma}_{v}\left(\mathbf{S}_{T}\right)$
is consistent for $\Sigma_{v}$ for all $\mathbf{S}_{T}\in\mathcal{S}$.
The convergence in the lemma occurs uniformly over all true parameters
$\beta$, $c$, $\gamma$, $\phi$ and over $\mathbf{S}_{T}\in\mathcal{S}$.
Thus, estimation based on any partition $\mathbf{S}_{T}\in\mathcal{S}$
leads to residuals $\left\{ \widehat{v}\left(\mathbf{S}_{T}\right)\right\} $
that can be used to construct a consistent estimate $\widehat{\Sigma}_{v}\left(\mathbf{S}_{T}\right)$
for $\Sigma_{v}$ under weak instruments.

For some partitions $\mathbf{S},\mathbf{S'}\subseteq(0,1]$, partition
$Q\left(\mathbf{S},\mathbf{S}'\right)$ conformably with 
\[
w(\mathbf{S}_{T})'w(\mathbf{S}_{T}')=\begin{bmatrix}Z(C_{T})'Z(C_{T}') & Z(C_{T})'X\\
X'Z(C_{T}') & X'X
\end{bmatrix},
\]
where $Z(C_{T})=C_{T}Z$, so that 
\begin{align*}
Q\left(\mathbf{S},\mathbf{S}'\right) & =\begin{bmatrix}Q_{11}\left(\mathbf{S},\mathbf{S}'\right) & Q_{12}\left(\mathbf{S}\right)\\
Q_{21}\left(\mathbf{S}'\right) & Q_{22}
\end{bmatrix},
\end{align*}
and let $Q(\mathbf{S})=Q\left(\mathbf{S},\mathbf{S}\right)$. In addition,
note that 
\begin{align}
\Sigma_{N_{1}}\left(\mathbf{S},\mathbf{S}'\right) & =J\left(\mathbf{S}\right)B_{0}\Psi\left(\mathbf{S},\mathbf{S}'\right)B'_{0}J\left(\mathbf{S}'\right)',\quad\Sigma_{N_{1}N_{2}}\left(\mathbf{S},\mathbf{S}'\right)=J\left(\mathbf{S}\right)A_{0}\Psi\left(\mathbf{S},\mathbf{S}'\right)B'_{0}J\left(\mathbf{S}'\right)',\nonumber \\
\Sigma_{N_{2}}^{*}\left(\mathbf{S},\mathbf{S}'\right) & =J\left(\mathbf{S}\right)A_{0}\Psi\left(\mathbf{S},\mathbf{S}'\right)A'_{0}J\left(\mathbf{S}'\right)'\label{Variance equivalence}
\end{align}
for $J\left(\mathbf{S}\right)=\left[I_{q}:-Q_{12}\left(\mathbf{S}\right)Q_{22}^{-1}\right],$
$B_{0}=\left(b'_{0}\otimes I_{q+p}\right)$ and $A_{0}=\left(\Sigma_{v}^{-1}a_{0,\beta}\right)'\otimes I_{q+p}$.\footnote{See \eqref{Eq. (14.32) AMS} and \eqref{Eq. (14.35) AMS} for details.} 

Finally, let $N_{1,\infty}\left(\cdot\right)$ and $N_{2,\infty}\left(\cdot\right)$
be independent $q$-dimensional Gaussian processes indexed by $\mathbf{S}\subseteq(0,1]$
with 
\begin{align}
N_{1,\infty}\left(\mathbf{S}\right) & \sim\mathscr{N}\left(\Sigma_{N_{1}}^{-1/2}\left(\mathbf{S}\right)\Sigma_{\overline{Z}}(\mathbf{S},\mathbf{S}_{0})ca'_{\beta}b_{0},\,I_{q}\right),\label{Eq. (9.10) AMS}\\
N_{2,\infty}\left(\mathbf{S}\right) & \sim\mathscr{N}\left(\Sigma_{N_{2}}^{-1/2}\left(\mathbf{S}\right)\left(\Sigma_{\overline{Z}}(\mathbf{S},\mathbf{S}_{0})ca'_{\beta}\Sigma_{v}^{-1}a_{0,\beta}-\Sigma_{N_{1}N_{2}}\left(\mathbf{S}\right)\Sigma_{N_{1}}^{-1}\left(\mathbf{S}\right)\Sigma_{\overline{Z}}(\mathbf{S},\mathbf{S}_{0})ca'_{\beta}b_{0}\right),\,I_{q}\right),\nonumber 
\end{align}
where $\Sigma_{\overline{Z}}(\mathbf{S},\mathbf{S}')=Q_{11}\left(\mathbf{S},\mathbf{S}'\right)-Q_{12}\left(\mathbf{S}\right)Q_{22}^{-1}Q_{21}\left(\mathbf{S}'\right)$.

\begin{lem}
\label{Lemma: Theorem 7 AMS}Let Assumptions \ref{Assumption 1 AMS}-\ref{Assumption 2nd Moment Stability}
hold and suppose $\theta=c/T^{1/2}$ for some nonstochastic $c\in\mathbb{R}^{q}$.
Then, for $\mathbf{S}_{T}\in\mathcal{S}$ and $\mathbf{S}=\lim_{T\rightarrow\infty}T^{-1}\mathbf{S}_{T}$,
we have $(N_{1,T}\left(\mathbf{S}_{T}\right),\,N_{2,T}\left(\mathbf{S}_{T}\right))\Rightarrow(N_{1,\infty}\left(\mathbf{S}\right),\,N_{2,\infty}\left(\mathbf{S}\right))$.
\end{lem}
\noindent\textit{Proof of Lemma }\ref{Lemma: Theorem 7 AMS}. By
Assumption \ref{Assumption 1 AMS}, 
\begin{align}
T^{-1}\overline{Z}\left(C_{T}\right)'\overline{Z}\left(C_{T}'\right) & =T^{-1}Z(C_{T})'Z(C_{T}')-T^{-1}Z(C_{T})'P_{X}Z(C_{T}')\overset{\mathbb{P}}{\rightarrow}\Sigma_{\overline{Z}}(\mathbf{S},\mathbf{S}'),\label{Eq. (14.31) AMS}
\end{align}
uniformly in $\mathbf{S}_{T},\mathbf{S}_{T}'\in\mathcal{S}$. By Assumptions
\ref{Assumption 1 AMS} and \ref{Assumption 3 AMS}, we have uniformly
in $\mathbf{S}_{T}\in\mathcal{S}$,
\begin{align}
T^{-1/2}\overline{Z}\left(C_{T}\right)'vb_{0} & =T^{-1/2}\left(Z\left(C_{T}\right)-P_{X}Z\left(C_{T}\right)\right)'vb_{0}\label{Eq. (14.32) AMS}\\
 & =T^{-1/2}\left(Z\left(C_{T}\right)-XQ_{22}^{-1}Q_{21}\left(\mathbf{S}_{T}\right)\right)'vb_{0}+o_{\mathbb{P}}\left(1\right)\nonumber \\
 & =\left[I_{q}:-Q_{12}\left(\mathbf{S}_{T}\right)Q_{22}^{-1}\right]T^{-1/2}w\left(\mathbf{S}_{T}\right)'vb_{0}+o_{\mathbb{P}}\left(1\right)\nonumber \\
 & =\left[I_{q}:-Q_{12}\left(\mathbf{S}_{T}\right)Q_{22}^{-1}\right]\left(b'_{0}\otimes I_{q+p}\right)T^{-1/2}\mathrm{vec}\left(w(\mathbf{S}_{T})'v\right),\nonumber \\
 & \Rightarrow J(\mathbf{S})B_{0}\mathscr{G}(\mathbf{S}).\nonumber 
\end{align}
Using \eqref{Variance equivalence}, \eqref{Eq. (14.31) AMS}, \eqref{Eq. (14.32) AMS}
and Assumption \ref{Assumption Uniform Consistent Covariance Matrix},
\begin{align}
N_{1,T} & \left(\mathbf{S}_{T}\right)=\widehat{\Sigma}_{N_{1}}^{-1/2}\left(\mathbf{S}_{T}\right)T^{-1/2}\overline{Z}\left(C_{T}\right)'\left(T^{-1/2}\overline{Z}\left(C_{0,T}\right)ca'_{\beta}+v\right)b_{0}\label{Eq. (14.44)}\\
 & \Rightarrow\Sigma_{N_{1}}^{-1/2}\left(\mathbf{S}\right)\Sigma_{\overline{Z}}(\mathbf{S},\mathbf{S}_{0})ca_{\beta}'b_{0}+\Sigma_{N_{1}}^{-1/2}\left(\mathbf{S}\right)J(\mathbf{S})B_{0}\mathscr{G}(\mathbf{S})\sim N_{1,\infty}\left(\mathbf{S}\right)\nonumber 
\end{align}
since $J(\mathbf{S})B_{0}\mathscr{G}(\mathbf{S})\sim\mathscr{N}(0,\Sigma_{N_{1}}(\mathbf{S}))$.
Similarly, using Lemma \ref{Lemma 5 AMS Weak IV}, Assumptions \ref{Assumption 1 AMS},
\ref{Assumption 3 AMS} and \ref{Assumption Uniform Consistent Covariance Matrix},
\eqref{Variance equivalence} and \eqref{Eq. (14.44)}, we have 
\begin{align}
N_{2,T} & \left(\mathbf{S}_{T}\right)\label{Eq. (N2_hat)}\\
 & =\widehat{\Sigma}_{N_{2}}^{-1/2}\left(\mathbf{S}_{T}\right)\nonumber \\
 & \quad\times\left(T^{-1/2}\overline{Z}\left(C_{T}\right)'\left(T^{-1/2}\overline{Z}\left(C_{0,T}\right)ca'_{\beta}+v\right)\widehat{\Sigma}_{v}^{-1}\left(\mathbf{S}_{T}\right)a_{0,\beta}-\widehat{\Sigma}_{N_{1}N_{2}}\left(\mathbf{S}_{T}\right)\widehat{\Sigma}_{N_{1}}^{-1/2}\left(\mathbf{S}_{T}\right)N_{1,T}\left(\mathbf{S}_{T}\right)\right)\nonumber \\
 & =\widehat{\Sigma}_{N_{2}}^{-1/2}\left(\mathbf{S}_{T}\right)\nonumber \\
 & \quad\times\left(\left(T^{-1}\overline{Z}\left(C_{T}\right)'\overline{Z}\left(C_{0,T}\right)ca'_{\beta}+T^{-1/2}\overline{Z}\left(C_{T}\right)'v\right)\Sigma_{v}^{-1}a_{0,\beta}-\widehat{\Sigma}_{N_{1}N_{2}}\left(\mathbf{S}_{T}\right)\widehat{\Sigma}_{N_{1}}^{-1/2}\left(\mathbf{S}_{T}\right)N_{1,T}\left(\mathbf{S}_{T}\right)\right)+o_{\mathbb{P}}\left(1\right)\nonumber \\
 & \Rightarrow\Sigma_{N_{2}}^{-1/2}\left(\mathbf{S}\right)\left(\Sigma_{\overline{Z}}(\mathbf{S},\mathbf{S}_{0})ca'_{\beta}\Sigma_{v}^{-1}a_{0,\beta}\right)+\Sigma_{N_{2}}^{-1/2}\left(\mathbf{S}\right)J\left(\mathbf{S}\right)A_{0}\mathscr{G}(\mathbf{S})\nonumber \\
 & \quad-\Sigma_{N_{2}}^{-1/2}\left(\mathbf{S}\right)\left(\Sigma_{N_{1}N_{2}}\left(\mathbf{S}\right)\Sigma_{N_{1}}^{-1}\left(\mathbf{S}\right)\Sigma_{\overline{Z}}(\mathbf{S},\mathbf{S}_{0})ca'_{\beta}b_{0}\right)\nonumber \\
 & \quad-\Sigma_{N_{2}}^{-1/2}\left(\mathbf{S}\right)\left(\Sigma_{N_{1}N_{2}}\left(\mathbf{S}\right)\Sigma_{N_{1}}^{-1}\left(\mathbf{S}\right)J\left(\mathbf{S}\right)B_{0}\mathscr{G}(\mathbf{S})\right)\sim N_{2,\infty}\left(\mathbf{S}\right),\nonumber 
\end{align}
where $T^{-1}\overline{Z}\left(C_{T}\right)'\overline{Z}\left(C_{0,T}\right)\overset{\mathbb{P}}{\rightarrow}\Sigma_{\overline{Z}}(\mathbf{S},\mathbf{S}_{0})$
uniformly over $\mathbf{S}_{T}\in\mathcal{S}$ by \eqref{Eq. (14.31) AMS}
and 
\begin{align}
T^{-1/2}\overline{Z}\left(C_{T}\right)'v\Sigma_{v}^{-1}a_{0,\beta} & \Rightarrow\left[I_{q}:-Q_{12}\left(\mathbf{S}\right)Q_{22}^{-1}\right]\left(a'_{0,\beta}\Sigma_{v}^{-1}\otimes I_{q+p}\right)\mathscr{G}(\mathbf{S})\label{Eq. (14.35) AMS}
\end{align}
in analogy with the arguments that show \eqref{Eq. (14.32) AMS}.
The distributional equivalence of the limit in \eqref{Eq. (N2_hat)}
can be seen after noting 
\begin{align}
\mathrm{Var} & \left(J\left(\mathbf{S}\right)A_{0}\mathscr{G}(\mathbf{S})-\Sigma_{N_{1}N_{2}}\left(\mathbf{S}\right)\Sigma_{N_{1}}^{-1}\left(\mathbf{S}\right)J\left(\mathbf{S}\right)B_{0}\mathscr{G}(\mathbf{S})\right)\label{Eq. (14.46) AMS}\\
 & =J\left(\mathbf{S}\right)A_{0}\Psi\left(\mathbf{S},\mathbf{S})\right)A'_{0}J\left(\mathbf{S}\right)'-J\left(\mathbf{S}\right)A_{0}\Psi\left(\mathbf{S},\mathbf{S})\right)B'_{0}J\left(\mathbf{S}\right)'\Sigma_{N_{1}}^{-1}\left(\mathbf{S}\right)\Sigma_{N_{1}N_{2}}\left(\mathbf{S}\right)'\nonumber \\
 & \quad-\Sigma_{N_{1}N_{2}}\left(\mathbf{S}\right)\Sigma_{N_{1}}^{-1}\left(\mathbf{S}\right)J\left(\mathbf{S}\right)B{}_{0}\Psi\left(\mathbf{S},\mathbf{S})\right)A'_{0}J\left(\mathbf{S}\right)'\nonumber \\
 & \quad+\Sigma_{N_{1}N_{2}}\left(\mathbf{S}\right)\Sigma_{N_{1}}^{-1}\left(\mathbf{S}\right)J\left(\mathbf{S}\right)B{}_{0}\Psi\left(\mathbf{S},\mathbf{S})\right)B'_{0}J\left(\mathbf{S}\right)'\Sigma_{N_{1}}^{-1}\left(\mathbf{S}\right)\Sigma_{N_{1}N_{2}}\left(\mathbf{S}\right)'\\
 & =\Sigma_{N_{2}}^{*}\left(\mathbf{S}\right)-\Sigma_{N_{1}N_{2}}\left(\mathbf{S}\right)\Sigma_{N_{1}}^{-1}\left(\mathbf{S}\right)\Sigma_{N_{1}N_{2}}\left(\mathbf{S}\right)'-\Sigma_{N_{1}N_{2}}\left(\mathbf{S}\right)\Sigma_{N_{1}}^{-1}\left(\mathbf{S}\right)\Sigma_{N_{1}N_{2}}\left(\mathbf{S}\right)'\nonumber \\
 & \quad+\Sigma_{N_{1}N_{2}}\left(\mathbf{S}\right)\Sigma_{N_{1}}^{-1}\left(\mathbf{S}\right)\Sigma_{N_{1}}\left(\mathbf{S}\right)\Sigma_{N_{1}}^{-1}\left(\mathbf{S}\right)\Sigma_{N_{1}N_{2}}\left(\mathbf{S}\right)'\nonumber \\
 & =\Sigma_{N_{2}}^{*}\left(\mathbf{S}\right)-\Sigma_{N_{1}N_{2}}\left(\mathbf{S}\right)\Sigma_{N_{1}}^{-1}\left(\mathbf{S}\right)\Sigma_{N_{1}N_{2}}\left(\mathbf{S}\right)'=\Sigma_{N_{2}}\left(\mathbf{S}\right).\nonumber 
\end{align}
The weak convergence in \eqref{Eq. (14.32) AMS} occurs jointly with
that in \eqref{Eq. (N2_hat)} since $N_{1,T}(\cdot)$ and $N_{2,T}(\cdot)$
are functions of the same data. And finally, they are asymptotically
independent since they are asymptotically Gaussian and 
\begin{align*}
\mathrm{Cov} & \left(\Sigma_{N_{1}}^{-1/2}\left(\mathbf{S}\right)J(\mathbf{S})B_{0}\mathscr{G}(\mathbf{S}),\Sigma_{N_{2}}^{-1/2}\left(\mathbf{S}'\right)\left(J\left(\mathbf{S}'\right)A_{0}-\Sigma_{N_{1}N_{2}}\left(\mathbf{S}'\right)\Sigma_{N_{1}}^{-1}\left(\mathbf{S}'\right)J\left(\mathbf{S}'\right)B_{0}\right)\mathscr{G}(\mathbf{S}')\right)\\
 & =\Sigma_{N_{1}}^{-1/2}\left(\mathbf{S}\right)J(\mathbf{S})B_{0}\Psi(\mathbf{S},\mathbf{S}')\left(A_{0}'J\left(\mathbf{S}'\right)'-B_{0}'J\left(\mathbf{S}'\right)'\Sigma_{N_{1}}^{-1}\left(\mathbf{S}'\right)\Sigma_{N_{1}N_{2}}\left(\mathbf{S}'\right)'\right)\Sigma_{N_{2}}^{-1/2}\left(\mathbf{S}'\right)\\
 & =\Sigma_{N_{1}}^{-1/2}\left(\mathbf{S}\right)\left(\Sigma_{N_{1}N_{2}}\left(\mathbf{S}',\mathbf{S}\right)'-\Sigma_{N_{1}}\left(\mathbf{S},\mathbf{S}'\right)\Sigma_{N_{1}}^{-1}\left(\mathbf{S}'\right)\Sigma_{N_{1}N_{2}}\left(\mathbf{S}'\right)'\right)\Sigma_{N_{2}}^{-1/2}\left(\mathbf{S}'\right)\\
 & =\pi(\mathbf{S}\cap\mathbf{S}')\Sigma_{N_{1}}^{-1/2}\left(\mathbf{S}\right)\left(\Sigma_{N_{1}N_{2}}'-\Sigma_{N_{1}}\Sigma_{N_{1}}^{-1}\Sigma_{N_{1}N_{2}}'\right)\Sigma_{N_{2}}^{-1/2}\left(\mathbf{S}'\right)=0
\end{align*}
for any $\mathbf{S},\,\mathbf{S}'\subseteq(0,1]$ by Assumption \ref{Assumption 2nd Moment Stability}.
$\square$

\medskip{}

Inspection of the  proof shows that the results hold uniformly over
compact sets of true values of $\beta$ and $c$ (including the zero
vector) and over arbitrary sets of true $\gamma$ and $\phi$ values.

\medskip{}

\noindent\textit{Proof of Theorem }\ref{Theorem 7 AMS HAR}. Let
\begin{align}
M_{\infty}\left(\mathbf{S}\right) & =\left[N_{1,\infty}\left(\mathbf{S}\right):N_{2,\infty}\left(\mathbf{S}\right)\right]'\left[N_{1,\infty}\left(\mathbf{S}\right)\colon N_{2,\infty}\left(\mathbf{S}\right)\right],\label{Eq. (9.5) AMS}\\
\overline{M}_{1,\infty}\left(\mathbf{S}\right) & =\left(N_{1,\infty}\left(\mathbf{S}\right)'N_{1,\infty}\left(\mathbf{S}\right),\,N{}_{1,\infty}\left(\mathbf{S}\right)'N_{2,\infty}\left(\mathbf{S}\right)\right)',\nonumber \\
M_{2,\infty}\left(\mathbf{S}\right) & =N_{2,\infty}\left(\mathbf{S}\right)'N_{2,\infty}\left(\mathbf{S}\right),\quad M_{1,2,\infty}\left(\mathbf{S}\right)=N_{1,\infty}\left(\mathbf{S}\right)'N_{2,\infty}\left(\mathbf{S}\right),\quad M_{1,\infty}\left(\mathbf{S}\right)=N_{1,\infty}\left(\mathbf{S}\right)'N_{1,\infty}\left(\mathbf{S}\right).\nonumber 
\end{align}
By Lemma \ref{Lemma: Theorem 7 AMS} $N_{1,T}\left(\cdot\right)$
and $N_{2,T}\left(\cdot\right)$ are asymptotically independent. $\widehat{\mathbf{S}}_{T}$
depends on $N_{2,T}\left(\cdot\right)$ only. Thus, $N_{1,\infty}(\cdot)$
and $\widehat{\mathbf{S}}_{T}$ are asymptotically independent. Using
Lemma \ref{Lemma: Theorem 7 AMS} we yield that under $H_{0}$ $N_{1,\infty}(\widehat{\mathbf{S}}_{T})$
is Gaussian with zero mean and variance $I_{q}$. Thus, $M_{1,\infty}\left(\mathbf{S}_{T}\right)\sim M_{1,\infty}$
for all $\mathbf{S}_{T}\in\mathcal{S}$. Part (i) follows by using
the continuous mapping theorem.

For part (ii), note that conditional on $N_{2,\infty}(\cdot)$ Lemma
\ref{Lemma: Theorem 7 AMS} implies that $M_{1,2,\infty}(\widehat{\mathbf{S}}_{T})$
is Gaussian with zero mean and variance $N_{2,\infty}(\widehat{\mathbf{S}}_{T})'N_{2,\infty}(\widehat{\mathbf{S}}_{T})$.
The result then follows from the continuous mapping theorem.

We now move to part (iii). By Lemma \ref{Lemma: Theorem 7 AMS} $N_{1,T}(\cdot)$
and $N_{2,T}(\cdot)$ are asymptotically independent and $\widehat{\mathbf{S}}_{T}$
depends on $N_{2,T}(\cdot)$ only. $\widehat{\mathbf{S}}_{T}$ is
inconsistent and converges in distribution to a random variable $\mathbf{S}_{\infty}.$
Thus, $LR_{T}(\widehat{\mathbf{S}}_{T})$ has asymptotically the same
distribution as 
\begin{align*}
CLR_{\infty} & \left(M_{1,\infty}(\mathbf{S}_{\infty}),\,M_{2,\infty}(\mathbf{S}_{\infty}),\,\Sigma_{v},\,\beta_{0}\right)\\
 & =\frac{1}{2}\left(M_{1,\infty}(\mathbf{S}_{\infty})-M_{2,\infty}(\mathbf{S}_{\infty})+\sqrt{\left(M_{1,\infty}(\mathbf{S}_{\infty})-M_{2,\infty}(\mathbf{S}_{\infty})\right)^{2}+4M_{1,2,\infty}^{2}(\mathbf{S}_{\infty})}\right).
\end{align*}
Conditional on $N_{2,T}(\cdot)$ $\widehat{\mathbf{S}}_{T}$ is fixed.
Define $\kappa_{CLR,\alpha}(M_{1,\infty}(\mathbf{S}_{\infty}),\,m_{2},\,\Sigma_{v},\,\beta_{0})$
to be the $1-\alpha$ quantile of the null distribution of $CLR_{\infty}(M_{1,\infty}(\mathbf{S}_{\infty}),\,m_{2},$
$\Sigma_{v},\,\beta_{0})$. Since under $H_{0}$ $\kappa_{\mathrm{\mathit{CLR}},\alpha}(\widehat{\mathbf{S}}_{T})$
has asymptotically the same distribution as $\kappa_{\mathrm{CLR},\alpha}(M_{1,\infty}(\mathbf{S}_{\infty}),\,M_{2,\infty}(\mathbf{S}_{\infty}),\,\Sigma_{v},\,\beta_{0})$,
we have that the distribution of $LR_{T}(\widehat{\mathbf{S}}_{T})-\kappa_{\mathrm{\mathit{CLR}},\alpha}(\widehat{\mathbf{S}}_{T})$
is asymptotically the same distribution as 
\begin{align*}
CLR_{\infty} & \left(M_{1,\infty}(\mathbf{S}_{\infty}),\,M_{2,\infty}(\mathbf{S}_{\infty}),\,\Sigma_{v},\,\beta_{0}\right)-\kappa_{CLR,\alpha}(M_{1,\infty}(\mathbf{S}_{\infty}),\,M_{2,\infty}(\mathbf{S}_{\infty}),\,\Sigma_{v},\,\beta_{0}).
\end{align*}
Conditional on $N_{2,\infty}(\cdot)$, $N_{1,\infty}(\widehat{\mathbf{S}}_{T})$$\sim\mathscr{N}\left(0,\,I_{q}\right)$
under $H_{0}$. This implies that the conditional null distribution
of $CLR_{\infty}$ given $N_{2,\infty}(\cdot)$ does not depend on
$\theta$ or $c$. Thus, the test that rejects $H_{0}$ when $LR_{T}(\widehat{\mathbf{S}}_{T})-\kappa_{\mathrm{\mathit{CLR}},\alpha}(\widehat{\mathbf{S}}_{T})>0$
is similar at significance level $\alpha.$ $\square$

\subsubsection{Proof of Proposition \ref{Lemma 1 AMS}}

The distribution of $y$ is multivariate normal with 
\begin{align}
\mathbb{E}\left(y\right) & =\overline{Z}(C_{0,T})\theta a'_{\beta}+X\eta,\label{Eq. (S.62) AMS}
\end{align}
independence across rows, and covariance matrix $\Sigma_{v}$ for
each row. Thus, the density of $y$ is 
\begin{align}
\left(2\pi\right)^{-T/2}\left|\Sigma_{v}\right|^{-T/2} & \exp\left(-\frac{1}{2}\sum_{t=1}^{T}\left(y_{t}-a_{\beta}\theta'\overline{Z}_{t}(C_{0,T})-\eta'X_{t}\right)'\Sigma_{v}^{-1}\left(y_{t}-a_{\beta}\theta'\overline{Z}_{t}(C_{0,T})-\eta'X_{t}\right)\right)\label{Eq. (S6.3) AMS}\\
 & =\left(2\pi\right)^{-T/2}\left|\Sigma_{v}\right|^{-T/2}\exp\Biggl(-\frac{1}{2}\Biggl[\sum_{t=1}^{T}y'_{t}\Sigma_{v}^{-1}y_{t}-2\theta'\left(\sum_{t=1}^{T}\overline{Z}_{t}(C_{0,T})y'_{t}\right)\Sigma_{v}^{-1}a_{\beta}\nonumber \\
 & \quad-2\mathrm{Tr}\left(\left(\sum_{t=1}^{T}X{}_{t}y'_{t}\right)\Sigma_{v}^{-1}\eta'\right)+\sum_{t=1}^{T}\left(a_{\beta}\theta'\overline{Z}_{t}(C_{0,T})-\eta'X_{t}\right)'\Sigma_{v}^{-1}\left(a_{\beta}\theta'\overline{Z}_{t}(C_{0,T})-\eta'X_{t}\right)\Biggr]\Biggr).\nonumber 
\end{align}
By the Fisher\textendash Neyman factorization theorem $N\left(y\right)$
is a sufficient statistic for $\psi=\left(\beta,\,\theta',\,\gamma',\,\phi'\right)'$
if and only if the density $\mathscr{L}\left(y;\beta,\,\theta,\,\gamma,\,\phi,\,\mathbf{S}_{0,T}\right)$
can be factorized as $\mathscr{L}\left(y;\beta,\,\theta,\,\gamma,\,\phi,\,\mathbf{S}_{0,T}\right)=f_{\psi}\left(N\left(y\right)\right)h\left(y\right)$
for nonnegative functions $f_{\psi}\left(\cdot\right)$ and $h\left(\cdot\right)$.
Note that $\overline{Z}(C_{0,T})=M_{X}C_{0,T}Z\neq M_{X}Z$ if $\pi_{0}<1$
and so $\mathscr{L}\left(y;\beta,\,\theta,\,\gamma,\,\phi,\,\mathbf{S}_{0,T}\right)$
cannot be factorized as above for $N(y)=[y'M_{X}Z:y'X]$. Thus, $Z'M_{X}y$
and $X'y$ are not sufficient statistics for $\psi$ if $\pi_{0}<1$.
Therefore when $\pi_{0}<1$, $Z'M_{X}y$ cannot be sufficient for
$\left(\beta,\,\theta'\right)'$ if (i) $Z'M_{X}y$ and $X'y$ are
independent, (ii) the distribution of $X'y$ does not depend on $\left(\beta,\,\theta'\right)'$
and (iii) the distribution of $Z'M_{X}y$ does not depend on $\left(\gamma',\,\phi'\right)'$.

To complete the proof, we verify that (i)\textendash (iii) above hold.
For (i), note that $Z'M_{X}y$ and $X'y$ are (jointly) multivariate
normal random matrices and for any $b_{1},\,b_{2}\in\mathbb{R}^{2}$,
we have  
\begin{gather*}
\mathrm{Cov}\left({Z}'M_{X}yb_{1},\,X'yb_{2}\right)=Z'M_{X}\mathrm{Cov}(yb_{1},yb_{2})X=Z'M_{X}b_{1}'\Sigma_{v}b_{2}I_{T}X=b_{1}'\Sigma_{v}b_{2}Z'M_{X}X=0,
\end{gather*}
where the second equality uses the independence of the rows of $y$.
Lemma 1(c) of \citeReferencesSupp{andrews/moreira/stock:2006} implies
(ii) and for (iii), note that the normality of $Z'M_{X}y$ has 
\begin{align*}
\mathbb{E}\left(Z'M_{X}y\right) & =Z'M_{X}\mathbb{E}\left(y\right)=Z'M_{X}\left(\overline{Z}(C_{0,T})\theta a'_{\beta}+X\eta\right)=Z'\overline{Z}(C_{0,T})\theta a'_{\beta},\\
\mathrm{Var}\left(Z'M_{X}yb\right) & =Z'M_{X}\mathrm{Var}\left(yb\right)M_{X}Z=Z'M_{X}b'\Sigma_{v}bI_{T}M_{X}Z,
\end{align*}
for any $b\in\mathbb{R}^{2}$, which do not depend on $\left(\gamma',\,\phi'\right)'$.
$\square$

\subsubsection{Proof of Proposition \ref{Lemma 1 AMS unknown subpopulation}}

The density of $y$ is given by $\mathscr{L}\left(y;\beta,\,\theta,\,\gamma,\,\phi,\,\mathbf{S}_{T}\right)$,
where $\mathscr{L}(\cdot)$ is defined in \eqref{Eq. (S6.3) AMS}
and $\mathbf{S}_{T}$ is an unknown parameter. Using the same logic
as in the proof of Proposition \ref{Lemma 1 AMS}, by the Fisher\textendash Neyman
factorization theorem $\{\overline{Z}\left(C_{T}\right)'y\}_{\mathbf{S}_{T}\in\mathcal{S}}$
is a sufficient statistic for $\left(\beta,\,\theta'\right)'$ if
(i) $\{\overline{Z}\left(C_{T}\right)'y\}_{\mathbf{S}_{T}\in\mathcal{S}}$
and $X'y$ are independent, (ii) the distribution of $X'y$ does not
depend on $\left(\beta,\,\theta'\right)'$ and (iii) the distribution
of $\{\overline{Z}\left(C_{T}\right)'y\}_{\mathbf{S}_{T}\in\mathcal{S}}$
does not depend on $\left(\gamma',\,\phi'\right)'$. For (i), note
that $\overline{Z}\left(C_{T}\right)'y$ and $X'y$ are (jointly)
multivariate normal random matrices and for any $b_{1},\,b_{2}\in\mathbb{R}^{2}$,
we have 
\begin{gather*}
\mathrm{Cov}\left(\overline{Z}\left(C_{T}\right)'yb_{1},\,X'yb_{2}\right)=Z'C'_{T}M_{X}\mathrm{Cov}(yb_{1},yb_{2})X=Z'C'_{T}M_{X}b_{1}'\Sigma_{v}b_{2}I_{T}X=b_{1}'\Sigma_{v}b_{2}Z'M_{X}X=0,
\end{gather*}
where the second equality uses the independence of the rows of $y$.
Lemma 1(c) of \citeReferencesSupp{andrews/moreira/stock:2006} implies
(ii) and for (iii), note that 
\begin{align*}
\mathbb{E}\left(Z'C'_{T}M_{X}y\right) & =Z'C'_{T}M_{X}\mathbb{E}\left(y\right)=Z'C'_{T}M_{X}\left(\overline{Z}(C_{0,T})\theta a'+X\eta\right)=Z'C'_{T}\overline{Z}(C_{0,T})\theta a'_{\beta},\\
\mathrm{Var}\left(Z'C'_{T}M_{X}yb\right) & =Z'C'_{T}M_{X}\mathrm{Var}\left(yb\right)M_{X}C_{T}Z=Z'C'_{T}M_{X}b'\Sigma_{v}bI_{T}M_{X}C{}_{T}Z,
\end{align*}
for any $b\in\mathbb{R}^{2}$, which does not depend on $\left(\gamma',\,\phi'\right)'$.
$\square$

\subsubsection{Proof of Proposition \ref{MLE for subpopulation}}

Under the conditions of the proposition, the log-likelihood of $y$
is given (up to a constant) by 
\begin{align}
\mathscr{\ell} & \left(\beta,\,\theta,\,\gamma,\,\phi\,,\mathbf{S}_{T}\right)\label{Eq. (30) M=00003D000026M}\\
 & =-\frac{1}{2}\mathrm{Tr}\left(\Sigma_{v}^{-1}\left(\left(y-\overline{Z}\left(C_{T}\right)\theta a'_{\beta}-X\eta\right)'\left(y-\overline{Z}\left(C_{T}\right)\theta a'_{\beta}-X\eta\right)\right)\right)\nonumber \\
 & =\mathrm{Tr}\left(\Sigma_{v}^{-1}a_{\beta}\theta'\overline{Z}\left(C_{T}\right)'y\right)-\frac{1}{2}\mathrm{Tr}\left(\Sigma_{v}^{-1}a_{\beta}\theta'\overline{Z}\left(C_{T}\right)'\overline{Z}\left(C_{T}\right)\theta a'_{\beta}\right)-\frac{1}{2}\mathrm{Tr}\left(\Sigma_{v}^{-1}(y-X\eta)'(y-X\eta)\right).\nonumber 
\end{align}
Maximizing this log-likelihood with respect to $\theta$ under $H_{0}$
yields 
\[
\widetilde{\theta}(C_{T})=\left(\overline{Z}\left(C_{T}\right)'\overline{Z}\left(C_{T}\right)\right)^{-1}\overline{Z}\left(C_{T}\right)'y\Sigma_{v}^{-1}a_{0,\beta}(a_{0,\beta}'\Sigma_{v}^{-1}a_{0,\beta})^{-1},
\]
so that the concentrated likelihood function under $H_{0}$ is 
\begin{align}
\ell & \left(y;\beta_{0},\,\widetilde{\theta}(C_{T}),\,\gamma,\,\phi\,,\mathbf{S}_{T}\right)\label{Eq concentrated lhd}\\
 & =(a_{0,\beta}'\Sigma_{v}^{-1}a_{0,\beta})^{-1}\mathrm{Tr}\left(\Sigma_{v}^{-1}a_{0,\beta}a_{0,\beta}'\Sigma_{v}^{-1}y'\overline{Z}\left(C_{T}\right)\left(\overline{Z}\left(C_{T}\right)'\overline{Z}\left(C_{T}\right)\right)^{-1}\overline{Z}\left(C_{T}\right)'y\right)\nonumber \\
 & \quad-\frac{1}{2}(a_{0,\beta}'\Sigma_{v}^{-1}a_{0,\beta})^{-2}\mathrm{Tr}\left(\Sigma_{v}^{-1}a_{0,\beta}a_{0,\beta}'\Sigma_{v}^{-1}y'\overline{Z}\left(C_{T}\right)\left(\overline{Z}\left(C_{T}\right)'\overline{Z}\left(C_{T}\right)\right)^{-1}\overline{Z}\left(C_{T}\right)'y\Sigma_{v}^{-1}a_{0,\beta}a_{0,\beta}'\right)\nonumber \\
 & \quad-\frac{1}{2}\mathrm{Tr}\left(\Sigma_{v}^{-1}(y-X\eta)'(y-X\eta)\right)\nonumber \\
 & =\frac{1}{2}(a_{0,\beta}'\Sigma_{v}^{-1}a_{0,\beta})^{-1}a_{0,\beta}'\Sigma_{v}^{-1}y'\overline{Z}\left(C_{T}\right)\left(\overline{Z}\left(C_{T}\right)'\overline{Z}\left(C_{T}\right)\right)^{-1}\overline{Z}\left(C_{T}\right)'y\Sigma_{v}^{-1}a_{0,\beta}\nonumber \\
 & \quad-\frac{1}{2}\mathrm{Tr}\left(\Sigma_{v}^{-1}(y-X\eta)'(y-X\eta)\right).\nonumber 
\end{align}
Maximizing \eqref{Eq concentrated lhd} with respect to $\mathbf{S}_{T}$
is equivalent to maximizing 
\begin{equation}
(a_{0,\beta}'\Sigma_{v}^{-1}a_{0,\beta})^{-1}a_{0,\beta}'\Sigma_{v}^{-1}y'\overline{Z}\left(C_{T}\right)\left(\overline{Z}\left(C_{T}\right)'\overline{Z}\left(C_{T}\right)\right)^{-1}\overline{Z}\left(C_{T}\right)'y\Sigma_{v}^{-1}a_{0,\beta},\label{Eq Subpopulation MLE}
\end{equation}
making the MLE of $\mathbf{S}_{T}$ equal to the maximizer of \eqref{Eq Subpopulation MLE}
over $\mathbf{S}_{T}\in\mathcal{S}$.

To complete the proof, we show that maximizing $M_{2,T}(\mathbf{S}_{T})={N}_{2,T}\left(\mathbf{S}_{T}\right)^{\prime}{N}_{2,T}\left(\mathbf{S}_{T}\right)$
over $\mathbf{S}_{T}\in\mathcal{S}$ is asymptotically equivalent
to maximizing \eqref{Eq Subpopulation MLE} over $\mathbf{S}_{T}\in\mathcal{S}$
under the conditions of the proposition. To see this, first note that
\begin{align}
\Sigma_{N_{1}N_{2}}(\mathbf{S}) & =\lim_{T\rightarrow\infty}T^{-1}\sum_{t=1}^{T}\mathbb{E}\left[v_{t}b_{0}\overline{Z}_{t}\left(C_{T}\right)'\overline{Z}_{t}\left(C_{T}\right)a_{0,\beta}'\Sigma_{v}^{-1}v_{t}'\right]\label{Asy Indep Cov}\\
 & =\lim_{T\rightarrow\infty}T^{-1}\sum_{t=1}^{T}\overline{Z}_{t}\left(C_{T}\right)'\overline{Z}_{t}\left(C_{T}\right)a_{0,\beta}'\Sigma_{v}^{-1}\mathbb{E}\left[v_{t}'v_{t}\right]b_{0}\nonumber \\
 & =\lim_{T\rightarrow\infty}T^{-1}\sum_{t=1}^{T}\overline{Z}_{t}\left(C_{T}\right)'\overline{Z}_{t}\left(C_{T}\right)a_{0,\beta}'b_{0}=0\nonumber 
\end{align}
for $\mathbf{S}=\lim_{T\rightarrow\infty}T^{-1}\mathbf{S}_{T}$, where
the first equality follows from the i.i.d. assumption, the second
from the assumption of fixed regressors, the third from $\mathbb{E}\left[v_{t}'v_{t}\right]=\Sigma_{v}$
and the fourth from $a_{0,\beta}'b_{0}=0$. Thus, applying Lemma \ref{Lemma 5 AMS Weak IV},
${N}_{2,T}\left(\mathbf{S}_{T}\right)$ is asymptotically equivalent
to ${\Sigma}_{N_{2}}^{-1/2}\left(\mathbf{S}_{T}\right)T^{-1/2}\overline{Z}\left(C_{T}\right)'y{\Sigma}_{v}^{-1}a_{0,\beta}$
under Assumption \ref{Assumption Uniform Consistent Covariance Matrix},
implying that $M_{2,T}(\mathbf{S}_{T})$ is asymptotically equivalent
to $T^{-1}a_{0,\beta}'\Sigma_{v}^{-1}y'\overline{Z}\left(C_{T}\right){\Sigma}_{N_{2}}^{-1}\left(\mathbf{S}_{T}\right)\overline{Z}\left(C_{T}\right)'y\Sigma_{v}^{-1}a_{0,\beta}.$
The result then follows from
\begin{align*}
\Sigma_{N_{2}}(\mathbf{S}) & =\lim_{T\rightarrow\infty}T^{-1}\sum_{t=1}^{T}\mathbb{E}\left[v_{t}\Sigma_{v}^{-1}a_{0,\beta}\overline{Z}_{t}\left(C_{T}\right)'\overline{Z}_{t}\left(C_{T}\right)a_{0,\beta}'\Sigma_{v}^{-1}v_{t}'\right]\\
 & =\lim_{T\rightarrow\infty}T^{-1}\sum_{t=1}^{T}\overline{Z}_{t}\left(C_{T}\right)'\overline{Z}_{t}\left(C_{T}\right)a_{0,\beta}'\Sigma_{v}^{-1}\mathbb{E}\left[v_{t}'v_{t}\right]\Sigma_{v}^{-1}a_{0,\beta}=\lim_{T\rightarrow\infty}T^{-1}\overline{Z}\left(C_{T}\right)'\overline{Z}\left(C_{T}\right)a_{0,\beta}'\Sigma_{v}^{-1}a_{0,\beta}
\end{align*}
for $\mathbf{S}=\lim_{T\rightarrow\infty}T^{-1}\mathbf{S}_{T}$, in
analogy with \eqref{Asy Indep Cov}. $\square$

\begin{singlespace}
{\small\bibliographystyleReferencesSupp{econometrica}  
\bibliographyReferencesSupp{References_Supp}}{\small\par}
\end{singlespace}


\end{document}